\documentclass[final]{agujournal2018}
\usepackage{apacite}
\usepackage{url} 
\usepackage{lineno}
\usepackage{amsmath}
\usepackage{amssymb}
\usepackage{lscape}
\usepackage{pdflscape}

\journalname{Earth's Future}

\begin{document}

%
%


\title{A flood damage allowance framework for coastal protection with deep uncertainty in sea-level rise}

%
%




\authors{D.J. Rasmussen\affil{1}, Maya K. Buchanan\affil{2}, Robert E. Kopp\affil{3}, Michael Oppenheimer\affil{4}}

\affiliation{1}{Woodrow Wilson School of Public and International Affairs, Princeton University, Princeton, NJ, USA}
\affiliation{2}{Climate Central, Princeton, NJ, USA}
\affiliation{3}{Department of Earth \& Planetary Sciences, Rutgers Energy Institute and Institute of Earth, Ocean, and Atmospheric Sciences, Rutgers University, New Brunswick, NJ, USA}
\affiliation{4}{Department of Geosciences and Woodrow Wilson School of Public and International Affairs, Princeton University, Princeton, NJ, USA}

\correspondingauthor{D.J. Rasmussen}{dj.rasmussen@princeton.edu}

\begin{keypoints}
\item Sea level allowances may underestimate the vertical adjustment needed to maintain the average flood damage under uncertain sea-level rise 
\item Damage allowances are the heights of flood protection strategies to maintain the average flood damage in a given year 
\item Damage allowances can be integrated into popular financial decision-making tools, like benefit-cost analysis
\item Under high greenhouse gas emission scenarios, damage allowances beyond 2050 are strongly dependent on subjective beliefs of Antarctic ice sheet behavior
\end{keypoints}

%
%

\begin{abstract}

Future projections of Antarctic ice sheet (AIS) mass loss remain characterized by deep uncertainty (i.e., behavior is not well understood or widely agreed upon by experts). This complicates decisions on long-lived projects involving the height of coastal flood protection strategies that seek to reduce damages from rising sea levels. If a prescribed margin of safety does not properly account for sea-level rise and its uncertainties, the effectiveness of flood protection will decrease over time, potentially putting lives and property at greater risk. We develop a flood damage allowance framework for calculating the height of a flood protection strategy needed to ensure that a given level of financial risk is maintained (i.e., the average flood damage in a given year). The damage allowance framework considers decision-maker preferences such as planning horizons, preferred protection strategies (storm surge barrier, levee, elevation, and coastal retreat), and subjective views of AIS stability. We use Manhattan (New York City)\textemdash with the distribution of buildings, populations, and infrastructure fixed in time\textemdash as an example to show how our framework could be used to calculate a range of damage allowances based on multiple plausible AIS outcomes. Assumptions regarding future AIS stability more strongly influence damage allowances under high greenhouse gas emissions (Representative Concentration Pathway [RCP] 8.5) compared to those that assume strong emissions reductions (RCP2.6). Design tools that specify financial risk targets, such as the average flood damage in a given year, allow for the calculation of avoided flood damages (i.e., benefits) that can be combined with estimates of construction cost and then integrated into existing financial decision-making tools, like benefit-cost or cost-effectiveness analyses. 

\end{abstract}

\noindent \textbf{Plain Language Summary:} The current level of knowledge of how Antarctica will respond to a warming planet remains limited. As such, future projections of sea-level rise are strongly dependent upon personal beliefs of how much and how quickly Antarctica will melt. This complicates the decision of how high to build a levee in order to reduce coastal flood damages from rising sea levels (or any other flood protection method, such as raising a building on stilts). Generally, the higher the levee (or other flood protection method), the greater the margin of safety afforded. If the prescribed margin of safety does not properly account for sea-level rise and its uncertainties, the effectiveness of the flood protection will decrease over time. This could potentially waste money and put lives and property at greater risk. We create a tool for flood protection designers that allows them to calculate the height of various flood protection strategies based on their beliefs of future Antarctic melt. Our tool is open-source and freely available, and it is compatible with existing decision making protocols such as those used by the U.S. Army Corps of Engineers.

\section{Introduction}


Rising sea levels increase the probability of a given extreme sea level (ESL) event and the average number of a given ESL event experienced in a given year \citep{Buchanan2017a,Vitousek2017a,Wahl2017a,Vousdoukas2018a}. For example, a 0.5 m increase in mean sea level at the Battery in lower Manhattan (New York City, USA) is anticipated to increase the frequency of the local 100-year ESL event from once every 100 years, on average, to about once every 20 years, on average \citep{Rasmussen2018a}. This poses a challenge to designers of coastal flood protection strategies that seek to maintain a given margin of safety over time, such as ESLs that have a 1\% chance of being equaled or exceeded in a given year. Without accounting for sea-level rise (SLR) in the design of coastal flood protection, the originally selected margin of safety will decrease, potentially leading to increased flood damages and greater numbers of people at risk.

A hazard or SLR `allowance' is the vertical distance by which something needs to be adjusted in order to ensure that the average number of ESL events is kept constant under uncertain sea level change \citep{Hunter2012a,Slangen2017a,Buchanan2016a}. Hazard allowances only consider the heights of physical water levels and not their damages. If risk is characterized by the probability of a hazard and its consequence \citep{Kaplan1981a}, then the assessment of the benefits of coastal risk reduction measures requires consideration of both the probability of an ESL event and its associated damage. An allowance that maintains financial risk over time (e.g., the annual average loss [AAL] due to flooding) rather than a physical hazard (e.g., ESLs) directly quantifies reductions in damage which could inform financial decision-making (e.g., benefit-cost analysis). Additionally, future projections of GMSL are characterized by `deep uncertainty' \citep[i.e., mechanisms and associated parameters are not known or are disagreed on by experts;][]{Lempert2002a}, in part due to incomplete understanding of the physical processes that govern the behavior (or response of) the Antarctic ice sheet (AIS) under climate forcing. These deep uncertainties inhibit the characterization of a single, unambiguous probability distribution of future GMSL \citep{Bakker2017a,Kopp2017a,LeCozannet2017a}. Considerably different probabilistic projections can result from differing physical modeling approaches. Therefore, flood allowance estimation should accommodate multiple plausible distributions of AIS melt \citep[e.g.,][]{Slangen2017a}. In this study, we address these two limitations of traditional flood allowances by linking ESLs to financial loss using a simple, time-invariant damage function for Manhattan that is modified based on the flood protection strategy used (Sec. \ref{sec:stabiliz}) and also employ future probabilistic projections of local SLR that accommodate multiple subjective beliefs regarding future AIS behavior (Sec. \ref{sec:ais}).

Flood mitigation strategies can produce significant benefits to society by reducing damage to buildings and infrastructure from ESLs and can potentially save lives \citep{Lincke2018a,Scussolini2017a,Aerts2014a}. Without additional investments in adaptation measures, by the end of the century direct damages from coastal floods on the global scale could reach into the trillions of U.S. dollars per year \citep{Hinkel2014a,Diaz2016a,Jevrejeva2018a}, with most of the losses occurring in highly exposed coastal cities \citep{Hanson2011a,Hallegatte2013a}. In response, many of these urban areas are exploring or implementing strategies that aim to offset current and future losses from floods \citep[e.g.,][]{Pirazzoli2006a,Merrell2011a,EA2012a,SIRR2013a}. These strategies can generally be characterized as accommodation, defense, advance, and retreat \citep{CZMS1990a,Klein2001a}. Accommodation aims to reduce damage when inundation occurs (e.g., elevating structures or other flood-proofing measures), while defense seeks to prevent inundation using structural measures such as levees and storm surge barriers. Advance creates new lands by building out into the sea and retreat permanently moves assets and populations away from the coastline. The height of a protection strategy (i.e., the design height) is a key variable as it is generally consistent with the reduction in the AAL due to flooding; the greater the height, the larger the reduction (all else being equal). A poorly selected design height could lead to greater residual risks in terms of public safety and damage.

Formal decision tools can aid in the appraisal of flood protection strategies, including calculating the design height of flood protection. For example, expected benefit-cost analysis (BCA) is commonly used by government agencies to economically optimize design heights by balancing incremental reductions in risk with incremental investments in greater levels of safety \citep[e.g.,][]{vanDantzig1956a,Fankhauser1995a,Ramm2017a,Kanyama2019a}. Different decision tools have different variables that can be prescribed by users (e.g., cost of construction, discount rates). For BCA, the margin of safety is not a parameter that can be specified (e.g., protection against the 100-year ESL event), but rather a variable to be solved for (e.g., by maximizing net present value). This approach may be insufficient if a specific margin of safety is desired. On the other hand, flood allowance frameworks specify a margin of safety \citep{Buchanan2016a}, but design heights are calculated without regard to cost. Between these two approaches lies a risk-constrained, cost-effectiveness position that prescribes a margin of safety, but also allows for the appraisal of various strategies based on their relative costs and benefits. Coastal localities require a tailor-made response to meet their flood protection needs. Flood protection strategies are all associated with unique financial and political resource demands \citep{Hinkel2018a}, residual risks due to potential failure (e.g., levee overtopping and breaching), and are vulnerable to less-than-perfect levels of compliance (e.g., the fraction of structures that actually elevate or retreat relative to the prescription). Appraisals of flood protection strategies to meet a prescribed margin of safety should consider these factors in addition to their relative costs and benefits.



In this study, we develop a flood `damage allowance' framework, a risk-constrained approach to calculate the design height of coastal protection strategies that maintain a user-defined AAL while considering deep uncertainty in AIS mass loss. Our approach is benefit-only as it does not consider costs. We specifically assess the strategies of elevation, coastal retreat, a levee, and a storm-surge barrier. Advance is not considered due to complexities associated with projecting property growth in new lands. For each strategy, a range of design heights are calculated under multiple assumptions of future AIS mass loss. The damage allowances provide a first-order measure of benefits that could be readily input into a BCA or cost-effectiveness framework \citep[e.g.,][]{Aerts2014a,Scussolini2017a} where each protection strategy is associated with a given cost (not quantified here). Additionally, more complex modeling could subsequently be used to investigate preferred strategies \citep[e.g.,][]{Aerts2014a,Fischbach2017a}. We use Manhattan to illustrate our damage allowance framework, as it and the rest of New York City is currently exploring multiple flood mitigation options \citep{USACE2016a,USACE2019a,SIRR2013a,NYC2019a,NYC2019b}.

\section{Framework} \label{sec:methods}

The flow and sources of information used in our flood damage allowance framework is shown in Fig. \ref{Sfig:methods}, and more details are given in the Supporting Information. First, we employ a flood damage model (Sec. \ref{Ssec:simpleMod}) to calculate Manhattan's AAL due to flooding. The damage model estimates the present-day probabilities of ESLs of various heights using both a long-term record of sea level observations at the Battery tide gauge in lower Manhattan and extreme value theory (Fig. \ref{fig:process}A; Sec. \ref{Ssec:ESL}). Damages are estimated using a simple, aggregate flood damage function constructed from property data and observed relationships between flood depth and building type (Fig. \ref{fig:process}B; Sec. \ref{Ssec:dmgmod}). The damage function assumes a ``frozen city'', in that the population and assets remain in place over time. The product of the ESL probability distribution and the damage function produces a probability distribution for damages (Fig. \ref{fig:process}C). Second, a flood damage mitigation strategy is chosen that modifies the shape of the simple damage function (either elevation, levee, storm surge barrier, or coastal retreat; Sec. \ref{sec:stabiliz}). Third, Manhattan's AAL due to flooding is projected into the future using probabilistic local SLR projections (Sec. \ref{Ssec:slr}) that have been adjusted by 1) subjectively weighting the likelihood of rapid AIS mass loss mechanisms and 2) specifying an upper limit to 2100 AIS contributions to GMSL (Sec. \ref{sec:ais}). As an example, the increase in the frequency of ESLs and damage events from a 0.5 m of local SLR is shown in Figs. \ref{fig:process}A and C, respectively. Finally, the design height of a given flood mitigation strategy is calculated such that the future AAL due to flooding under uncertain SLR equals a user-specified level of acceptable financial risk (e.g., the current AAL).

\subsection{A generalizable damage stabilizing model} \label{sec:stabiliz}

If $f(z^{*})$ is the current annual exceedance probability (AEP) of a given ESL event with surge height $z^{*}$ (e.g., the 100-year event), then the instantaneous hazard allowance that maintains the AEP under uncertain sea level change $\Delta$ can be expressed as:

\begin{linenomath*}
\begin{equation}
f(z^{*}) = \int_{\Delta}f(z^{*}-\Delta+A(z^{*}))P(\Delta)\, \text{d}\Delta,
\end{equation} \label{eq:elevation2}
\end{linenomath*}

\noindent where $f(z^{*}-\Delta)$ is the AEP of $z^{*}$ after including sea level change whose uncertainty is given by the probability distribution $P(\Delta)$, and $A(z^{*})$ is the vertical adjustment required to maintain the current AEP. For a given AEP, the hazard allowance can be interpreted as the horizontal distance between the historical and future ESL return curve (i.e., the curve that shows the average number of ESL events that meet or exceed a given ESL in a given year). If $\Delta$ is known, then $A = \Delta$, but if SLR projection uncertainty is considered, the hazard allowance will always be greater than the expected SLR due to an approximately log-linear relationship between the AEP of ESLs and water height \citep{Buchanan2016a}.

Two types of flood allowances have been proposed in the literature: 1) instantaneous and 2) average annual design life. Instantaneous flood allowances are designed for maintaining a target level of risk in the final year of a project design life (traditionally the expected number of ESL exceedances). In the preceding years, the average level of risk protection would be above the target. On the other hand, the average annual design life flood allowance (not explored in this study), maintains the average risk over the lifetime of a project by providing greater protection than prescribed during the early years of the design life and less protection than prescribed at the end of the design life \citep{Buchanan2016a}.

%
%
%
%

We extend the flood allowance concept to damage events by employing a simple, time-invariant damage function that describes the relationship between ESLs and direct physical damages. The damage allowance is the design height of a flood mitigation strategy needed to maintain the current frequency of damage events (i.e., the AAL due to flooding) under uncertain SLR. To offset additional damages due to SLR, we create a ``protected'' damage function using idealized representations of how the flood damage reduction strategies of elevation (Sec. \ref{sec:elevation}), a levee or storm surge barrier (Sec. \ref{sec:protect}), and coastal retreat (Sec. \ref{sec:retreat}) could impact the relationship between ESLs and damage in the ``unprotected'' damage function.

In order to stabilize the AAL resulting from uncertain SLR, the current AAL must equal the projected AAL that includes both an arbitrary sea level change ($\Delta$) plus an adjustment to offset the increase in damages resulting from SLR. This ``conservation of damage'' can be mathematically represented by:

\begin{linenomath*}
\begin{equation}
\int_{z_{min}}^{\infty} \int_{\Delta}D^{*}(z)f(z-\Delta)P(\Delta)\, \text{d}\Delta\, \text{d}z = \int_{z_{min}}^{\infty}D(z)f(z)\, \text{d}z,  \label{eq:balance1}
\end{equation}
\end{linenomath*}

\noindent where $z$ is the ESL height (relative to mean higher high water [MHHW])\footnote{The average level of high tide over the last 19-years}, $D^{*}(z)$ is a protected damage function that includes adjustments to mitigate additional flood damage resulting from SLR depending on the adaptation strategy taken (elevation, levee, storm surge barrier, or coastal retreat; Secs. \ref{sec:elevation} to \ref{sec:retreat}), and $D(z)$ is the unprotected damage function (Sec. \ref{Ssec:dmgmod}). The equation is integrated from the current protection height in Manhattan ($z_{min}$) to infinity. We assume that the island of Manhattan is currently protected by a bulkhead that is 1.0 m above MHHW, below which no damage occurs\footnote{This is approximately the height with which water begins to flow over the bulkhead at the Battery in lower Manhattan (retrieved from https://water.weather.gov/ahps; July, 2019). This is also a simplification, as the bulkhead height varies around the island of Manhattan.}. While all allowances traditionally aim to maintain the current level of risk (i.e., the historical AAL), we note that this could be replaced by any user-specified level of risk. 

%
%
%
%


\subsubsection{Damage reduction strategy: elevation} \label{sec:elevation}

Accommodation strategies, such as elevating buildings on columns, reduces the vulnerability to populations and the built environment. Elevation allows for floods to occur below the design height without incurring significant damage to the structure. We model an elevation strategy that raises all structures by the vertical height $A$ at or below the land elevation $A$ (Fig. \ref{fig:dmgfunc}A). We acknowledge the impracticality of elevating high-rises structures in Manhattan. However, roughly 88\% of all buildings in Manhattan sited at $\le$ 8 m in elevation (relative to MHHW) have less than six floors above ground level \citep{NYCpluto2018a}\footnote{However, these buildings tend to be older and the relation to the total fraction of damage is unclear.}. Additionally, our approach is also consistent with wet-proofing basements and subsequently building new floors on top of the existing structure. To estimate the height $A$ that structures need to be elevated to maintain the current AAL under uncertain SLR, $D^{*}(z)$ is substituted in Eqn. \ref{eq:balance1} with the following:

\begin{equation}
D^{*}_{e}(z,A) = \underbrace{\phi(z-A) \int_{e_{min}}^{A} p(e) \, \text{d}e }_{\text{Damage to elevated structures}}  + \underbrace{\int_{A}^{z} p(e) \cdot  \phi(z-e)\, \text{d}e,}_{\text{Damage to non-elevated structures}} \label{eq:elevation1}
\end{equation} 

\noindent where $z$ is the ESL height (above MHHW), $A$ is the damage allowance (above the current protection level, $z_{min}$), $p(e)$ is the amount of property at elevation $e$ and $\phi(z-e)$ is an inundation depth-damage function for NYC that relates the flood height (i.e., $z-e$) to damage as a fraction of the total property value (see Sec. \ref{Ssec:depthdamage}). Equation \ref{eq:elevation1} is plotted in Fig. \ref{fig:dmgfunc}B assuming $A$ = 0.75 m. Note that no flood damage occurs to structures when $z \le A$, but when $z > A$, $D^{*}_{e}(z,A)$ begins to converge to the unprotected damage function $D(z)$. While not used for Manhattan, a method for elevating all structures within a damage function is given in the Supporting Information (Sec. \ref{Ssec:elevAll}).

\subsubsection{Damage reduction strategies: levee and storm surge barrier} \label{sec:protect}

Flood defenses physically block the inland advance of ESLs and include both hard defenses (e.g., levees and storm surge barriers) as well as soft defenses (e.g., beach-dune systems). Levees (also known as dikes) are stationary embankments strategically used to prevent areas from flooding. Storm surge barriers are gates placed within bodies of water (usually tidally-influenced rivers or estuaries) that remain open to allow for tidal flushing and maritime navigation, but close during forecasted ESLs (e.g., coastal storms). Storm surge barriers are often placed within levee systems \citep{Mooyaart2017a}.

While levees and surge barriers can be public good solutions that do not require private actors to undertake action (e.g., elevation or retreat), both have the disadvantage of maximal loss occurrence by either being overtopped (e.g., by experiencing EWLs outside of the designed margin of safety) or by suffering structural failure. We model the probability of levee and storm surge barrier failure ($p_{f}$) conditional on the ESL $z$ (i.e., the structural load), the targeted design height $A$, the failure rate for the water load at the design height ($tol_{A}$; assumed to be 0.10, which is the minimum threshold for the probability of a loading breach for Dutch flood defenses \citep{TAW1998a}), and the height of the the freeboard $F_{b}$ (i.e., the protection above $A$): $p_{f} = P(fail\,|\, z, A, tol_{A}, F_{b})$. According to \citet{Wolff2008a}, for a well designed levee, the probability of failure below the design height should be ``unlikely'' and then increase rapidly until it reaches unity at the targeted design height plus the specified freeboard. We model $p_{f}$ using an exponential relationship that is a function of $z$, $p_{f} = a \cdot \text{exp}(b \cdot z)$, where $b = ln(1/tol_{A})/ F_{b}$ and $a = \text{exp}(-b \cdot A+ln(tol_{A}))$ (Fig. \ref{Sfig:failrate}). For levees, $D^{*}$ is substituted in Eq. \ref{eq:balance1} with:

\begin{equation}
 D^{*}_{l}(z) = p_{f}(z,A)D(z).
 \end{equation} \label{eq:levee}
 
\noindent In other words, total damage scales with the probability of levee failure until overtopping occurs (Figs. \ref{fig:dmgfunc}B and \ref{Sfig:failrate}). 

%

Due to increased chances of mechanical failure, environmental impacts, and shipping disruption, the frequency of storm surge barrier gate closure is often restricted \citep{SustainableSolutionsLab2018a}. This prevents storm surge barriers from protecting against more frequent events that can cause minor flooding, such as extreme tidal events \citep{Sweet2016a}. Some of the largest storm surge barriers are designed to close approximately once every 10 years, such as the Maeslant Barrier that protects Rotterdam, while smaller to medium sized barriers, such as the Thames Barrier in London, have historically been designed to close no more than two to three times per year \citep{SustainableSolutionsLab2018a}. We assume that a surge barrier designed to protect Manhattan would be similar to the Maeslant Barrier, designed to close no more than once every 10 years on average under current mean sea levels. This corresponds to an ESL closure threshold ($z_{close}$) of roughly 1.0 m above MHHW, as estimated from the ESL return curve (Fig. \ref{fig:process}A). For storm surge barriers, $D^{*}$ is substituted in Eq. \ref{eq:balance1} with:

\begin{equation}
 D^{*}_{b}(z) = H_{b}[z]D(z).
 \end{equation} \label{eq:levee}
 
\noindent where $H_{b}$ is the discrete form of the Heaviside function,

\begin{equation}
H_{b}[z]=\left\{\begin{matrix}
\begin{aligned}
    &p_{f}(z,A), && z \ge z_{close}, \\
    &1, && z < z_{close}.  \label{eq:hretreat}
\end{aligned}
\end{matrix}\right. 
\end{equation}

\subsubsection{Damage reduction strategy: coastal retreat} \label{sec:retreat}

Coastal retreat can be described as the reduction in exposure to ESLs through the removal of populations and the built environment below an elevation $A$ in order to reduce expected flood damages. This could be voluntary migration, forced displacement, or planned relocation. Retreat is the only strategy that completely eliminates residual risks if there is perfect compliance. Mathematically, coastal retreat modifies the unprotected damage function $D(z)$ by subtracting the damages above $A$ (i.e., $D(A)$) and removing damages below $A$ (Fig. \ref{fig:dmgfunc}B). The protected damage function is, 

\begin{equation}
D^{*}_{r}(z)=D(z)H_{r}[z],
\end{equation}

\noindent where $H_{r}$ is the discrete form of the Heaviside function,  

\begin{equation}
H_{r}[z]=\left\{\begin{matrix}
\begin{aligned}
    &1-\alpha, && z < A, \\
    &1-\alpha \frac{D(A)}{D(z)}, && z > A, \label{eq:hretreat}
\end{aligned}
\end{matrix}\right. 
\end{equation}

\noindent and $\alpha$ is the fraction of assets [0,1] below $A$ that have retreated (i.e., the retreat compliance). Perfect compliance should not be expected, especially if retreat is voluntary. Some risk targets may not be achievable using coastal retreat if compliance is not high enough. As retreat compliance decreases, $D^{*}_{r}(z)$ approaches the unprotected damage function (Fig. \ref{Sfig:retreat_compli}).

\subsection{Uncertainty in Antarctic ice sheet collapse} \label{sec:ais}

Beyond mid century, the dynamic response of the AIS to warming is a key uncertainty in protecting future sea levels. Several different plausible estimates of continental-scale AIS melt exist \citep[e.g.,][]{Golledge2019a,Edwards2019a,Deconto2016a,Ritz2015a,Levermann2014a,Bamber2013a,Little2013a}, but there is currently not an agreed upon full range of outcomes and likelihoods necessary for risk assessment using a single SLR probability distribution. Some investigators find significant collapse of marine-based sectors of the AIS under strong climate forcing before 2100 using physical models that include dynamical ice sheet mechanisms \citep[e.g.,][]{Golledge2019a,Edwards2019a,Deconto2016a}, while others assume or suggest greater stability \citep[e.g.,][]{Ritz2015a,Little2013a,Little2013b}. An implication of this ambiguity is that the probability distribution of SLR in the second half of the century remains strongly dependent upon subjective assessment of potential AIS contributions. A single probabilistic projection can only represent a fraction of the true total uncertainty. Employing multiple probability distributions is one method to better illustrate deep uncertainty \citep[e.g.,][]{Sriver2018a,Kopp2017a,Slangen2017a,Wong2017b,Wong2017a}.

Imprecise probability methods can be used in cases where multiple probability distributions cannot be reduced to a single distribution. For example, a probability box (or `p-box') can be used to express SLR incertitude by constraining plausible cumulative probability distribution functions (CDFs) of SLR within a defined space \citep{Baudrit2007a,LeCozannet2017a}. If it is assumed that the upper (i.e., right edge) and lower (i.e., left edge) limits of the p-box contain the unknown distribution, then the true probability of exceeding a given amount of SLR lies within the bounds of the CDFs. A p-box is shown in Fig. \ref{fig:deep}A for 2100 local SLR for Manhattan. Here, we limit the upper p-box boundary with local probabilistic SLR projections from \citet{Kopp2017a}, which employ the fast ice loss AIS projections from \citet{Deconto2016a}, and the lower p-box boundary with SLR projections from \citet{Kopp2014a}, which include more sluggish AIS mass loss based on a combination of the Intergovernmental Panel on Climate Change's (IPCC) Fifth Assessment Report (AR5) and expert elicitation of total ice sheet mass loss from \citet{Bamber2013a}. These SLR projections were chosen for illustrative purposes only. For instance, a more credible approach could employ the more recent projections from \citet{Bamber2019a} that include new expert elicitation of AIS behavior. End-of-century SLR projections from \citet{Bamber2019a} under RCP8.5 largely fall in between those from \citet{Kopp2014a} and \citet{Kopp2017a}, but have a higher probability of SLR outcomes $>$ 3 m (Fig. \ref{Sfig:slrtail}).

Subjective beliefs regarding future AIS behavior are used to generate an ``effective'' SLR distribution (Sec. \ref{Ssec:pbox}) within the p-box. The ``effective'' distribution is generated by 1) selecting an upper limit to 2100 AIS contributions ($\text{AIS}_{max}$) and 2) averaging the upper and lower bounds of the p-box using a weight that reflects the subjective likelihood of AIS collapse initiation before 2100 ($\beta_{c} \in [0,1]$). Larger values of $\beta_{c}$ imply a higher likelihood of AIS collapse initiation before 2100, while lower values of  $\beta_{c}$ imply a lower likelihood. More details regarding the SLR projections and the p-box construction are given in the Supporting Information (Secs. \ref{Ssec:slr} and \ref{Ssec:pbox}, respectively). Similar approaches that weigh worst-case outcomes for SLR and other climate variables have been used for decision making under deep uncertainty \citep{Buchanan2016a,McInerney2012a}.

\subsection{Example of damage allowance framework} \label{sec:framework}

\begin{figure}[htb]
\centering
\includegraphics[width=.99\textwidth]{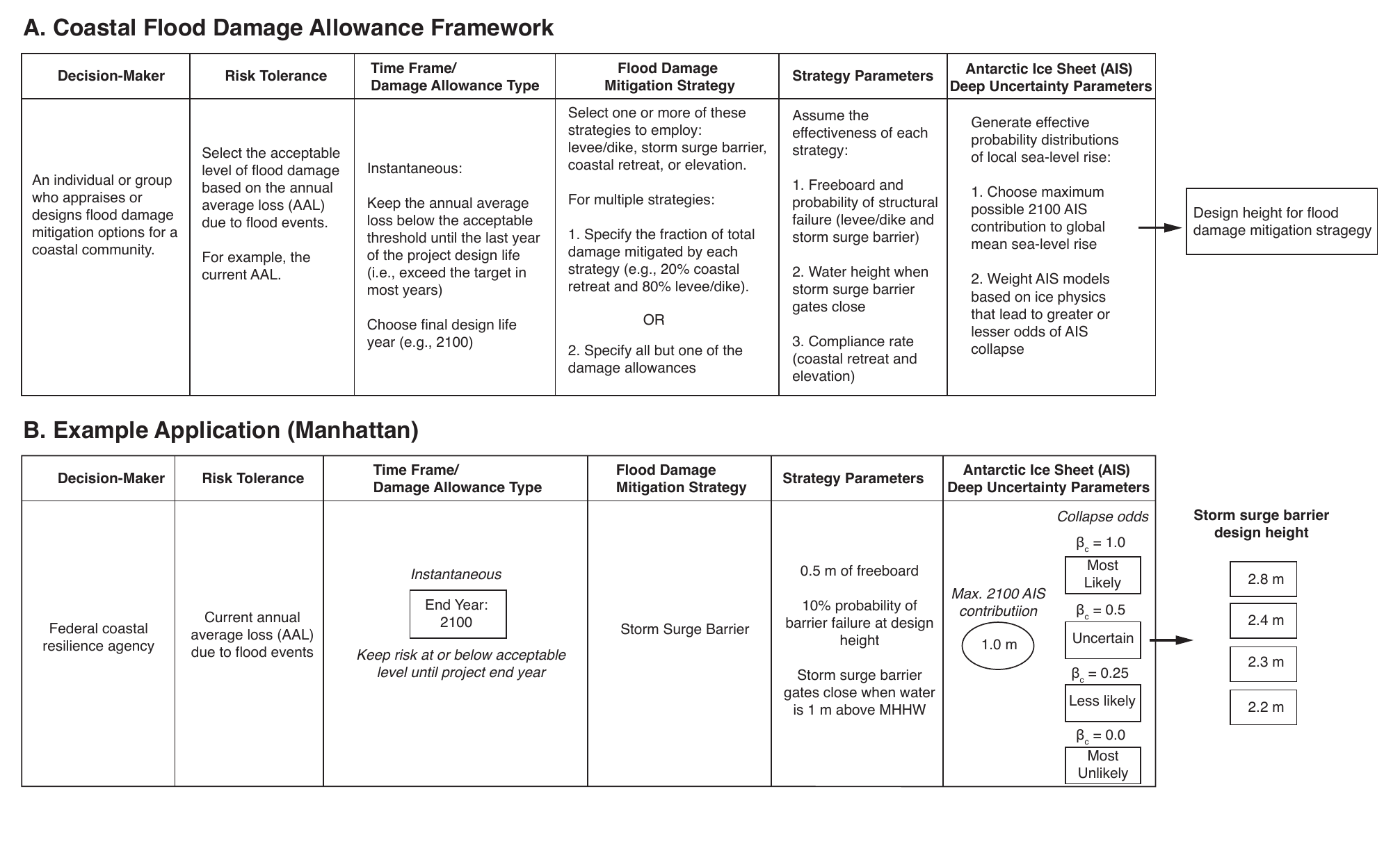}
\caption{ \textbf{A.} Flow chart illustrating how to apply the coastal flood damage allowance framework \textbf{B.} An example application for Manhattan seeking to maintain the current annual average loss from flood damages using a storm surge barrier.}\label{fig:flowchart}
\end{figure}

The flow of the damage allowance framework is presented in Fig. \ref{fig:flowchart}A. First, an acceptable AAL risk target is selected. This could be the current AAL to maintain the current level of flood risk (i.e., the traditional allowance definition), or any value greater than zero. Second, both the time frame and the approach for meeting the risk target are chosen. In this paper, we only consider the instantaneous allowance (i.e., risk is kept below the target through the duration of the project and is met in the final year), however, this could alternatively be a period of years over which the mean AAL equals the risk target \citep[earlier years are below the risk target and later years are above it; i.e., the average annual design life allowance;][]{Buchanan2016a}. Third, a flood damage mitigation strategy is selected. Finally, an upper limit to 2100 AIS contributions ($\text{AIS}_{max}$) is chosen, as well as a value of $\beta_{c}$ that reflects a level of confidence that the AIS will begin to collapse before 2100 (Sec \ref{sec:ais}). An application of the damage allowance framework is shown in Fig. \ref{fig:flowchart}B. In this example, a federal coastal resilience agency chooses to maintain Manhattan's current AAL due to flooding through 2100 using a storm surge barrier. The barrier has a 10\% probability of failure at the design height and the gates close when the ESL is $\ge$ 1.0 m (the assumed current bulkhead height around Manhattan). The agency believes that the 2100 AIS mass loss contributions to GMSL is limited to 1.0 m (relative to 2000) and that collapse initiation of the AIS is ``less likely'' ($\beta_{c}$ = 0.25). According to the framework, the storm surge barrier should be built to a height of 2.3 m above the current protection level. This design height increases by 0.5 m if the AIS collapse odds are instead believed to be ``most likely'' ($\beta_{c}$ = 1.0).

\section{Results} \label{sec:results}

\subsection{Impact of flood protection strategy on damage event frequency}

Fig. \ref{fig:process}A shows the ESL hazard return curve for Manhattan. Sea-level rise increases the frequency of all ESL events. The hazard allowance that maintains the historical frequency of ESL events is 0.50 m under 0.50 m of known SLR and is 0.86 m when considering the entire probability distribution of projected 2070 SLR ($\text{AIS}_{max}$ = 1.75 m, $\beta_{c}$ = 1.0). However, the hazard allowance only maintains the number of ESL events. It does not assure that the number of damage events will also coincidentally be held constant over time. An ESL event may or may not lead to damage depending on the existing local flood protection. An unprotected damage function for Manhattan is shown in Fig. \ref{fig:process}B that connects ESLs to direct financial loss. A 1.0 m high bulkhead around Manhattan eliminates damages $<$ $\sim$\$0.5 billion (2017 USD)\footnote{All monetary values in this paper are given in 2017 U.S. dollars (USD)}. Also shown is a 1.70 m levee on top of the current protection level (i.e., the 1.0 m bulkhead) that modifies the unprotected damage function, reducing the damage for some ESLs (Fig. \ref{fig:process}B). In the absence of additional protection measures, SLR increases the frequency of all damage events and the AAL (Fig. \ref{fig:process}C). For instance, a 0.5 m of SLR is estimated to increase the current AAL for Manhattan from roughly \$0.1 billion/yr to roughly \$0.7 billion/yr, and projected 2070 SLR is expected to increase the current AAL by more than a factor of 10, roughly \$1.6 billion/yr ($\text{AIS}_{max}$ = 1.75 m, $\beta_{c}$ = 1.0). Under 2070 SLR, a 1.70 m levee reduces the AAL with SLR to the current AAL (i.e., the damage allowance). The damage allowance for the levee that is needed to maintain the current AAL is nearly two times greater than the ESL hazard allowance needed to maintain the current frequency of ESL events (1.70 m vs. 0.86 m) and more than twice as large as the expected amount of SLR in 2070 (0.80 m; $\text{AIS}_{max}$ = 1.75 m, $\beta_{c}$ = 1.0; Table \ref{Stab:slr}).

\begin{figure}[htb]
\centering
    \includegraphics[width=.999\textwidth]{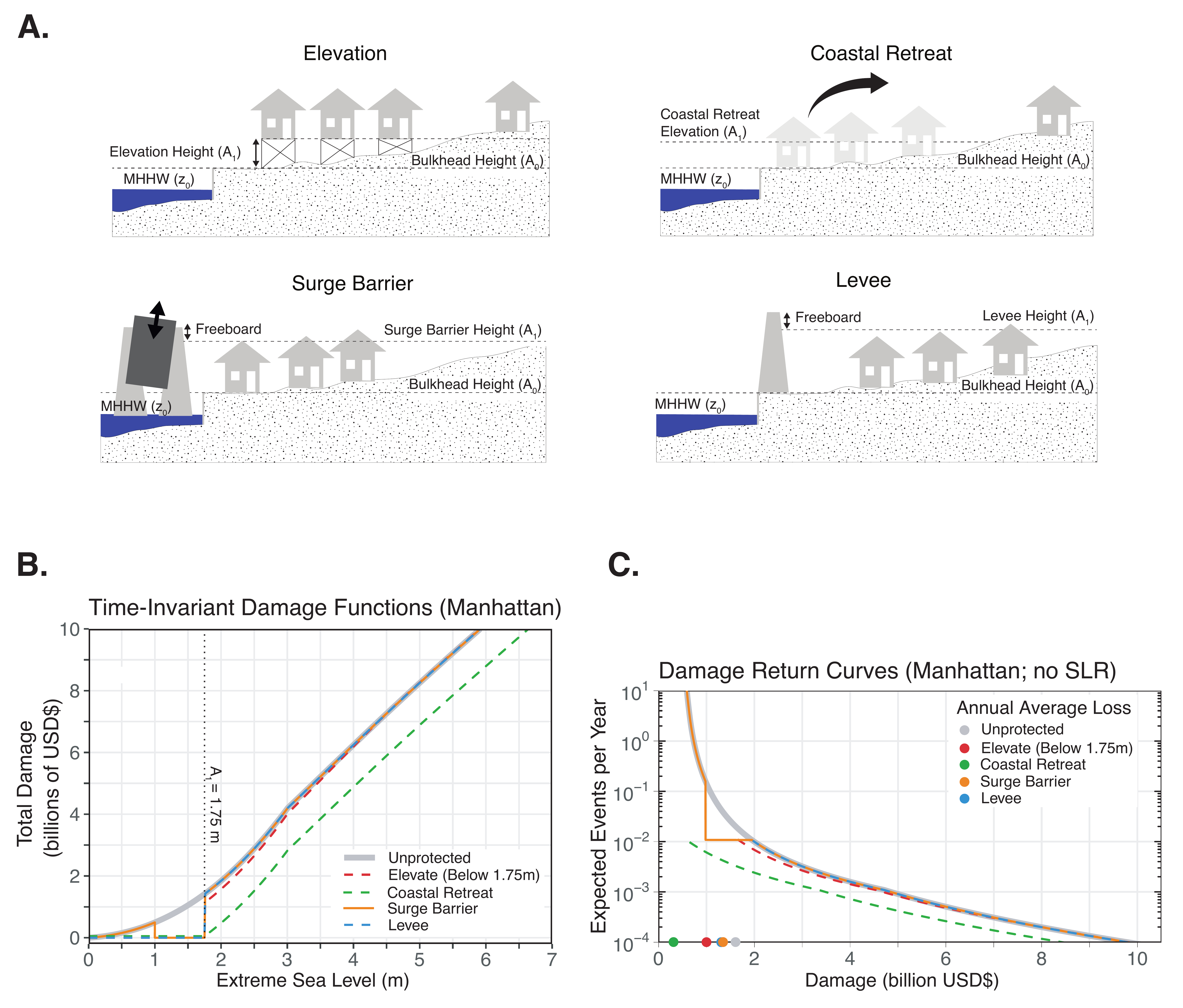}
  \caption{\textbf{A.} Schematics illustrating each flood protection strategy for an arbitrary protection height (A$_{1}$). \textbf{B.} Time-invariant damage functions for Manhattan that relate extreme sea level (ESL; meters) to total direct damage due to flooding (billions of USD\$; Sec. \ref{Ssec:dmgmod}). The thick grey line is the unprotected damage function (i.e., no flood protection strategy) that assumes no existing bulkhead around Manhattan; the dashed red line is the damage function for elevating all structures by the design height below 1.75 m in elevation (relative to MHHW); the dashed green line is the damage function for coastal retreat of all structures below 1.75 m of elevation (relative to MHHW); the solid orange line is the damage function for a storm surge barrier with a protection height of 1.75 m (relative to MHHW) with gates that close when the ESL is $>$ 1.0 m and has zero probability of structural failure and no freeboard; the dashed blue line is the damage function for a levee with a protection height of 1.75 m (relative to MHHW) that has zero probability of structural failure and no freeboard. \textbf{C.} Damage event return curves under no sea-level rise (SLR) showing the expected number of flood damage events per year (billions of 2017 USD\$) with no protection strategy (thick grey curve) and under the flood protection strategies of elevation (dashed red curve), coastal retreat (dashed green curve), a storm surge barrier (solid orange curve), and a levee (dashed blue curve). All cases assume  that Manhattan's distribution of buildings, people, and infrastructure remains constant in time. For illustrative purposes, all cases assume no bulkhead around Manhattan and the storm surge barrier and levee strategies assume no possibility of structural failure and no freeboard. The discontinuity occurs for the storm surge barrier due to protection being limited to a range of ESLs (here 1.0 m to 1.75 m). The annual average loss under each protection strategy is plotted on the x-axis with a filled circle.} \label{fig:dmgfunc}
\end{figure}

Fig. \ref{fig:dmgfunc}C shows the impact of a levee on the current return periods of damage events without SLR, as well as for elevation, a storm surge barrier, and coastal retreat. For illustrative purposes, the 1.0 m bulkhead around Manhattan is not considered here (but is in the damage allowance calculations; Sec. \ref{sec:dmgresults}). With out considering both the bulkhead protection and SLR, the AAL is roughly \$1.6 billion/yr. A 1.75 m levee used to protect Manhattan would prevent all flood damage events $<$ \$2 billion (assuming zero probability of levee breaching) and would lower the AAL to \$1.2 billion/yr (structural failure of a levee is not depicted in Fig. \ref{fig:dmgfunc}B or Fig. \ref{fig:dmgfunc}C, but considered in the damage allowance calculation; Sec. \ref{sec:protect}). Raising all structures below 1.75 m in elevation to 1.75 m has nearly the same effect as implementing a levee, but in this study, failure associated with structural elevation is not considered. Compared to the levee, the elevation strategy reduces the AAL more, to \$1.0 billion/yr. The effect of a storm surge barrier is similar to the levee, but because barrier gates remain open below a specified threshold ESL (here 1.0 m), flood damage events $<$ \$1 billion can occur. When the gates close, the barrier behaves like the 1.75 m high levee and damage events are prevented until ESLs are $>$ 1.75 m (over-topping) or structural failure occurs (not depicted in Fig. \ref{fig:dmgfunc}B or \ref{fig:dmgfunc}C, but considered in the damage allowance calculation; Sec. \ref{sec:protect}). A surge barrier would lower the current AAL to \$1.3 billion/yr, slightly higher than the levee because the barrier allows ESL events to occur below the closure threshold (here 1.0 m). Coastal retreat not only eliminates damage events below the targeted protection height, like levees and structure elevation, it also removes the possibility of damage to structures below the protection height from higher ESLs (i.e., overtopping). Of all strategies explored in this study, coastal retreat reduces the AAL the most. If all structures below 1.75 m in elevation retreat, the AAL is reduced from \$1.6 billion/yr to \$0.3 billion/yr. We also consider the effects of multiple flood protection strategies (Sec. \ref{Ssec:multi}).

\subsection{Characterizing deep uncertainty with sea-level rise and extreme sea level return periods}
\begin{figure}[htb]
\centering
    \includegraphics[width=.999\textwidth]{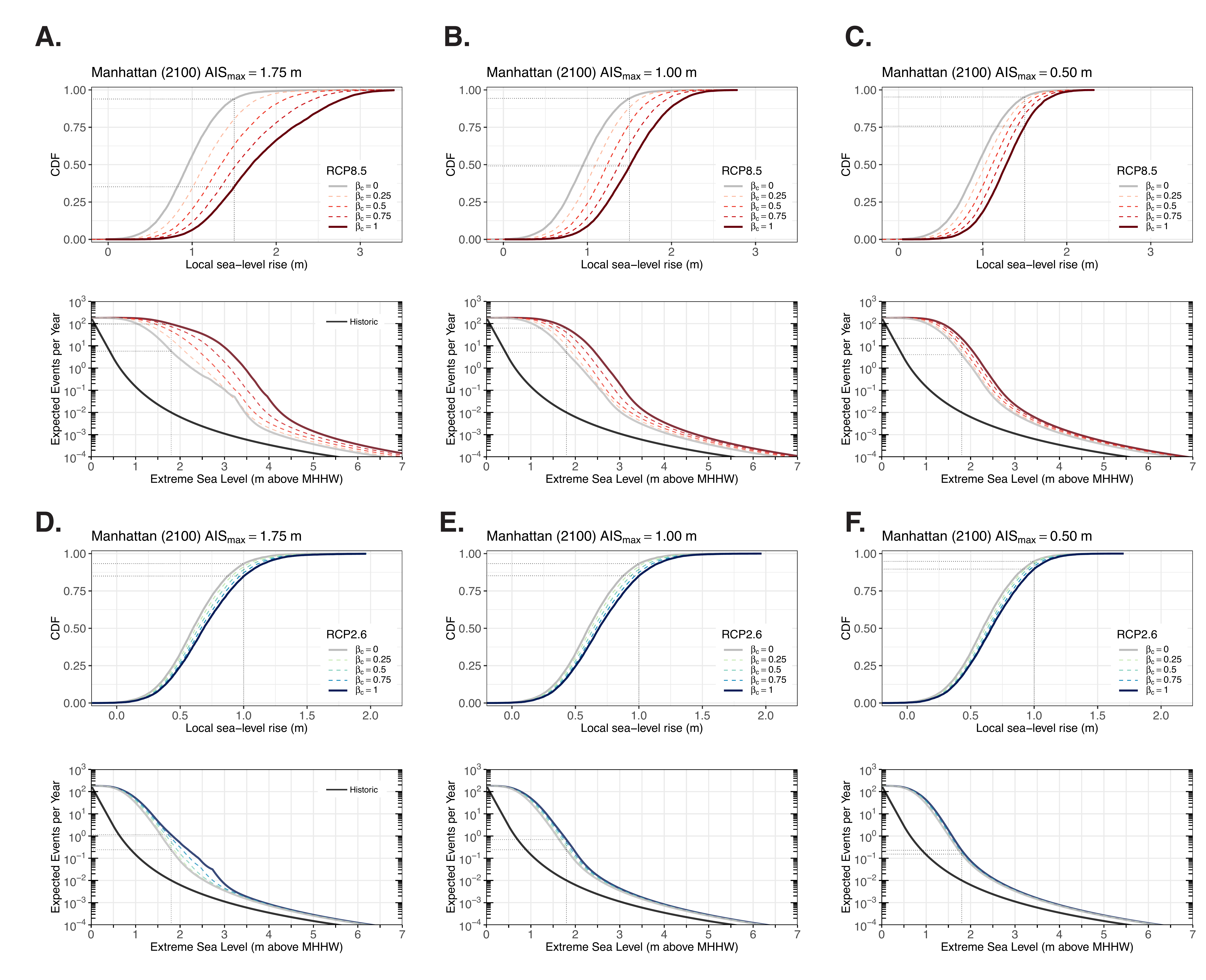}
  \caption{\textbf{A.} Top Row: probability boxes ('p-boxes'; solid lines) for 2100 local sea-level rise (SLR) in Manhattan (located at the Battery tide gauge) under the representative concentration pathway (RCP) 8.5 climate forcing scenario. Effective cumulative distribution functions (CDFs) of local SLR (dashed lines) are generated within each p-box by averaging the edges using weights ($\beta_{c}$ $\in$ [0,1]) that reflect a user's belief of AIS collapse initiation within the 21st century (higher values reflect higher likelihood of collapse) and by constraining the maximum possible 2100 Antarctic ice sheet (AIS) melt (AIS$_{max}$, relative to 2000; here, 1.75 m; Sec. \ref{sec:ais}). The black dotted lines highlight the cumulative probability of 1.0 m of local SLR under different assumptions of AIS collapse initiation (i.e., values of $\beta_{c}$). Bottom Row: extreme sea level (ESL) return curves for Manhattan showing the relationship between the expected number of ESLs per year and ESL height (meters above mean higher high water [MHHW]) for: 1) historical sea levels (black curve) and 2) year 2100 (RCP8.5) for different values of $\beta_{c}$. All curves incorporate generalized Pareto distribution (GPD) parameter uncertainty (Sec. \ref{Ssec:ESL}) and the future return curves additionally incorporate local SLR projection uncertainty by integrating across the entire local SLR probability distribution. The black dotted lines highlight the annual expected number of historically experienced 100-yr ESL events under different values of $\beta_{c}$. \textbf{B.} As for A, but for AIS$_{max}$ = 1.0 m. \textbf{C.} As for A, but for AIS$_{max}$ = 0.5 m. \textbf{D.} As for A, but for RCP2.6. \textbf{E.} As for A, but for RCP2.6 and AIS$_{max}$ = 1.0 m. \textbf{F.} As for A, but for RCP2.6 and AIS$_{max}$ = 0.5 m.} \label{fig:deep}
\end{figure}

In the absence of additional coastal flood protection, local SLR is expected to increase the frequency of all Manhattan flood damage events (Fig. \ref{fig:process}C), but late century projections of SLR remain deeply uncertain. In Fig. \ref{fig:deep}, we illustrate deep uncertainties associated with 2100 local SLR projections and ESL return curves at the Battery tide gauge in lower Manhattan under maximum possible AIS melt contributions ($\text{AIS}_{max}$) of 1.75 m, 1.0 m, and 0.5 m (relative to 2000) for both RCP8.5 and RCP2.6. For each plot, deep uncertainties are explored within the p-box ($\beta_{c} \in$ [0,1]) using multiple assumptions regarding the likelihood of collapse of the AIS. Under RCP8.5 and the largest AIS melt contribution threshold considered ($\text{AIS}_{max}$ = 1.75 m; Fig. \ref{fig:deep}A), the probability of local SLR exceeding 1.5 m is 6\% under the least likely AIS collapse assumption ($\beta_{c}$ = 0), but is 65\% under the most likely AIS collapse assumption ($\beta_{c}$ = 1). The spread between these probabilities is a reflection of the deep uncertainty present, and the spread decreases as $\text{AIS}_{max}$ decreases. For instance, in the case of $\text{AIS}_{max}$ = 1.0 m (Fig. \ref{fig:deep}B), the probability of local SLR exceeding 1.0 m is 6\% ($\beta_{c}$ = 0) and 51\% ($\beta_{c}$ = 1), while in the case of $\text{AIS}_{max}$ = 0.5 m (Fig. \ref{fig:deep}C), the probability of local SLR exceeding 1.5 m is 5\% ($\beta_{c}$ = 0) and 24\% ($\beta_{c}$ = 1). Deep uncertainties for the AIS are the greatest under high-emission future scenarios (i.e., RCP8.5). After mid century, under RCP2.6, median SLR projections are lower compared to RCP8.5 (Table \ref{Stab:slr}), and assumptions regarding future AIS behavior do not significantly impact SLR projections. For instance, for $\text{AIS}_{max}$ = 1.75 m, the probability of local SLR exceeding 1.0 m is 7\% under the least likely AIS collapse assumption ($\beta_{c}$ = 0), but is 14\% under the most likely AIS collapse assumption ($\beta_{c}$ = 1; Fig. \ref{fig:deep}D). For $\text{AIS}_{max}$ = 0.5 m, these probabilities reduce slightly to 5\% and 13\%, respectively (Fig. \ref{fig:deep}F). Additional p-boxes for local SLR projections under other values of $\text{AIS}_{max}$ and for the years 2050 and 2070 are given in the Supporting Information (Figs. \ref{Sfig:pbox1750} to \ref{Sfig:pbox250}). 

Deep uncertainty associated with the AIS is also reflected in the expected number of ESL events experienced at the Battery tide gauge. For instance, under RCP8.5 and assuming $\text{AIS}_{max}$ = 1.75 m, the expected number of historical 100-yr ESL events in 2100 is expected to increase from 0.01/yr, on average, to between roughly 6/yr ($\beta_{c}$ = 0) and 100/yr ($\beta_{c}$ = 1), on average (Fig. \ref{fig:deep}A). When truncating the AIS melt distribution at 1.75 m (i.e., $\text{AIS}_{max}$ = 1.75 m), ``kinks'' appear in the return curves (e.g., the $\beta_{c}$ = 0 return curve in Fig. \ref{fig:deep}A and the $\beta_{c}$ = 1 return curve in Fig. \ref{fig:deep}D) as a result of extreme sea level samples in the upper tail of the distribution causing ESL frequencies to saturate at 182.6/yr. The saturation subsequently increases the expected value. Both the positioning and the presence of the kinks are sensitive to the truncation of the upper tail of the mean sea level distribution. For example, the kinks disappear in the return curves for $\text{AIS}_{max}$ $\le$ 1.0 m (Fig. \ref{fig:deep}B,C,E, and F). As with local SLR projections, reducing $\text{AIS}_{max}$ also decreases the spread of ESL projections under differing assumptions of AIS collapse likelihood. Under RCP8.5, for $\text{AIS}_{max}$ = 0.5 m, the expected number of historical 100-yr ESL events experienced at the Battery tide gauge is roughly 4/yr ($\beta_{c}$ = 0) and 20/yr ($\beta_{c}$ = 1), on average (Fig. \ref{fig:deep}C). Under RCP2.6, the number of expected ESL events per year dramatically decreases relative to RCP8.5, owing to lower SLR projections, and the spread in results under different AIS collapse assumptions decreases relative to RCP8.5. Compared to RCP8.5, the assumption of strong greenhouse gas mitigation also results in ESL return curves that are less sensitive to the truncation of $\text{AIS}_{max}$. For instance, assuming an AIS melt threshold of $\text{AIS}_{max}$ = 1.75 m, the number of expected 100-yr ESL events is roughly between 0.2/yr ($\beta_{c}$ = 0) and 1/yr ($\beta_{c}$ = 1), on average (Fig. \ref{fig:deep}D). Applying an AIS melt truncation of $\text{AIS}_{max}$ = 1.0 m under RCP2.6 slightly decreases the number of expected 100-yr ESL events from 0.01/yr, on average, to roughly 0.2/yr ($\beta_{c}$ = 0) and 0.7/yr ($\beta_{c}$ = 1), on average (Fig. \ref{fig:deep}E). For $\text{AIS}_{max}$ = 0.5 m, there's little to no difference in the number of expected 100-yr flood events based on the perceived likelihood of AIS collapse; the number of expected 100-yr flood events for both $\beta_{c}$ = 0 and $\beta_{c}$ = 1 is roughly 0.2/yr, on average (Fig. \ref{fig:deep}F). Additional ESL return curves under other values of $\text{AIS}_{max}$ and for the years 2050 and 2070 are given in the Supporting Information (Figs. \ref{Sfig:pbox1750} to \ref{Sfig:pbox250}). 

\subsection{Flood damage allowances: surge barrier and coastal retreat}\label{sec:dmgresults}

Instantaneous flood damage allowances for a levee and coastal retreat that maintain the current AAL for Manhattan under different assumptions of AIS behavior for the climate forcing scenarios of RCP8.5 and RCP2.6 are given in Fig. \ref{fig:allwnc}. These allowances are also tabulated for 2050, 2070, and 2100 in Tables \ref{tab:DmgAllwncLevee} and \ref{tab:DmgAllwncRetreat}, respectively. The damage allowances for the levee include 0.5 m of freeboard and the levee is prescribed a 10\% probability of failure when the ESL is at the height of the damage allowance (see failure rate discussion in Sec. \ref{sec:protect}). The damage allowances are relative to the assumed current protection level (a 1.0 m bulkhead). As sea levels rise, the damage allowances increase over time. The allowances for the levee increase faster than for coastal retreat because as the levee height increases, more property stands to be impacted in the event of a levee failure (over-topping or breaching). This must be compensated by an increase in levee height. Coastal retreat removes property above and below the protection height and so more greatly modifies the damage function. The damage allowances for the levee and coastal retreat increase at rates similar to that of local SLR. For instance, the expected levee damage allowances for 2050, 2070, and 2100 are 0.9 m, 1.6 m, and 3.4 m, while the corresponding expected local SLR is 0.4 m, 0.8 m, and 1.8 m (RCP8.5; $\text{AIS}_{max}$ = 1.75 m and $\beta_{c}$ = 0, Table \ref{Stab:slr}).

For both RCP scenarios, the assumptions regarding AIS behavior (i.e., maximum 2100 melt contribution [$\text{AIS}_{max}$] and collapse likelihood [$\beta_{c}$]) produce little to no difference in the allowances before mid century for a given strategy. However, after mid century, under strong climate forcing (i.e., RCP8.5), allowances may differ by up to 1.4 m depending on the truncation of the AIS distributions (1.75, 1.5, 1.0, 0.5, or 0.25 m) and the perceived likelihood of AIS collapse ($\beta_{c} \in [0,1]$; Fig. \ref{fig:allwnc}A and Table \ref{tab:DmgAllwncLevee}). Under RCP2.6, these differences are $\le$ 0.6 m and only occur after 2080 under higher likelihoods of AIS collapse ($\beta_{c} >$ 0.75). For RCP8.5, the time period at which the impact of the $\text{AIS}_{max}$ assumptions begin to matter occurs earlier for higher collapse likelihoods ($\beta_{c} >$ 0.75) compared to lower collapse likelihoods ($\beta_{c} <$ 0.5). For instance, under the least likely AIS collapse assumption ($\beta_{c}$ = 0), damage allowances for both the levee and coastal retreat are not strongly dependent on the truncation of the AIS melt distribution until roughly 2080, but under the most likely AIS collapse assumption ($\beta_{c}$ = 1), damage allowances for both the levee and coastal retreat begin to diverge just after 2050 for AIS truncation assumptions of $\text{AIS}_{max}$ = 0.25 m and $\text{AIS}_{max}$ = 1.75 m. The size of the spread between the levee and coastal retreat damage allowances by 2100 under different 2100 AIS melt cut-offs is also dependent on the assumed likelihood of AIS collapse. For example, for low likelihood collapse odds, the range of levee damage allowances conditional on the possible truncations of AIS melt is 1.8--2.6 m ($\beta_{c}$ = 0) and for high likelihood collapse odds is 2.0--3.4 m ($\beta_{c}$ = 1). The sensitivity of allowances based on assumptions of AIS behavior is consistent with previous findings \citep{Slangen2017a}. Damage allowances for a storm surge barrier and elevation are given in time series figures and tables in the Supporting Information (Figs. \ref{Sfig:timeseries_surge} and \ref{Sfig:timeseries_elev2}; Tables \ref{Stab:DmgAllwncSurgeB} and \ref{Stab:DmgAllwncElevBelowA}).

Damage allowances consider the entire probability distribution of SLR. If the probability distribution is long-tailed, then low probability, extreme outcomes can greatly increase expected values beyond those that might result when considering a single high-end SLR outcome such as the 95th percentile. To illustrate, damage allowances that only consider the 5/95th percentile 2100 SLR projections are also given in the margins of each time series plot (Fig. \ref{fig:allwnc}; SLR projections given in Table \ref{Stab:slr}). In the case of $\text{AIS}_{max}$ = 1.75 m and $\beta_{c}$ = 0 (RCP8.5), the levee and coastal retreat damage allowances that consider only the 95th percentile SLR projection are lower than the damage allowances that just considers the full distribution of SLR. This could be a result of the $\beta_{c}$ = 0 case more heavily weighing the probability distribution from \citet{Kopp2014a}, which has a longer tail compared to \citet{Kopp2017a} (Fig. \ref{Sfig:slrtail}). If the underlying SLR probability distribution is long tailed, then illustrating allowance uncertainty based on a range of SLR percentiles may be more informative for some design applications. Heavy-tailed probability distributions could lead to low probability outcomes dominating the expected BCA calculation and all subsequent decisions made based on the expected BCA.

\section{Discussion and Conclusion} \label{sec:discuss} 

A number of different formal decision making frameworks exist for designing strategies that mitigate flood damages \citep{Walker2013a}. These are generally driven by economic objectives, such as choosing courses of action where monetized discounted benefits exceed discounted costs (i.e., BCA). Flood damage allowances are a new approach for determining the design heights of coastal protection strategies in this class of financial-decision making tools. They follow a `decision-centered' approach \citep[e.g.,][]{Ranger2013a} because their outcomes are dependent on more than just SLR projections alone; users specify a tolerable level of risk (annual average loss), a protection strategy type (elevation, levee, storm surge barrier, and coastal retreat), and assumptions regarding future AIS behavior (maximum AIS melt contribution and AIS stability).

\begin{figure}[htb]
\centering
\includegraphics[width=.99\textwidth]{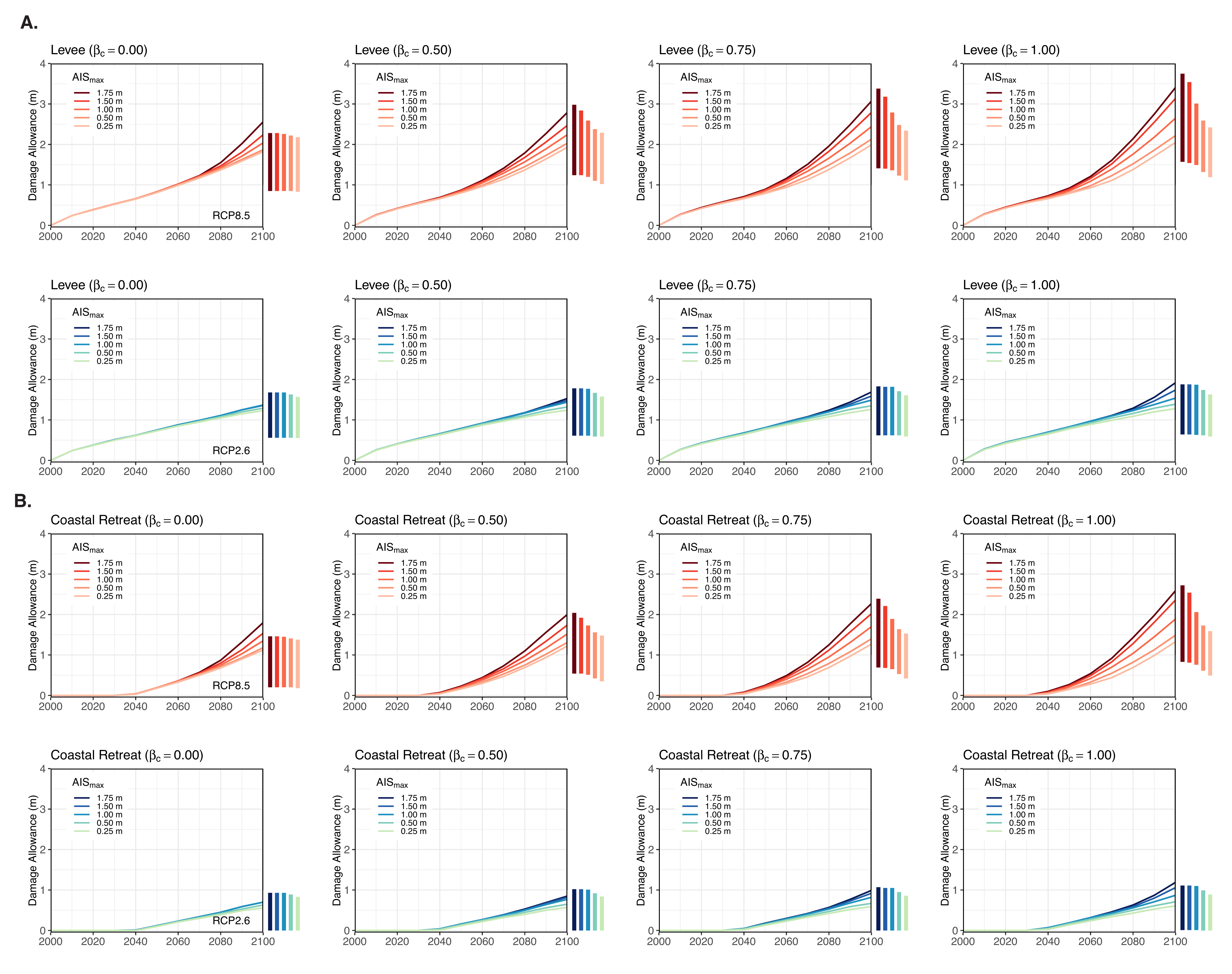}
\caption{ \textbf{A.} Top Row: Levee damage allowances (meters above the current protection height) over time (2000--2100) for protecting Manhattan under different maximum 2100 Antarctic ice sheet (AIS) contribution thresholds (AIS$_{max}$, relative to 2000), different subjectively perceived likelihoods of AIS collapse ($\beta_{c}$; 0 being `most unlikely' and 1 being `most likely'), and for the representative concentration pathway (RCP) 8.5 climate forcing scenario. The colored bars in the margins of each plot show the 2100 damage allowances using only the 5/95th percentile local sea-level rise projections. The levee allowances include 0.5 m of freeboard and have a 10\% probability of failure at the design height. Bottom Row: As for Top Row, but for RCP2.6. \textbf{B.} As for A, but for coastal retreat (assuming perfect compliance).}\label{fig:allwnc}
\end{figure}

As shown in this paper and elsewhere, decision tools that employ future projections of SLR may be appropriate when uncertainty can confidently be represented with a single PDF (e.g., near mid century, when AIS melt uncertainty is well defined), but is insufficient when uncertainty is deep \citep[e.g., late century AIS melt;][]{Hall2007a}. Using multiple distributions may better illustrate deep uncertainty (Sec. \ref{sec:ais}). Coastal flood protection designs can accommodate this information ambiguity by employing a range of design heights (e.g., Table \ref{tab:DmgAllwncLevee} and Fig. \ref{fig:allwnc}), rather than an expected value that has been integrated-out from a single PDF that assumes specific future AIS behavior. The expected monetized benefits of the chosen coastal flood protection strategy (i.e., avoided damages) can then be integrated into a simple cost-effectiveness decision framework, the costs of which could be estimated using existing sources \citep[e.g.,][]{Jonkman2013a,Lenk2017a,Aerts2018a}. 

The U.S. Army Corps of Engineers (USACE), the principal agency tasked with designing and implementing coastal flood risk management projects in the U.S., has proposed storm surge barriers to protect various areas of New York City from SLR and ESLs. The heights of the barriers are based on the height of the historical 100-yr ESL event plus a single, deterministic, `intermediate' future SLR projection ($\sim$0.5 m of SLR by 2100) and an amount of freeboard \citep{USACE2019a}. According to the SLR projections in this study, there's $>$ 90\% chance that local SLR will exceed 0.5 m by 2100 under all $\text{AIS}_{max}$ assumptions under RCP8.5 and $>$ 60\% chance under all $\text{AIS}_{max}$ assumptions for RCP2.6. If the freeboard calculations by the USACE do not account for potentially under-estimated future SLR, this could lead to mis-priced monetized benefits from flood protection and potentially sub-optimal selection of flood protection strategies and design heights.

In addition to decision-making frameworks employing multiple priors to deal with deep uncertainty, flexible/adaptive approaches have also been promoted in the literature as a way to make decisions under deep uncertainty. Flexible/adaptive decision approaches commit to short-term actions in response to new information \citep[e.g.,][]{Haasnoot2013a,Wise2014a,Walker2013a}. They have the advantage of being less dependent on accurate projections of the future. An example could be flexible levee design that allows for heightening over time as risk tolerances change or as new information is learned about future sea-level rise. 

Despite their dependence on parameters characterized by deeply uncertainty, there are multiple political reasons why a multiple priors approach (like flood damage allowances) might be pursued over a flexible/adaptive framework. First, existing laws and government agency protocols may favor approaches that require future distributions of uncertain parameters, such as BCA or cost-effective analysis. For example, the USACE largely uses BCA \citep{Flood1936a} to assess and select coastal protection strategies that best ``contribute to national economic development consistent with protecting the Nation's environment'' \citep{USWRC1983a}. Second, financing for disaster preparedness is usually only available following a major disaster \citep{NRC2014a}, in part due to motivations of elected officials \citep{Healy2009a}. For flood protection, current revenue streams are inadequate for supporting either new construction or regular upgrades that may occur with an adaptive/flexible approach \citep{Knopman2017a,SustainableSolutionsLab2018b}. This financing arrangement may reinforce the use of prediction-first approaches that force a decision to be made now with no planned opportunity to revisit the course of action in the future. Third, adaptive/flexible approaches may be more expensive compared to engineering to a fixed life \citep[i.e., designing once,][]{Fankhauser1999a,Haasnoot2019a}, and they may involve more political overhead to change existing governance structures away from appraisals based on BCA \citep[e.g.,][]{Ramm2017a,Kanyama2019a}. This could delay making a decision regarding what to build during which additional flood damages may occur. Development times for building flood protection are already long. For example, experience with storm surge barriers has shown that it can take decades to design and build a multi-billion dollar project \citep{Morang2016a,SustainableSolutionsLab2018a}.

While simple models like flood damage allowances may make too many approximations for readily implementable final project designs, in early planning phases they can identify coastal protection strategies that are worth examining in greater detail with more complex models. Rather than being viewed as substitutes, reduced-form models can complement their more complex peers. For example, while more complex integrated assessment frameworks may better simulate reality, they demand high computational costs which places limits on the number of flood protection strategies that can be investigated simultaneously \citep[e.g.,][]{Fischbach2017a}. Simple models may be more useful in cases where the appraisal of several project designs is needed, such as robust strategy identification \citep[e.g.,][]{Lempert2003a,Sriver2018a}. Additionally, exploring interactions between multiple variables with complex models can make it challenging to understand how different flood protection strategies and SLR assumptions impact expected benefits from proposed solutions. 

There are multiple caveats associated with flood damage allowances. First, flood allowances assume a risk tolerance remains constant over the lifetime of the investment. There might be a desire to increase it during that period due to a preference for a higher margin of safety. Second, the effectiveness of flood protection is time-dependent given changes in the hazard (SLR, ground subsidence rates, coastal storm frequency and severity) and changes in the consequence (e.g., what is behind the levee and how vulnerable is it to flood damage). In this study, SLR is the only variable that changes over time. Design heights needed to meet risk tolerance targets could be higher or lower if there are changes in these and other variables that impact the damage function. For instance, it has been observed that well designed flood damage reduction strategies can lead to increased development in protected areas due to a greater sense of perceived safety \citep[the ``levee effect'';][]{White1945a}. New development can increase residual flood risk over time. Additionally, from an aesthetic perspective, elevation, levee and surge barrier construction, and retreating from the floodplain could all impact building amenity value. These impacts could be important, but are not considered in our simple, illustrative framework. Third, levees and storm surge barriers can produce storm surge funneling effects that further elevate the water surface. This, as well as the added impact of waves, are not included in our framework and could greatly increase damage allowances. Fourth, damages from permanent inundation are not accounted for. This could have a significant impact on the effectiveness of a specific flood protection strategy. For example, in the case of a storm surge barrier, after sea level has risen above the gate closure threshold, the barrier may have to remain permanently closed to be effective. Finally, if the damage allowances are used to produce benefits for a BCA or cost-effectiveness framework, the usual limitations may still apply \citep[e.g., those given in][]{Arrow1996a}, including that calculations based on human population exposure, rather than financial metrics (e.g., AAL due to flooding), could give greater weight to lower income groups.

Many decision tools exist for making long-lived decisions under deep uncertainty, but they all come with trade-offs, including incompatibilities with current governing systems. Flexible/adaptable solutions have been preferred by some flood risk managers \citep[e.g.,][]{NYC2019c,Ranger2013a}, however, challenges are anticipated with implementing these approaches, including those that are political in nature (see discussion above). On the other hand, a common limitation of expected BCA and cost-effectiveness analysis is that they require future projections of relevant variables that determine benefits of protection over time, some of which are deeply uncertain and cannot be fully represented with a single probability distribution (e.g., sea-level rise, economic development, climate forcing scenarios). To partly address this issue, our flood damage allowance framework accommodates multiple possible AIS melt contributions. While there is value in information to reduce deep uncertainties, until this occurs, approaches to flood protection design that rely on future distributions of relevant variables will require a multi-prior approach to more accurately depict true states of knowledge. 

\newpage
\clearpage

\begin{figure}[h]
\centering
    \includegraphics[width=0.32\textwidth]{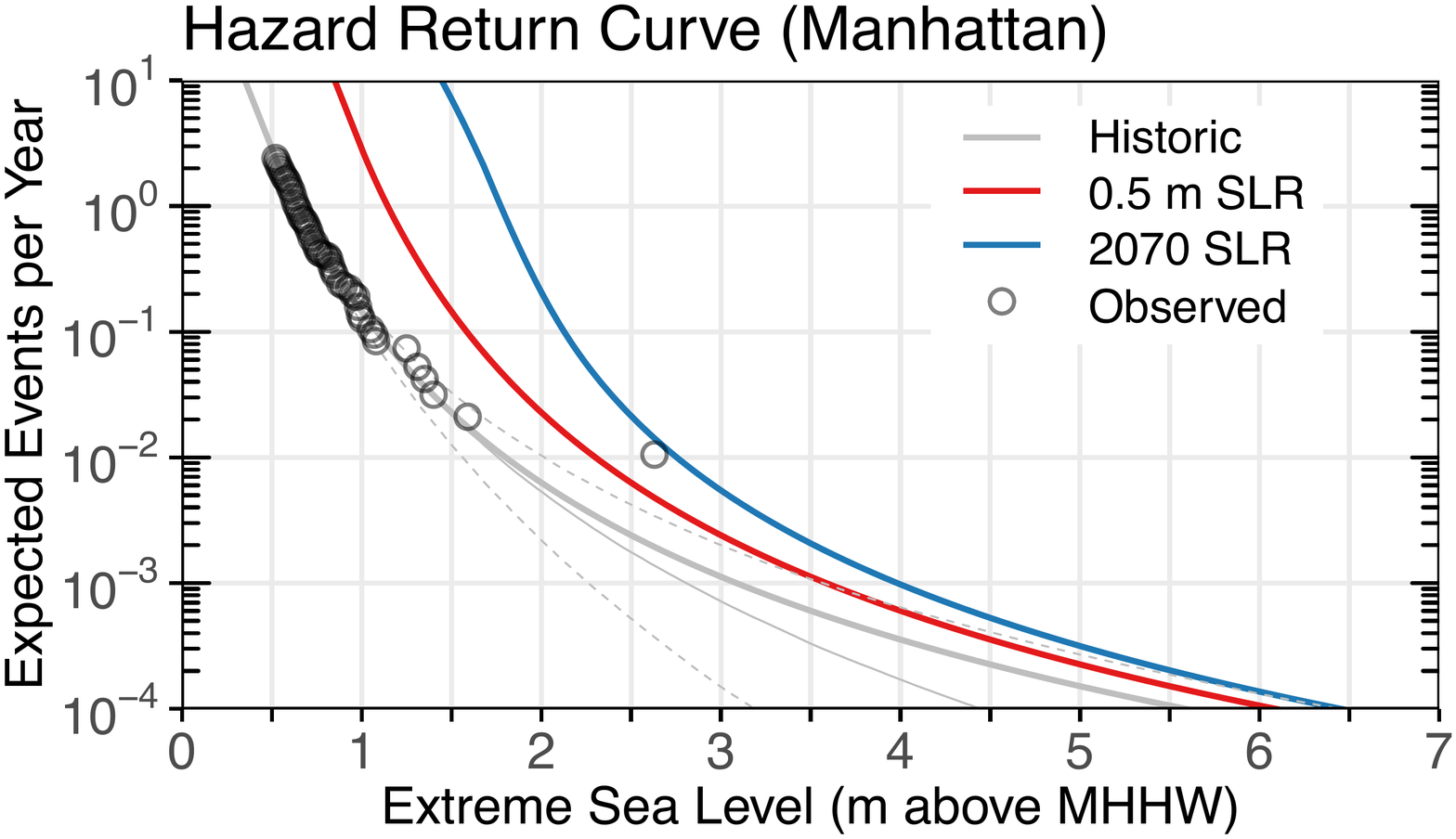}
    \includegraphics[width=0.32\textwidth]{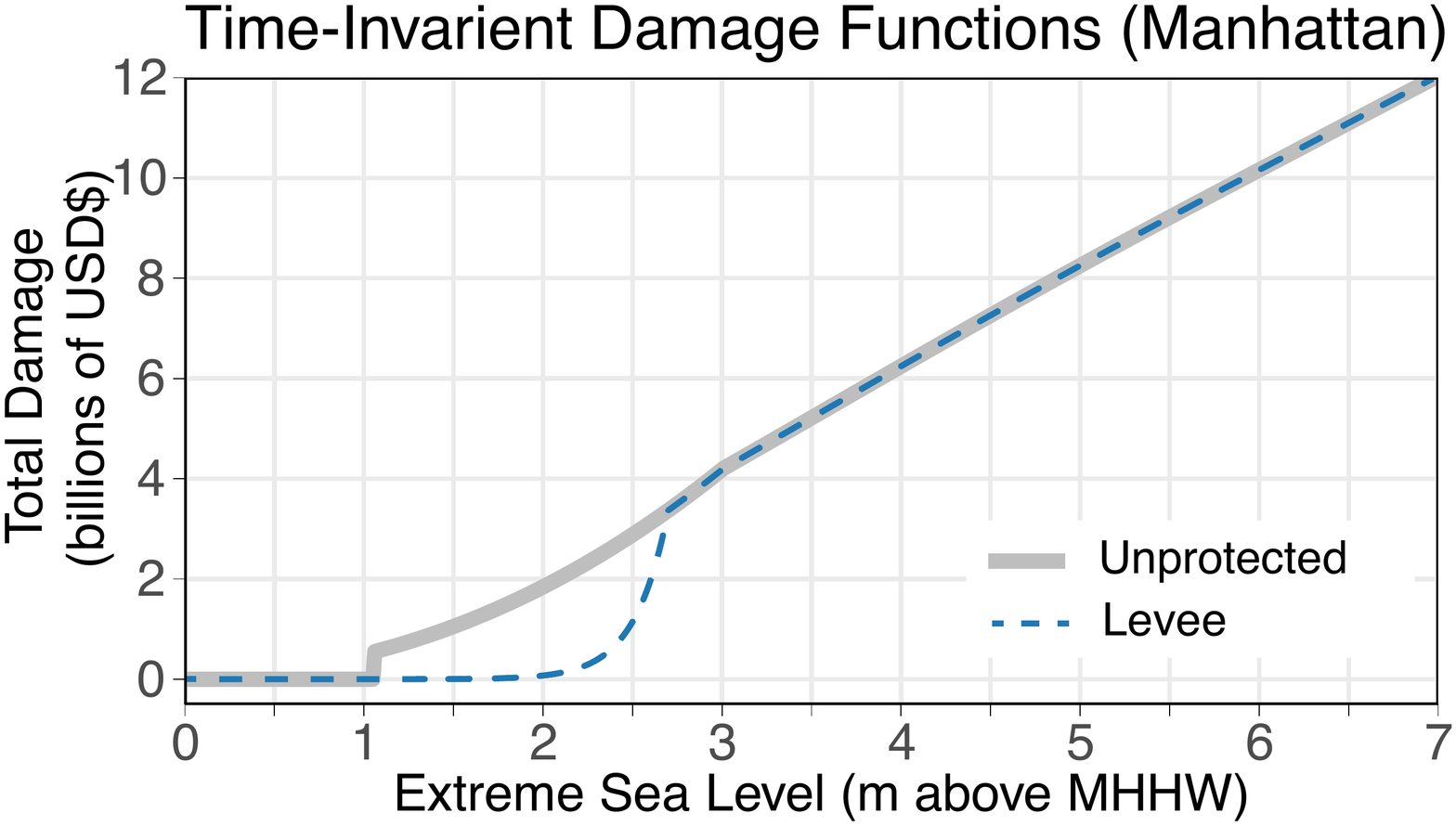}
    \includegraphics[width=0.32\textwidth]{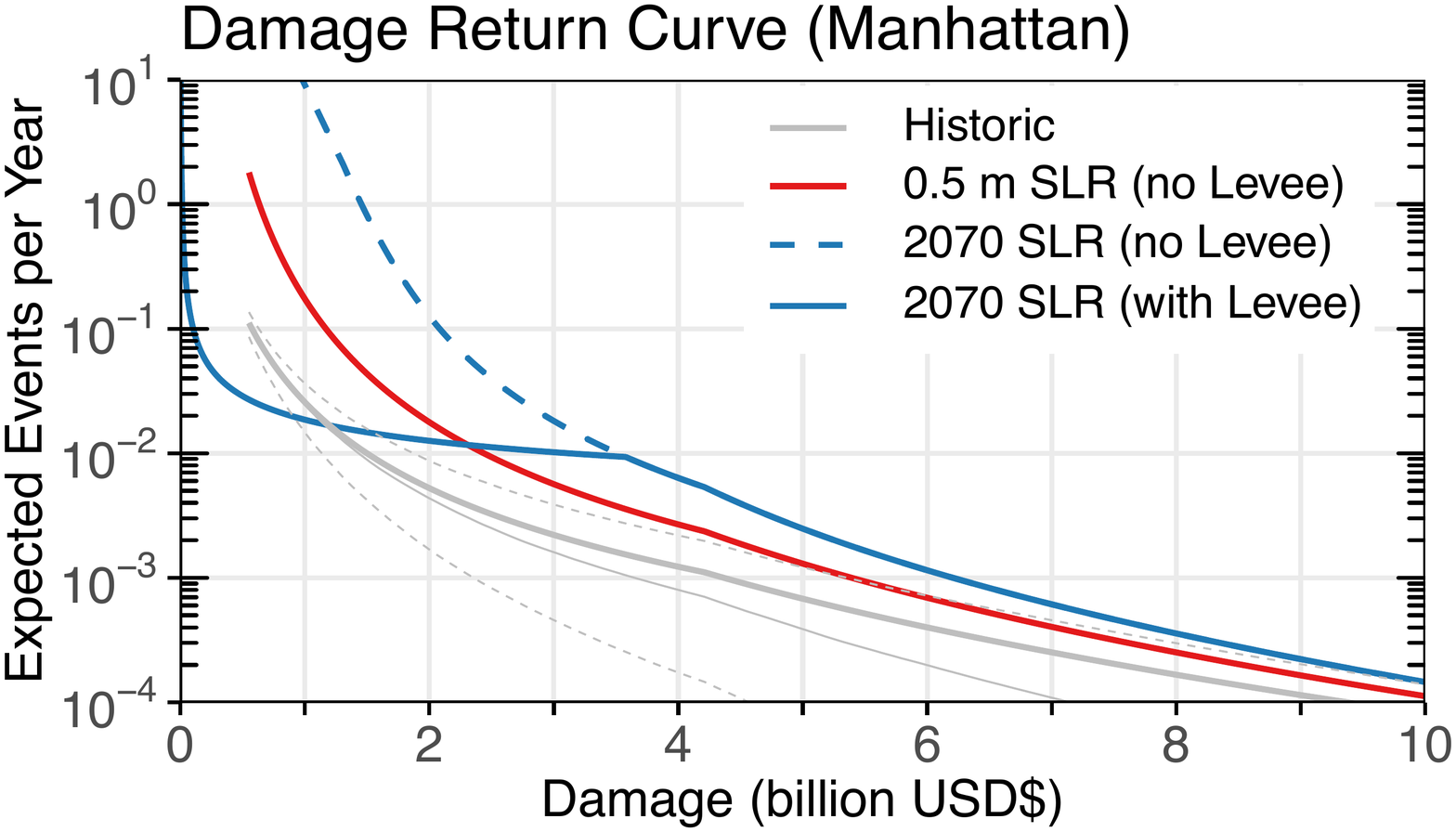}
\caption{\textbf{A.} Expected number of extreme sea level (ESL) events per year as a function of ESL height (m above mean higher high water [MHHW]) at the Battery (Manhattan, New York City) for historical mean sea level (grey lines), 0.5 m of sea-level rise (SLR; red line), and projected SLR in 2070 (blue line). Thin grey lines are the historical ESL height return curves for the 17/50/83 percentiles of the generalized Pareto distribution parameter uncertainty range (dotted/solid/dotted lines, respectively). Tide gauge observations (1920--2014) are plotted as open black circles. \textbf{B.} Direct physical flood damage in Manhattan (billions of 2017 USD\$) as a function of ESL height [meters above mean higher high water (MHHW)] as estimated by a time-invariant flood damage function that assumes a 1.0 m high bulkhead around Manhattan (grey curve; Sec. \ref{Ssec:dmgmod}) and a time-invariant flood damage function that assumes a 1.7 m high levee on top of the bulkhead protecting Manhattan that includes the probability of structural failure below the top of the levee (dashed blue curve; Sec. \ref{sec:protect}). \textbf{C.} As for A, but for the expected number of flood damage events per year (billions of 2017 USD\$) for historical mean sea level (grey lines), 0.5 m of sea-level rise and no added protection (red line), projected SLR in 2070 with no added protection (dash blue line), and projected SLR in 2070 with a 1.7 m levee that maintains the historical annual average loss due to flooding (solid blue line). The projected damages assume that Manhattan's distribution of buildings, people, and infrastructure remains constant in time.} \label{fig:process}
\end{figure}

\newpage
\clearpage

\newpage
\clearpage

\setlength{\tabcolsep}{3pt}
\begin{table}[htbp]
\centering
\setlength{\tabcolsep}{3pt}
\caption{Levee damage allowances (meters above the current protection height) for 2100, 2070, and 2050 under representative concentration pathway (RCP) 8.5 and RCP2.6 and for different assumptions regarding future Antarctic ice sheet (AIS) behavior (e.g, likelihood of AIS collapse [$\beta_{c}$] and maximum 2100 AIS contribution [AIS$_{max}$]). Levee damage allowances include 0.5 m of freeboard and have a 10\% probability of failure at the design height.}
{\small
\begin{tabular}{lcccccc|lccccccc}
\multicolumn{7}{l}{\textbf{Levee  Damage  Allowances}  (m)}  \\
RCP8.5 &     &  \multicolumn{5}{c}{AIS$_{max}$}     &     &  RCP2.6   &     &  \multicolumn{5}{c}{AIS$_{max}$}  \\
 &    $\beta_{c}$     &  1.75  m   &  1.5  m   &  1.0  m     &  0.5  m     &  0.25  m     &     &     &  $\beta_{c}$   &  1.75  m   &  1.5  m   &  1.0  m   &  0.5  m   &  0.25  m  \\
\hline
2100   &   &   &   &   &   &   &   &  2100   &   &   &   &   &   &  \\
 &  1.0   &3.4   &  3.1   &  2.6   &  2.2   &  2.0   &   &   &  1.0   &1.9   &  1.7   &  1.5   &  1.4   &  1.3  \\
 &  0.75   &3.1   &  2.8   &  2.4   &  2.1   &  2.0   &   &   &  0.75   &1.7   &  1.6   &  1.5   &  1.4   &  1.3  \\
 &  0.50   &2.8   &  2.5   &  2.2   &  2.0   &  1.9   &   & &  0.50   &1.5   &  1.5   &  1.4   &  1.3   &  1.2  \\
 &  0.25   &2.6   &  2.2   &  2.1   &  1.9   &  1.9   &   & &  0.25   &1.4   &  1.4   &  1.4   &  1.3   &  1.2  \\
 &  0.0   &2.6   &  2.2   &  2.0   &  1.9   &  1.8   &   & &  0.0   &1.4   &  1.4   &  1.4   &  1.3   &  1.2  \\
2070   &   &   &   &   &   &   &   &  2070   &   &   &   &   &   &  \\
 &  1.0   &1.6   &  1.5   &  1.4   &  1.2   &  1.1   &   &   &  1.0   &1.1   &  1.1   &  1.1   &  1.1   &  1.0  \\
 &  0.75   &1.5   &  1.4   &  1.3   &  1.2   &  1.1   &   &   &  0.75   &1.1   &  1.1   &  1.1   &  1.0   &  1.0  \\
 &  0.50   &1.4   &  1.4   &  1.3   &  1.2   &  1.1   &   & &  0.50   &1.1   &  1.1   &  1.0   &  1.0   &  1.0  \\
 &  0.25   &1.3   &  1.3   &  1.2   &  1.2   &  1.2   &   & &  0.25   &1.0   &  1.0   &  1.0   &  1.0   &  1.0  \\
 &  0.0   &1.2   &  1.2   &  1.2   &  1.2   &  1.2   &   & &  0.0   &1.0   &  1.0   &  1.0   &  1.0   &  0.9  \\
2050   &   &   &   &   &   &   &   &  2050   &   &   &   &   &   &  \\
 &  1.0   &0.9   &  0.9   &  0.8   &  0.8   &  0.8   &   &   &  1.0   &0.8   &  0.8   &  0.8   &  0.8   &  0.8  \\
 &  0.75   &0.9   &  0.9   &  0.8   &  0.8   &  0.8   &   &   &  0.75   &0.8   &  0.8   &  0.8   &  0.8   &  0.8  \\
 &  0.50   &0.9   &  0.9   &  0.8   &  0.8   &  0.8   &   & &  0.50   &0.8   &  0.8   &  0.8   &  0.8   &  0.8  \\
 &  0.25   &0.8   &  0.8   &  0.8   &  0.8   &  0.8   &   & &  0.25   &0.8   &  0.8   &  0.8   &  0.8   &  0.7  \\
 &  0.0   &0.8   &  0.8   &  0.8   &  0.8   &  0.8   &   & &  0.0   &0.8   &  0.8   &  0.8   &  0.7   &  0.7  \\
\hline
\end{tabular}}
\label{tab:DmgAllwncLevee}
\end{table}

\newpage

\begin{table}[htbp]
\centering
\setlength{\tabcolsep}{3pt}
\caption{Coastal retreat damage allowances (meters above the current protection height) for 2100, 2070, and 2050 under representative concentration pathway (RCP) 8.5 and RCP2.6 and for different assumptions regarding future Antarctic ice sheet (AIS) behavior (e.g, likelihood of AIS collapse [$\beta_{c}$] and maximum 2100 AIS contribution [AIS$_{max}$]). Assumes perfect compliance of coastal retreat (i.e., $\alpha$ =  1; Sec. \ref{sec:retreat}).}
{\small
\begin{tabular}{lcccccc|lccccccc}
\multicolumn{7}{l}{\textbf{Coastal  Retreat  Damage  Allowances}  (m)}  \\
RCP8.5 &     &  \multicolumn{5}{c}{AIS$_{max}$}     &     &  RCP2.6   &     &  \multicolumn{5}{c}{AIS$_{max}$}  \\
 &    $\beta_{c}$     &  1.75  m   &  1.5  m   &  1.0  m     &  0.5  m     &  0.25  m     &     &     &  $\beta_{c}$   &  1.75  m  &  1.5  m   &  1.0  m   &  0.5  m   &  0.25  m  \\
\hline
2100   &   &   &   &   &   &   &   &  2100   &   &   &   &   &   &  \\
 &  1.0   &2.6   &  2.4   &  1.9   &  1.5   &  1.3   &   &   &  1.0   &1.2   &  1.1   &  0.9   &  0.7   &  0.6  \\
 &  0.75   &2.3   &  2.0   &  1.7   &  1.4   &  1.3   &   &   &  0.75   &1.0   &  0.9   &  0.8   &  0.7   &  0.6  \\
 &  0.50   &2.0   &  1.7   &  1.5   &  1.3   &  1.2   &   & &  0.50   &0.8   &  0.8   &  0.8   &  0.7   &  0.6  \\
 &  0.25   &1.8   &  1.5   &  1.4   &  1.2   &  1.2   &   & &  0.25   &0.8   &  0.8   &  0.7   &  0.6   &  0.6  \\
 &  0.0   &1.8   &  1.5   &  1.4   &  1.2   &  1.1   &   & &  0.0   &0.7   &  0.7   &  0.7   &  0.6   &  0.6  \\
2070   &   &   &   &   &   &   &   &  2070   &   &   &   &   &   &  \\
 &  1.0   &0.9   &  0.8   &  0.7   &  0.6   &  0.4   &   &   &  1.0   &0.5   &  0.4   &  0.4   &  0.4   &  0.3  \\
 &  0.75   &0.8   &  0.7   &  0.7   &  0.5   &  0.5   &   &   &  0.75   &0.4   &  0.4   &  0.4   &  0.4   &  0.3  \\
 &  0.50   &0.7   &  0.7   &  0.6   &  0.5   &  0.5   &   & &  0.50   &0.4   &  0.4   &  0.4   &  0.3   &  0.3  \\
 &  0.25   &0.6   &  0.6   &  0.6   &  0.5   &  0.5   &   & &  0.25   &0.4   &  0.4   &  0.4   &  0.3   &  0.3  \\
 &  0.0   &0.6   &  0.6   &  0.5   &  0.5   &  0.5   &   & &  0.0   &0.3   &  0.3   &  0.3   &  0.3   &  0.3  \\
2050   &   &   &   &   &   &   &   &  2050   &   &   &   &   &   &  \\
 &  1.0   &0.3   &  0.2   &  0.2   &  0.2   &  0.1   &   &   &  1.0   &0.2   &  0.2   &  0.2   &  0.2   &  0.1  \\
 &  0.75   &0.2   &  0.2   &  0.2   &  0.2   &  0.2   &   &   &  0.75   &0.2   &  0.2   &  0.2   &  0.1   &  0.1  \\
 &  0.50   &0.2   &  0.2   &  0.2   &  0.2   &  0.2   &   & &  0.50   &0.2   &  0.2   &  0.2   &  0.1   &  0.1  \\
 &  0.25   &0.2   &  0.2   &  0.2   &  0.2   &  0.2   &   & &  0.25   &0.1   &  0.1   &  0.1   &  0.1   &  0.1  \\
 &  0.0   &0.2   &  0.2   &  0.2   &  0.2   &  0.2   &   & &  0.0   &0.1   &  0.1   &  0.1   &  0.1   &  0.1  \\
\hline
\end{tabular}}
\label{tab:DmgAllwncRetreat}
\end{table}

\clearpage

\appendix

\title{Supporting Information for ``A flood damage allowance framework for coastal protection with deep uncertainty in sea-level rise''}
%
%

%
%



\authors{D.J. Rasmussen\affil{1}, Maya K. Buchanan\affil{2}, Robert E. Kopp\affil{3}, Michael Oppenheimer\affil{4}}

\affiliation{1}{Woodrow Wilson School of Public and International Affairs, Princeton University, Princeton, NJ, USA}
\affiliation{2}{Climate Central, Princeton, NJ, USA}
\affiliation{3}{Department of Earth \& Planetary Sciences, Rutgers Energy Institute and Institute of Earth, Ocean, and Atmospheric Sciences, Rutgers University, New Brunswick, NJ, USA}
\affiliation{4}{Department of Geosciences and Woodrow Wilson School of Public and International Affairs, Princeton University, Princeton, NJ, USA}

\correspondingauthor{D.J. Rasmussen}{dj.rasmussen@princeton.edu}

\section{Supplementary Methods}

\begin{figure}[h]
\centering
    \includegraphics[width=.80\textwidth]{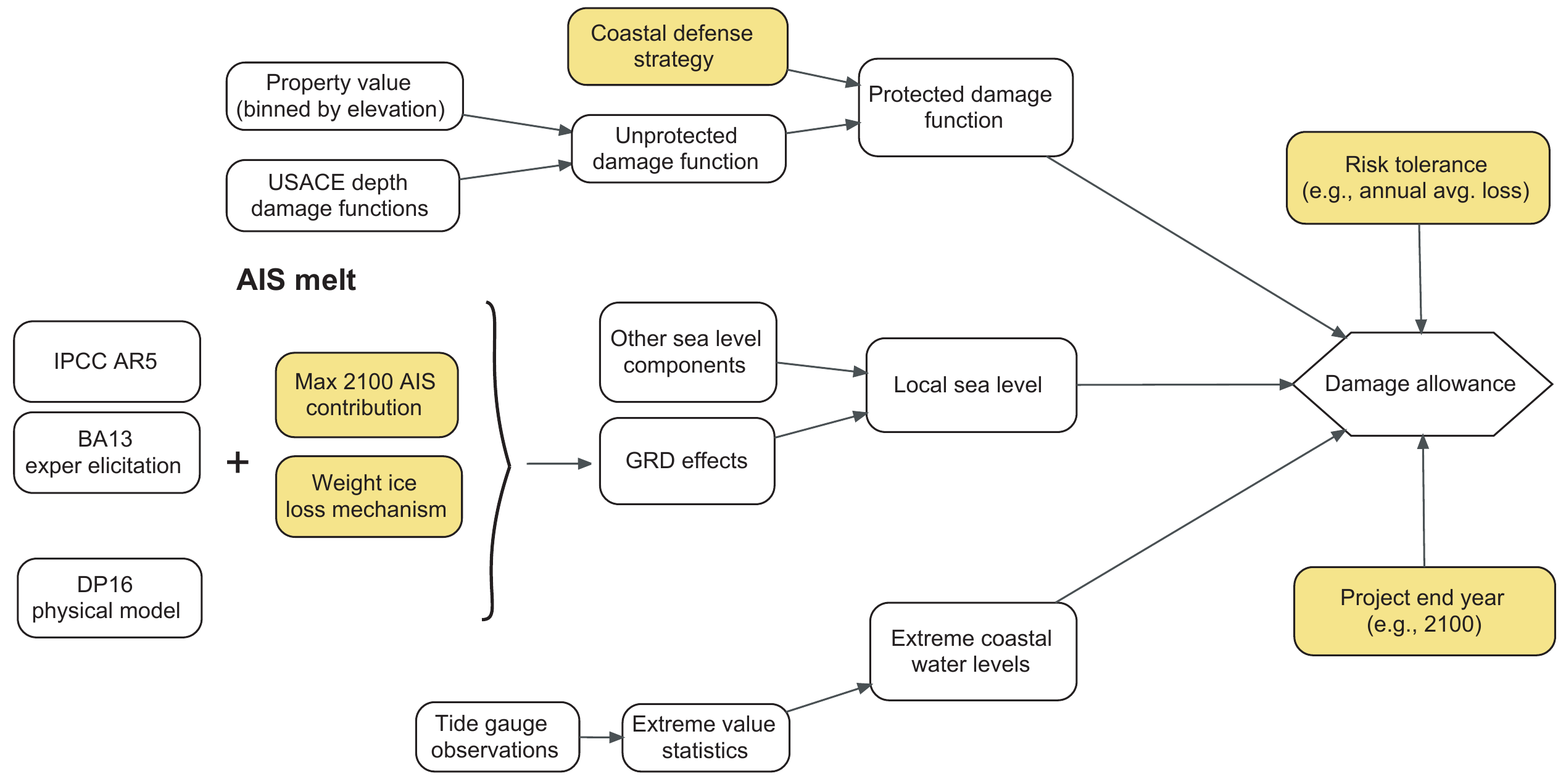}
\caption{Logical flow of sources of information used in damage allowance calculation. AIS is Antarctic Ice Sheet; USACE is U.S. Army Corps of Engineers; BA13 is \citet{Bamber2013a}; DP16 is \citet{Deconto2016a}; GRD are gravitational, rotational, and deformational effects; ``Other sea level components'' includes land water storage, Greenland ice sheet melt, glacier ice melt, oceanographic processes, and non-climatic background changes, such as glacial-isostatic adjustment (GIA).} \label{Sfig:methods}
\end{figure}

\subsection{Extreme sea level damage model}\label{Ssec:simpleMod}

We model the annual average loss (AAL) due to extreme sea level (ESL) damage as the average loss (insured and uninsured) of all modeled ESL damage events $D(z)$, weighted by the annual probability of occurrence $f(z)$. This can mathematically be written as,

\begin{equation}
\mathop{{\mathbb{E}}[D(z)]=\int_{z}}D(z)f(z)\, \text{d}z. \label{eq:aal}
\end{equation}

\noindent While this is a one-dimensional model (vertical direction only), we note that spatial variation in coastal protection and ESL event frequency could be accommodated using a 3-dimensional damage model where the ESL parameter (\textbf{z}) is a vector that various horizontally (i.e., $x$,$y$-direction) as well as vertically (i.e., $z$-direction).

\subsection{Modeling the frequency of extreme water levels} \label{Ssec:ESL}

Following \citet{Buchanan2016a} and \citet{Rasmussen2018a}, we estimate the probability of ESLs in Manhattan using extreme value theory and daily maximum sea levels calculated from quality-controlled tide gauge records from the University of Hawaii Sea Level Center\footnote{retrieved from: https://uhslc.soest.hawaii.edu, June 2017; \citet{Caldwell2015}}. While the frequency of extreme water levels varies though out the New York-New Jersey Harbor Estuary due to hydrological factors such as coastal bathymetry and topography, we simplify this spatial variation by using a single, long-standing tide gauge (1920--2014), located at the Battery in lower Manhattan, and assume it is representative of the ESL frequency experienced through out the entire borough. Tide gauge observations are de-trended to remove the effect of SLR and referenced to mean higher high water (MHHW)\footnote{Here defined as the average level of high tide over the last 19-years in the tide gauge record, which is different from the current U.S. National Tidal Datum Epoch of 1983--2001.}. The annual probability of ESLs $f(z)$ is modeled with a generalized Pareto distribution (GPD)\citep{Coles2001a,Coles2001b} given by:

\begin{equation}
f_{(\xi,\mu,\sigma)}(z) = \frac{1}{\sigma}\left(1 + \frac{\xi (z-\mu)}{\sigma}\right)^{\left(-\frac{1}{\xi} - 1\right)} \label{eq:freq}
\end{equation}

\noindent for $z \geqslant \mu$ when $\xi \geqslant 0$, and $\mu \leqslant z \leqslant \mu - \sigma /\xi$ when $\xi < 0$. The GPD parameters are the following: the shape parameter ($\xi$) governs the curvature and upward statistical limit of the ESL probability distribution function (PDF) and embodies the local coastal storm climate, the scale parameter ($\sigma$) characterizes the annual variability in the maxima of tides and storm surges, and the location parameter ($\mu$) is the threshold water-level above which return levels are estimated with the GPD\textemdash here the 99th percentile of daily maximum sea levels, which is generally above the highest seasonal tide, balances the bias-variance trade-off in the GPD parameter estimation \citep{Tebaldi2012a} and has been found to perform well at global scales \citep{Wahl2017a}. Daily maximum sea levels above 99th percentile are de-clustered to meet the statistical independence assumption of the GPD. The GPD parameters are estimated using the method of maximum likelihood. Uncertainty in the GPD parameters is calculated from their estimated covariance matrix and is sampled using Latin hypercube sampling of 1000 normally distributed GPD parameter pairs. The GPD parameters are given in Table \ref{Stab:gpd}. Events that occur outside of the support of the GPD (i.e., below $\mu$) are modeled with a Gumbel distribution (e.g., tidal floods). Other probability mixture approaches have been presented elsewhere \citep[e.g.,][]{Ghanbari2019a}. The historical flood return curve at the Battery tide gauge is presented in Fig. \ref{fig:process}A.

\begin{table}[h]
\begin{tabular}{lcccccccccc}
Site               & Lat  & Lon    & Uhawaii ID & Start & End  & Length (yrs) & $\lambda$ & $\mu$ (m)  & $\xi$    & $\sigma$            \\
\hline
Battery & 40.7 & -74.15 & 745a       & 1920  & 2014 & 95  & 2.63   & 0.51 & 0.19 (0.05, 0.33) & 0.13 (0.10, 0.15)
\end{tabular}
\caption{Generalized Pareto distribution (GPD) parameters estimated for the Battery tide gauge in Manhattan (New York City; Sec. \ref{Ssec:ESL}). The GPD threshold ($\mu$) is given as meters above mean higher high water (MHHW) Both the shape ($\xi$) and scale ($\sigma$) parameters of the GPD are given as 50th (5th/95th) percentiles. Parameters are estimated using the method of maximum likelihood.} \label{Stab:gpd}
\end{table}

\subsection{Modeling damages from extreme water levels}\label{Ssec:dmgmod}

\begin{figure}[h]
\centering
    \includegraphics[width=.48\textwidth]{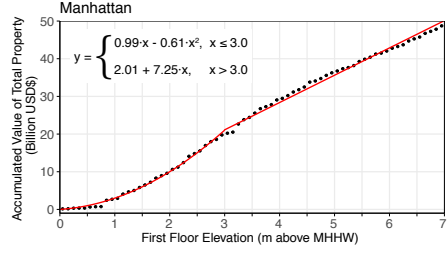}
\caption{Accumulated tax-assessed value of Manhattan property (building only, land excluded; billion USD\$) by first floor building elevation [meters above mean higher high water (MHHW); black circles] and piece-wise fit using a quadratic function below 3 m and linear function above 3 m (red line). Property data is from the New York City Department of City Planning \citep{NYCpluto2018a}.} \label{Sfig:accprop}
\end{figure}

Following the methodology from \citet{Diaz2016a}, we construct a 1-dimensional ($z$-direction), aggregate ESL damage function for Manhattan by integrating damages from the lowest unprotected elevation $e_{min}$ to an ESL height $z$ using:

\begin{equation}
\text{D(z)}=\int_{e_{min}}^{z}p(e) \cdot \phi (z-e) \text{d}e \label{eq:damage}
\end{equation}

\noindent where $p(e)$ is the total tax assessed value of all buildings at the estimated first floor elevation $e$ from the NYC Department of City Planning \citep[Fig. \ref{Sfig:accprop};][]{NYCpluto2018a}\footnote{Note that the NYC Department of City Planning makes available both tax assessed building value and combined tax assessed building and property value. We assume that floods only damage structures and not the land itself.}, $z$ is the ESL height, and $\phi(z-e)$ is an aggregate inundation depth-damage function for Manhattan that relates the flood height (i.e., $z-e$) to damage as a fraction of the total tax assessed building value (see Sec. \ref{Ssec:depthdamage}). The first floor structure elevation was estimated in 0.1 m increments starting from 0 m above the North American Vertical Datum of 1988 (NAVD88) using a 0.3 m vertical resolution LIDAR-derived digital elevation model (DEM) from the City of New York\footnote{https://data.cityofnewyork.us/City-Government/1-foot-Digital-Elevation-Model-DEM-/dpc8-z3jc}. Other covariates that may cause damage, such as wind gusts, waves, and precipitation, are not included. Also, not included in our damage accounting is the loss of human life, damage to infrastructure (both above and below ground), the value lost from permanently inundated lands, and indirect damage effects such as business interruption. Figure \ref{fig:process}B shows the damage function. The damage function is then multiplied by ESL probability distribution (Sec. \ref{Ssec:ESL}) to give a probability distribution of flood damages (Fig. \ref{fig:process}C).

\subsection{Depth-damage functions} \label{Ssec:depthdamage}

We employ structure-specific depth-damage functions constructed from expert elicitation \citep{USACE2015a} that relate the fraction of structure damage solely to the depth of the flood. We do not consider other potentially important flood characteristics that may cause more damage, such as flood velocity, wave height, and flood duration \citep{Merz2010a}. Manhattan (New York City) is comprised of several thousand classes of structures that may have distinct relationships between inundation depth and damage. New York City property tax assessments indicate that Manhattan is comprised of roughly 95 percent residential and 5 percent commercial \citep{NYC2018a}. Accordingly, we reduce the complexity of building type to three inundation depth-damage functions that represent these classes of buildings, high-rises with basements (95 percent of Manhattan) and two-story residences with basements (5 percent of Manhattan).

\begin{figure}[htb]
\begin{minipage}{.45\textwidth}
\centering
\includegraphics[width=.99\textwidth]{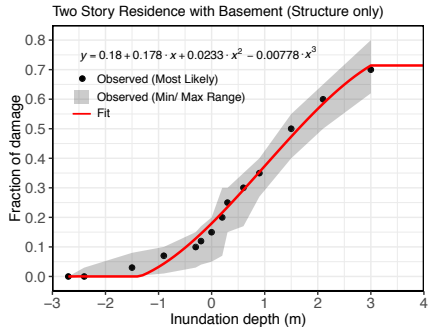}
\caption{Observed depth-damage relationship for two story residences with basements (structure only) from \citet{USACE2015a}. The contents of the structure are not included. A 3rd-order polynomial is fit through the observed (most likely) values (red line).}\label{Sfig:dd1}
\end{minipage}\hfill
\begin{minipage}{.55\textwidth}
    \centering 
    \begin{tabular}{cc|ccc}
    \multicolumn{5}{l}{\textbf{Two Story Residence with Basement (Structure only)} } \\
    \multicolumn{1}{c}{Depth (ft)} & \multicolumn{1}{c|}{Depth (m)} & \multicolumn{1}{c}{Min} & \multicolumn{1}{c}{Most Likely} & \multicolumn{1}{c}{Max} \\
    \hline
    -9.0  & -2.7  & 0.000 & 0.000 & 0.000 \\
    -8.0  & -2.4  & 0.000 & 0.000 & 0.030 \\
    -5.0  & -1.5  & 0.000 & 0.030 & 0.080 \\
    -3.0  & -0.9  & 0.010 & 0.070 & 0.100 \\
    -1.0  & -0.3  & 0.030 & 0.100 & 0.150 \\
    -0.5  & -0.2  & 0.040 & 0.120 & 0.170 \\
    0.0   & 0.0   & 0.050 & 0.150 & 0.200 \\
    0.5   & 0.2   & 0.070 & 0.200 & 0.300 \\
    1.0   & 0.3   & 0.150 & 0.250 & 0.300 \\
    2.0   & 0.6   & 0.170 & 0.300 & 0.350 \\
    3.0   & 0.9   & 0.270 & 0.350 & 0.400 \\
    5.0   & 1.5   & 0.400 & 0.500 & 0.550 \\
    7.0   & 2.1   & 0.500 & 0.600 & 0.650 \\
    10.0  & 3.0   & 0.620 & 0.700 & 0.800 \\
    \end{tabular}%
\end{minipage} 
\end{figure}

\begin{figure}[htb]
\begin{minipage}{.45\textwidth}
\centering
\includegraphics[width=.99\textwidth]{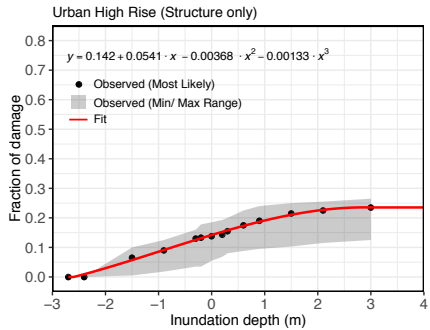}
\caption{Observed depth-damage relationship for an urban high-rise (structure only) from \citet{USACE2015a}. A 3rd-order polynomial is fit through the observed (most likely) values (red line).} \label{Sfig:dd2}
\end{minipage}\hfill
\begin{minipage}{.55\textwidth}
    \centering 
    \begin{tabular}{cc|ccc}
    \multicolumn{5}{l}{\textbf{Urban High Rise (Structure only)} } \\
    \multicolumn{1}{c}{Depth (ft)} & \multicolumn{1}{c|}{Depth (m)} & \multicolumn{1}{c}{Min} & \multicolumn{1}{c}{Most Likely} & \multicolumn{1}{c}{Max} \\
    \hline
    -9.0  & -2.7  & 0.000 & 0.000 & 0.000 \\
    -8.0  & -2.4  & 0.000 & 0.000 & 0.000 \\
    -5.0  & -1.5  & 0.005 & 0.065 & 0.100 \\
    -3.0  & -0.9  & 0.018 & 0.090 & 0.125 \\
    -1.0  & -0.3  & 0.035 & 0.130 & 0.160 \\
    -0.5  & -0.2  & 0.035 & 0.133 & 0.178 \\
    0.0   & 0.0   & 0.055 & 0.138 & 0.185 \\
    0.5   & 0.2   & 0.068 & 0.143 & 0.193 \\
    1.0   & 0.3   & 0.080 & 0.155 & 0.200 \\
    2.0   & 0.6   & 0.088 & 0.175 & 0.225 \\
    3.0   & 0.9   & 0.095 & 0.190 & 0.240 \\
    5.0   & 1.5   & 0.103 & 0.215 & 0.250 \\
    7.0   & 2.1   & 0.115 & 0.225 & 0.255 \\
    10.0  & 3.0   & 0.125 & 0.235 & 0.265 \\
    \end{tabular}%
\end{minipage} 
\end{figure}

The depth-damage function for an urban high-rise ($\phi_{hrise}(z-e)$) is shown in SI Fig. \ref{Sfig:dd2}, and the least-squares fit is given by:

\[
    \phi_{hrise}(z-e) = 
\begin{cases}
    0.142 + 0.0541\cdot (z-e) - 0.00368\cdot (z-e)^2 - 0.00133 \cdot (z-e)^3, & \text{if } z > e \\
    0,              & \text{otherwise}
\end{cases}
\]

\noindent where $z$ is the height of the extreme water level, $e$ is the first-floor elevation of the structure, and $z-e$ is the flood height. The depth-damage function for a two-story residential structure with a basement ($\phi_{res}(z-e)$) is shown in SI Fig. \ref{Sfig:dd1}, and the least-squares fit is given by,

\[
    \phi_{res}(z-e) = 
\begin{cases}
    0.18 + 0.178\cdot (z-e) + 0.0233\cdot (z-e)^2 - 0.00778 \cdot (z-e)^3, & \text{if } z > e \\
    0,              & \text{otherwise}
\end{cases}
\]

%
\noindent  An aggregate depth-damage function for Manhattan is constructed using the weighted average of the equations for each building class,


\[
\phi(z-e) = 0.95\cdot\phi_{hrise}(z-e)) + 0.05\cdot\phi_{res}(z-e).
\]

\newpage
\clearpage

\subsection{Elevation of all structures by same height} \label{Ssec:elevAll}

\begin{figure}[htb]
\centering
    \includegraphics[width=.66\textwidth]{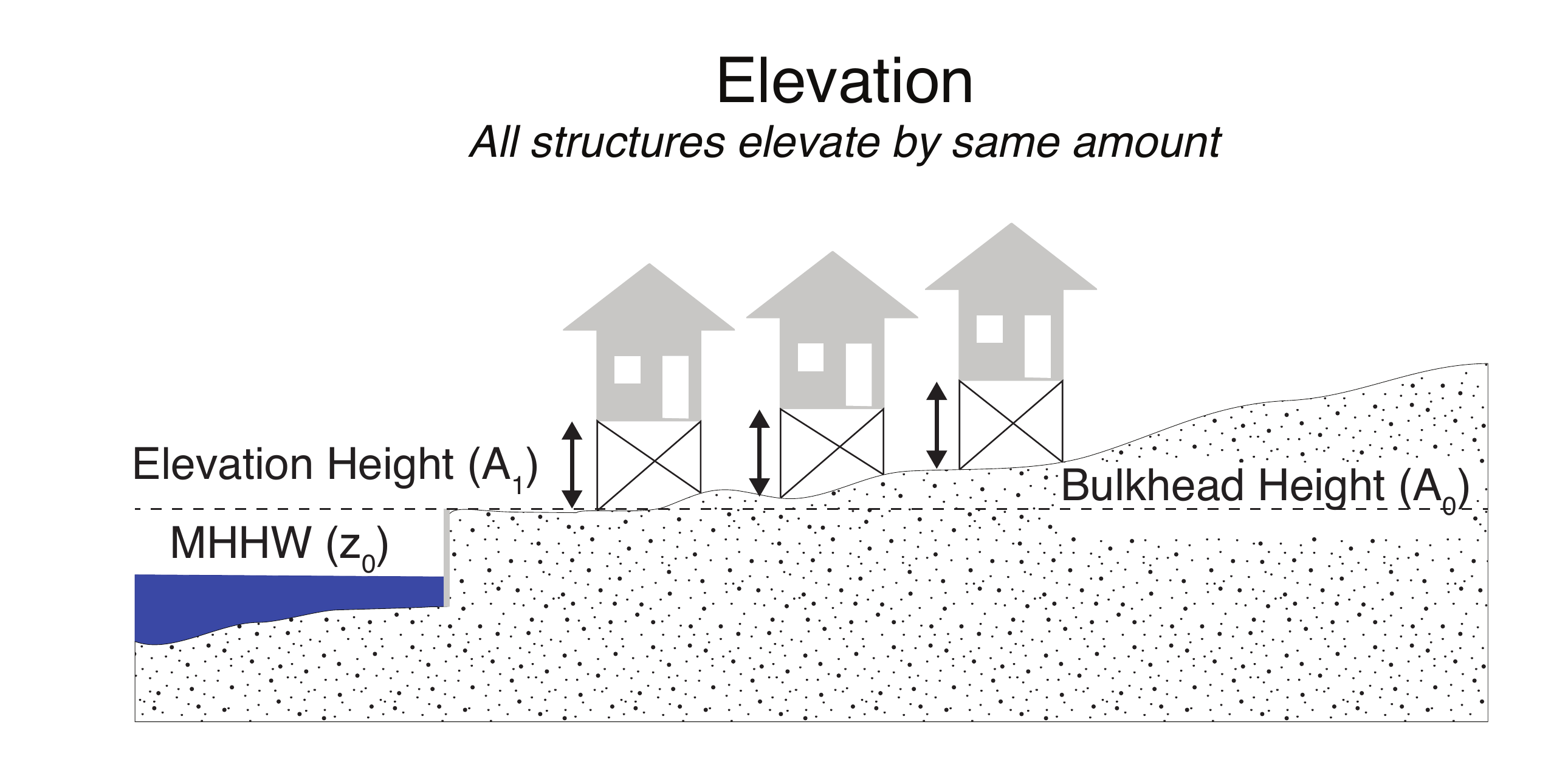} 
  \caption{Schematic illustrating an elevation flood defense strategy (all structures) for an arbitrary design height (A$_{1}$).} \label{Sfig:elevAll}
\end{figure}

While not used for Manhattan due to the impracticality of elevating high-rises, we present an method for elevation all structures within the damage function (Fig. \ref{Sfig:elevAll}). If $A$ is the vertical height that all structures would need to be elevated in order to maintain the current AAL under uncertain sea-level rise and if $\alpha$ is the fraction of assets [0,1] that have elevated by $A$ (i.e., the elevation compliance), then the protected damage function is:

\begin{equation}
D^{*}(z,A) = \underbrace{\alpha D(z-A)}_{\text{Damage to elevated structures}}  + \underbrace{(1-\alpha)D(A).}_{\text{Damage to non-elevated structures}} \label{eq:elevation2}
\end{equation} 

\noindent The elevation of all structures is mathematically represented as a horizontal shift of the ``unprotected'' damage function by $A$ to represent the uniform elevation of all assets by $A$.

\subsection{Combined flood protection strategy approach} \label{Ssec:multi}

A multi-strategy approach to flood defenses may provide an added level of safety through redundancy. For example, if a single levee fails, the area behind the levee is impacted. A second line of defense could compensate for failures of the first. If multiple strategies are employed, users could either assign a fraction of the total risk target to mitigate for each strategy or specify a damage allowance for all but one mitigation strategy and then solve for the unknown damage allowance. For example, if a user desires to maintain the current AAL using both coastal retreat and a levee, they may choose to retreat coastal assets below a pre-determined elevation $A_{1}$ that is also the base of the levee (e.g., $A_{1}$ = 1.0 m) and then solve for the height of the levee $A_{2}$ (Fig. \ref{Sfig:multi}A). This can mathematically be described by:

\begin{multline}
\underbrace{\int_{z_{min}}^{A_1}\int_{\Delta}D^{*}_{r}(z)f(z-\Delta)P(\Delta)\, \text{d}\Delta\, \text{d}z}_{\text{Damages below retreat elevation}}  + 
\underbrace{\int_{A_1}^{\infty}\int_{\Delta}D^{*}_{l}(z)f(z-\Delta)P(\Delta)\, \text{d}\Delta\, \text{d}z}_{\text{Damages from levee failure and overtopping}} = \underbrace{\int_{z_{min}}^{\infty}D(z)f(z)\, \text{d}z.}_{\text{Current AAL}}
\end{multline}

\newpage
\clearpage

\begin{figure}[htb]
\centering
    \includegraphics[width=.33\textwidth]{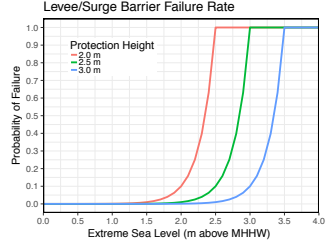} 
  \caption{Fragility curves showing the relationship between structural loading on a levee or storm surge barrier from extreme sea levels (meters above mean higher high water [MHHW]) and the conditional probability of structural failure of the levee or storm surge barrier for protection design heights of 2.0 m (red), 2.5 m (green), 3.0 m (blue), all with 0.5 m of freeboard above the design height. For all, the structural failure rate for extreme sea levels at the design height is 0.10.} \label{Sfig:failrate}
\end{figure}


\begin{figure}[htb]
\centering
    \includegraphics[width=.48\textwidth]{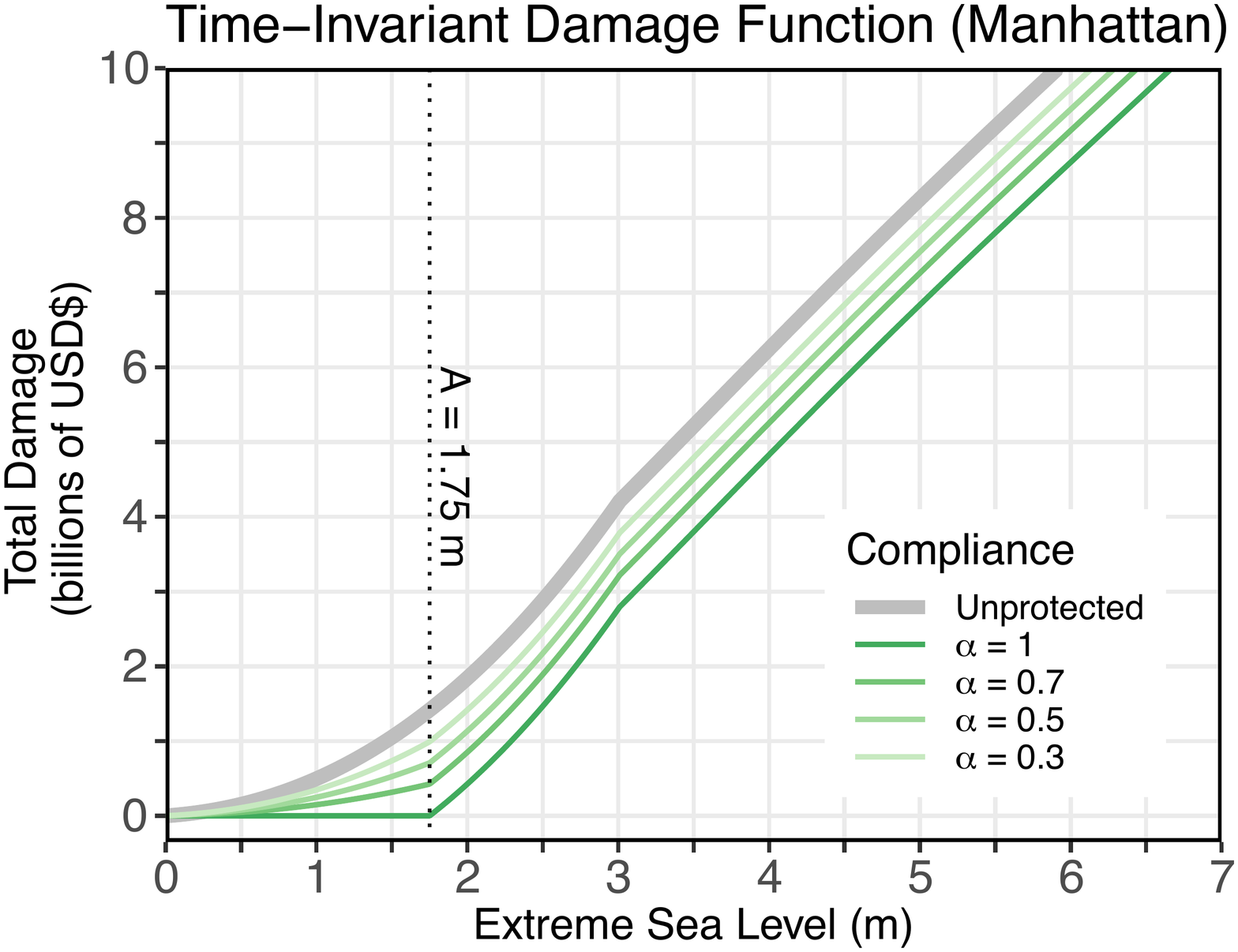} 
    \includegraphics[width=.48\textwidth]{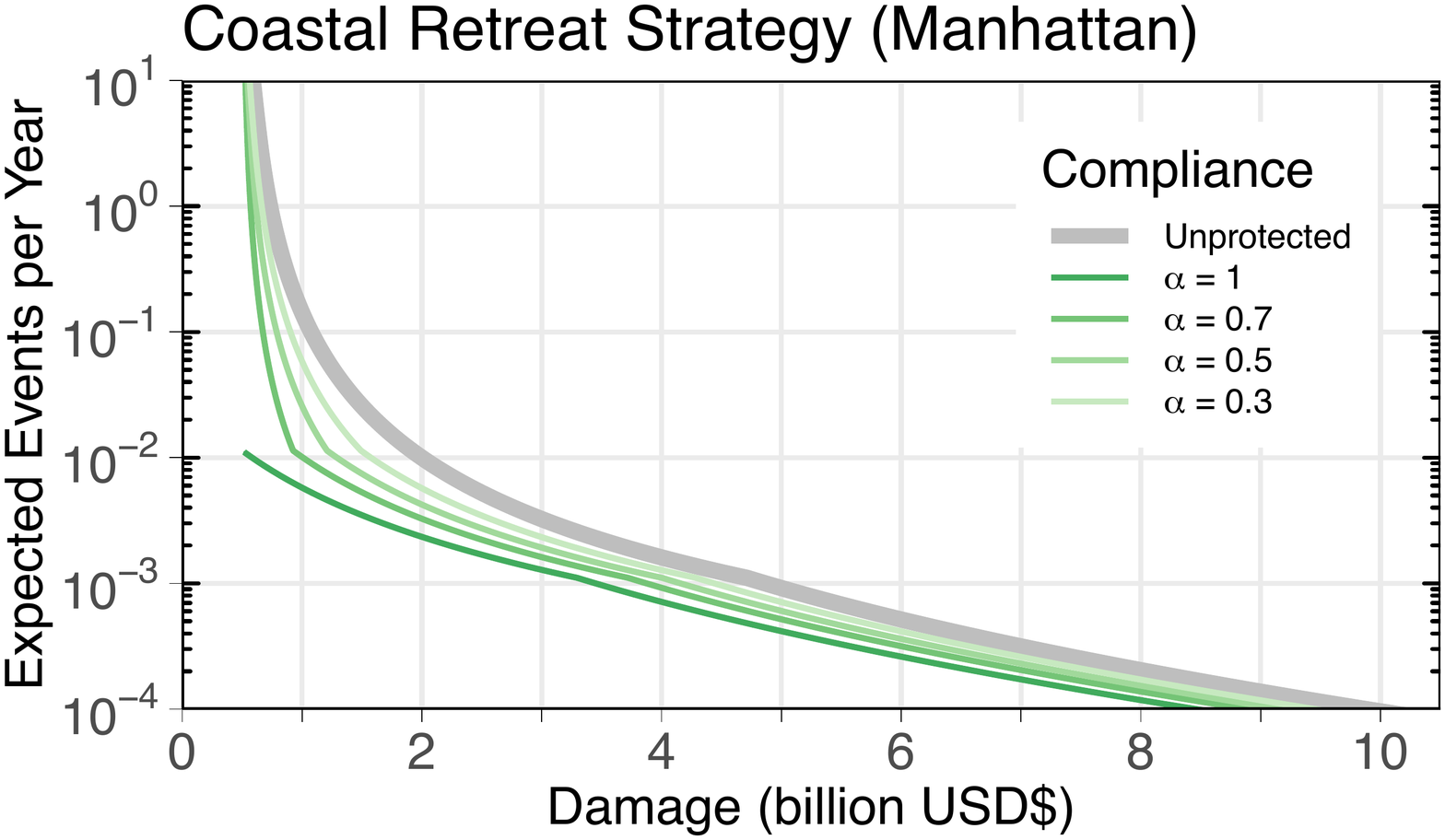} 
  \caption{\textbf{(Left)} Expected number of damage events per year for Manhattan assuming no change in protection strategy (grey line) and using an elevation strategy (1.75 m) with various levels of compliance ($\alpha$; red lines) \textbf{(Right)} As for Left, but for a coastal retreat strategy (1.75 m; green lines)}\label{Sfig:retreat_compli}
\end{figure}

\newpage
\clearpage

\subsection{Sea-level rise projections} \label{Ssec:slr}
Probabilistic, time-varying, local relative sea level (RSL) projections for Manhattan (modeled at the Battery tide gauge\footnote{https://tidesandcurrents.noaa.gov/stationhome.html?id=8518750}) are taken from the component-based studies of \citet{Kopp2014a} and \citet{Kopp2017a}. Both frameworks are identical, but differ in how they model Antarctic ice sheet (AIS) contributions. \citet{Kopp2014a} combines the Intergovernmental Panel on Climate Change's (IPCC) Fifth Assessment Report (AR5) projections of ice sheet dynamics and surface mass balance \citep[table 13.5 in][]{Church2013a} and expert elicitation of total ice sheet mass loss from \citet{Bamber2013a}, while \citet{Kopp2017a} implement a limited ensemble of physical AIS simulations from \citet{Deconto2016a}, which include two glaciological processes previously not accounted for in other continental scale models that can rapidly increase ice-sheet mass loss \citep[marine ice-sheet hydrofracturing and marine ice-cliff instability;][]{Pollard2015a}. The simulations from \citet{Deconto2016a} do not sample the full model parameter space, as such they do not provide a probabilistic assessment of future AIS behavior \citep{Kopp2017a,Edwards2019a}. Nonetheless, they previously have been implemented in probabilistic projection frameworks \citep{LeBars2017a,Bakker2017a}. Probability distributions of local RSL are produced using 10,000 Latin hypercube samples of individual sea level component contributions. Each probability distribution is conditional on either the high greenhouse gas (GHG) emission scenario of representative concentration pathway (RCP) 8.5 or the strong GHG reduction scenario of RCP2.6 \citep{VanVuuren2011a}.



\subsection{Probability box construction} \label{Ssec:pbox}
Our approach to constructing a probability box (`p-box') is presented in Section \ref{sec:ais}, but is expanded with more details here. In order to keep the p-box boundaries from overlapping, we arbitrarily truncate the maximum AIS contribution from \citet{Kopp2014a} at 1.75 m (relative to 2000), the highest predicted AIS contributions from \citet{Deconto2016a}. This limits the maximum 2100 GMSL projection below 3.5 m (relative to 2000). We note that this truncation is arbitrary and is used only for the purpose of illustrating the p-box approach to dealing with deep uncertainty. The truncation of the AIS contribution about the 1.75 m limit could impact results in a significant way, but is not investigated here. There currently is no consensus upper limit for 2100 GMSL or AIS contributions. The 5/95th percentile ranges of GMSL from \citet{Kopp2014a} and \citet{Kopp2017a} roughly bound either the 17/83 or 5/95 end-of-century ranges from current published RCP8.5 GMSL projections surveyed in \citet{Horton2018a}, but not those for RCP2.6. 

Flood allowances and ESL return curves that consider the full probability distribution of sea level projections have been shown to be sensitive to upper-bound estimates of AIS ice mass loss in the second half of the 21st century \citep{Buchanan2016a,Slangen2017a,Rasmussen2018a}. As such, we use a parameter that sets the truncation of the upper tail of the 2100 AIS contribution distribution (AIS$_{max}$). Specifically, we use AIS$_{max}$ values of 0.25, 0.5, and 1.0 m, which are in-line with the range of published end-of-century AIS melt estimates \cite[e.g., Table 2 in][]{LeCozannet2017a}, as well as limits of 1.5 m and 1.75 m, which are upper-end estimates from \citet{Deconto2016a}. An additional parameter weighs contributions from the projections that bound the p-box ($\beta_{c} \in [0,1]$). When there is greater confidence of AIS collapse (i.e., larger values of $\beta_{c}$), more weight is given to the \citet{Kopp2017a} projections, which include faster ice mass loss and greater AIS contributions to GMSL in the second half of the 21st century \citep[via marine ice-sheet hydrofracturing and ice-cliff collapse;][]{Pollard2015a}, relative to \citet{Kopp2014a}. Note that a value of zero for $\beta_{c}$ does not imply a scenario in which there is zero probability of AIS collapse initiation, nor does a value of one for $\beta_{c}$ imply certainty in AIS collapse initiation. $\beta_{c}$ simply corresponds to the relative likelihoods of AIS collapse initiation before 2100. The effective probability distribution $\tilde{P}$ at time $t$ is given by:

\begin{equation}
\tilde{P}(\beta_{c},\text{AIS}_{max},t) = \beta_{c} P_{high}(\text{AIS}_{max},t) + (1-\beta_{c})P_{low}(\text{AIS}_{max},t),
\end{equation}

\noindent where $P_{low}(\Delta,t)$ and $P_{high}(\Delta,t)$ are the minimum and maximum projections at each point in the CDFs from \citet{Kopp2014a} and \citet{Kopp2017a}. 

\newpage
\clearpage

\section{Supplemental Results, Figures and Tables}

\begin{figure}[h]
\centering
\includegraphics[width=.999\textwidth]{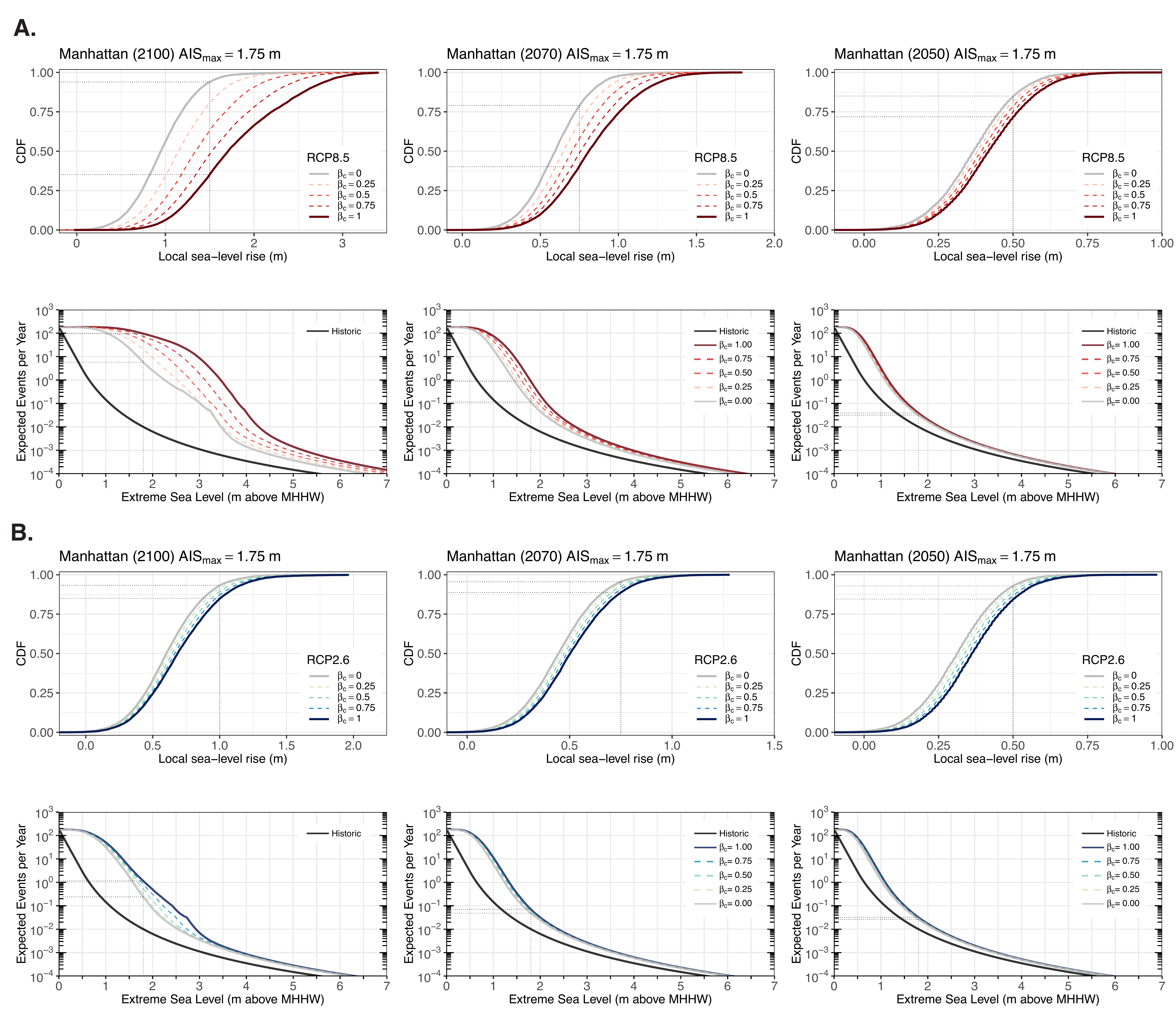}
\caption{\textbf{A.} Top: Probability boxes ('p-boxes'; solid lines) for 2100 (left), 2070 (center), and 2050 (right) local sea-level rise (SLR) in Manhattan (located at the Battery tide gauge) under the representative concentration pathway (RCP) 8.5 climate forcing scenario. Effective CDFs of local SLR (dashed lines) are generated within each p-box by averaging the edges using weights ($\beta_{c} \in$ [0,1]) that reflect a user's belief of AIS collapse initiation within the 21st century (higher values reflect higher likelihood of collapse) and by constraining the maximum possible 2100 Antarctic Ice Sheet (AIS) melt (AIS$_{max}$, relative to 2000; here, 1.75 m; Sec. \ref{sec:ais}). The black dotted lines highlight the cumulative probability of 1.0 m or 0.5 m of local SLR under different assumptions of AIS collapse initiation (i.e., values of $\beta_{c}$). Bottom: extreme sea level (ESL) event return curves for Manhattan showing the relationship between the expected number of ESL events per year and ESL height (meters above mean higher high water [MHHW]) for: 1) historical sea levels (black curve) and 2) the year 2100, 2070, and 2050 (RCP8.5) for different values of $\beta_{c}$. All curves incorporate generalized Pareto distribution (GPD) parameter uncertainty (Sec. \ref{Ssec:ESL}) and the future return curves additionally incorporate local SLR projection uncertainty by integrating across the entire local SLR probability distribution. The black dotted lines highlight the annual expected number of historically experienced 100-yr ESL events under different values of $\beta_{c}$. \textbf{B.} As for A, but for RCP2.6.}\label{Sfig:pbox1750}
\end{figure}

\begin{figure}[htb]
\centering
\includegraphics[width=.999\textwidth]{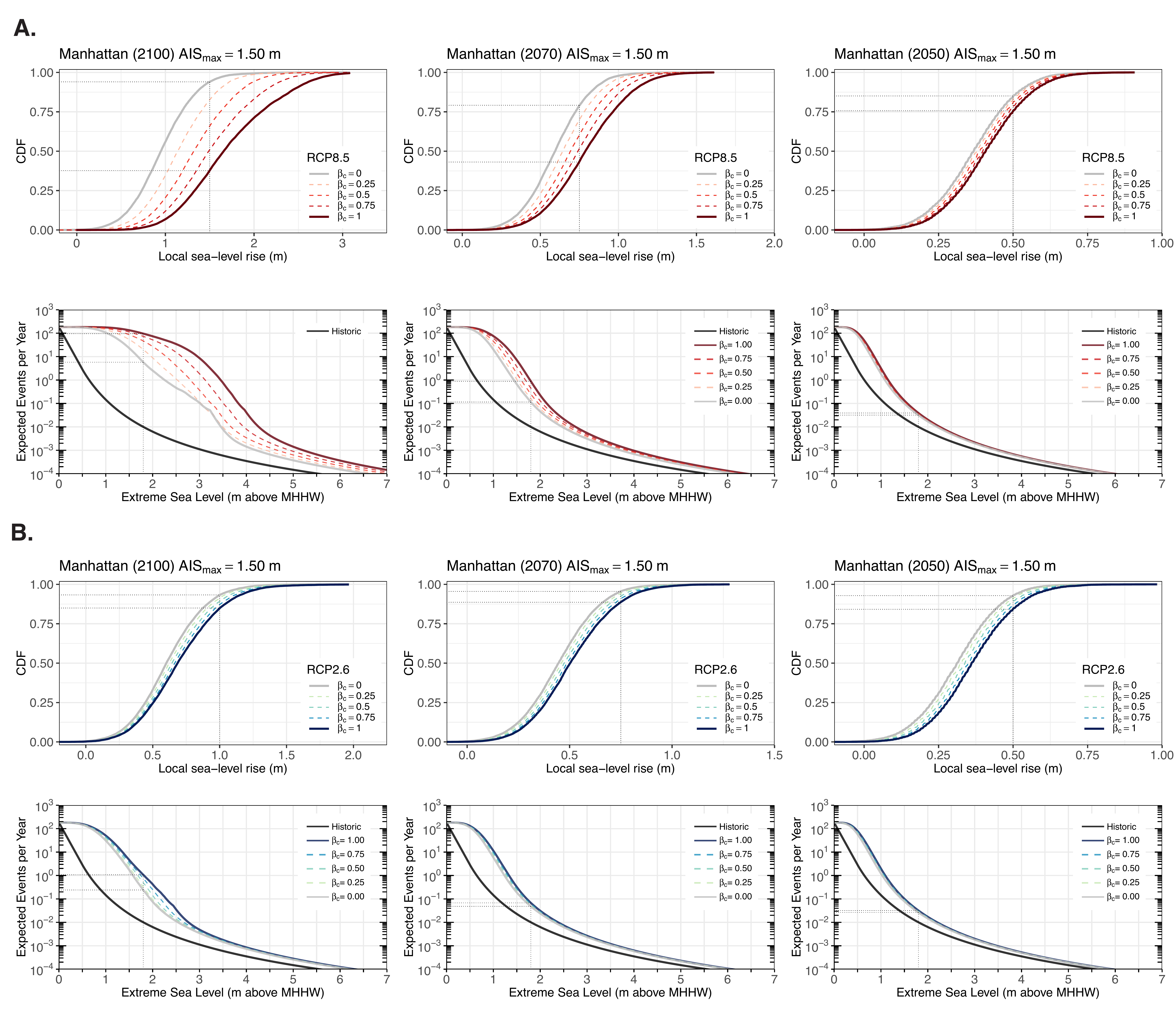}
\caption{\textbf{A.} Top: Probability boxes ('p-boxes'; solid lines) for 2100 (left), 2070 (center), and 2050 (right) local sea-level rise (SLR) in Manhattan (located at the Battery tide gauge) under the representative concentration pathway (RCP) 8.5 climate forcing scenario. Effective CDFs of local SLR (dashed lines) are generated within each p-box by averaging the edges using weights ($\beta_{c} \in$ [0,1]) that reflect a user's belief of AIS collapse initiation within the 21st century (higher values reflect higher likelihood of collapse) and by constraining the maximum possible 2100 Antarctic Ice Sheet (AIS) melt (AIS$_{max}$, relative to 2000; here, 1.5 m; Sec. \ref{sec:ais}). The black dotted lines highlight the cumulative probability of 1.0 m or 0.5 m of local SLR under different assumptions of AIS collapse initiation (i.e., values of $\beta_{c}$). Bottom: extreme sea level (ESL) event return curves for Manhattan showing the relationship between the expected number of ESL events per year and ESL height (meters above mean higher high water [MHHW]) for: 1) historical sea levels (black curve) and 2) the year 2100, 2070, and 2050 (RCP8.5) for different values of $\beta_{c}$. All curves incorporate generalized Pareto distribution (GPD) parameter uncertainty (Sec. \ref{Ssec:ESL}) and the future return curves additionally incorporate local SLR projection uncertainty by integrating across the entire local SLR probability distribution. The black dotted lines highlight the annual expected number of historically experienced 100-yr ESL events under different values of $\beta_{c}$. \textbf{B.} As for A, but for RCP2.6.}\label{Sfig:pbox1500}
\end{figure}

\begin{figure}[htb]
\centering
\includegraphics[width=.999\textwidth]{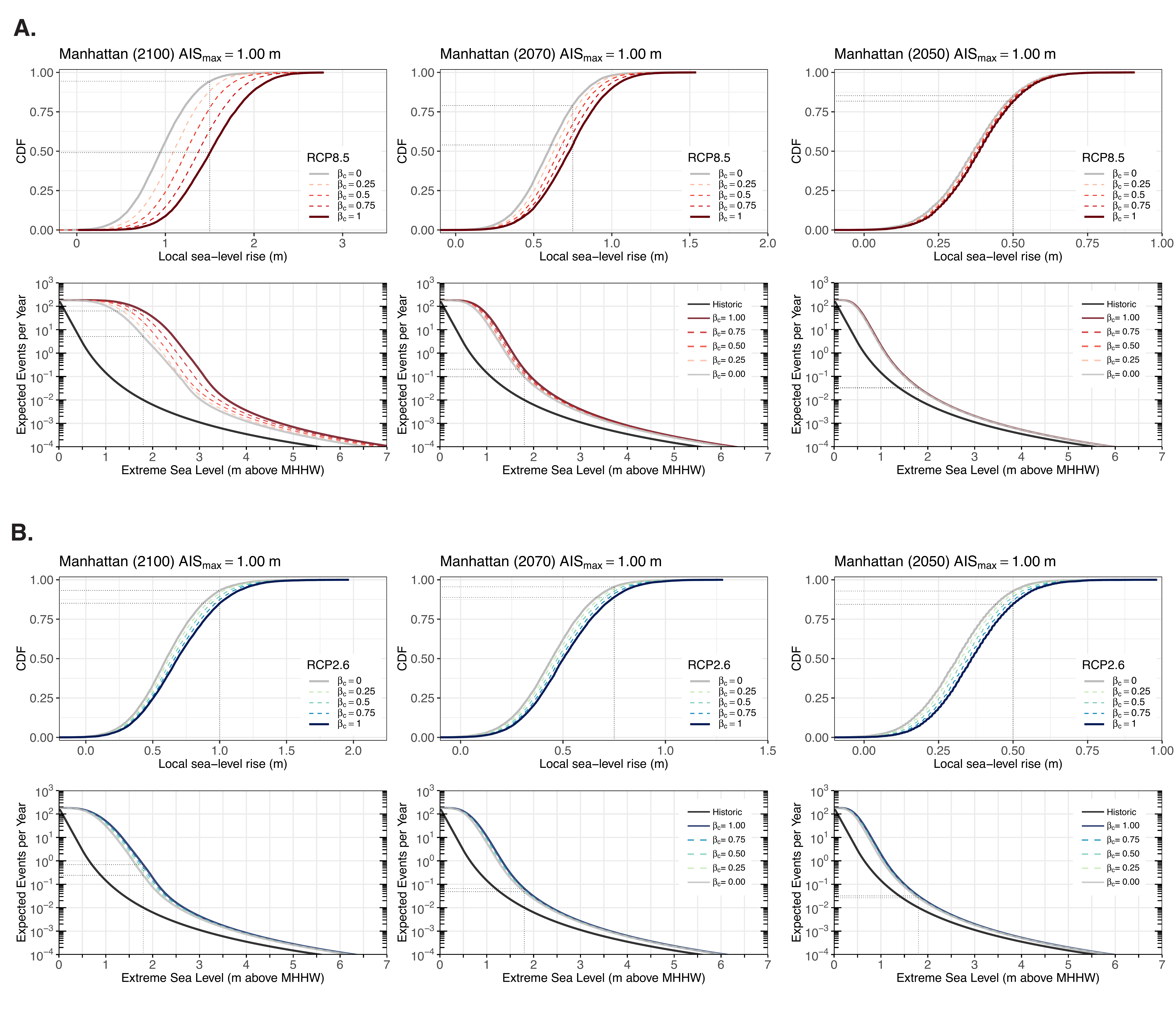}
\caption{\textbf{A.} Top: Probability boxes ('p-boxes'; solid lines) for 2100 (left), 2070 (center), and 2050 (right) local sea-level rise (SLR) in Manhattan (located at the Battery tide gauge) under the representative concentration pathway (RCP) 8.5 climate forcing scenario. Effective CDFs of local SLR (dashed lines) are generated within each p-box by averaging the edges using weights ($\beta_{c} \in$ [0,1]) that reflect a user's belief of AIS collapse initiation within the 21st century (higher values reflect higher likelihood of collapse) and by constraining the maximum possible 2100 Antarctic Ice Sheet (AIS) melt (AIS$_{max}$, relative to 2000; here, 1.0 m; Sec. \ref{sec:ais}). The black dotted lines highlight the cumulative probability of 1.0 m or 0.5 m of local SLR under different assumptions of AIS collapse initiation (i.e., values of $\beta_{c}$). Bottom: extreme sea level (ESL) event return curves for Manhattan showing the relationship between the expected number of ESL events per year and ESL height (meters above mean higher high water [MHHW]) for: 1) historical sea levels (black curve) and 2) the year 2100, 2070, and 2050 (RCP8.5) for different values of $\beta_{c}$. All curves incorporate generalized Pareto distribution (GPD) parameter uncertainty (Sec. \ref{Ssec:ESL}) and the future return curves additionally incorporate local SLR projection uncertainty by integrating across the entire local SLR probability distribution. The black dotted lines highlight the annual expected number of historically experienced 100-yr ESL events under different values of $\beta_{c}$. \textbf{B.} As for A, but for RCP2.6.}\label{Sfig:pbox1000}
\end{figure}

\begin{figure}[htb]
\centering
\includegraphics[width=.999\textwidth]{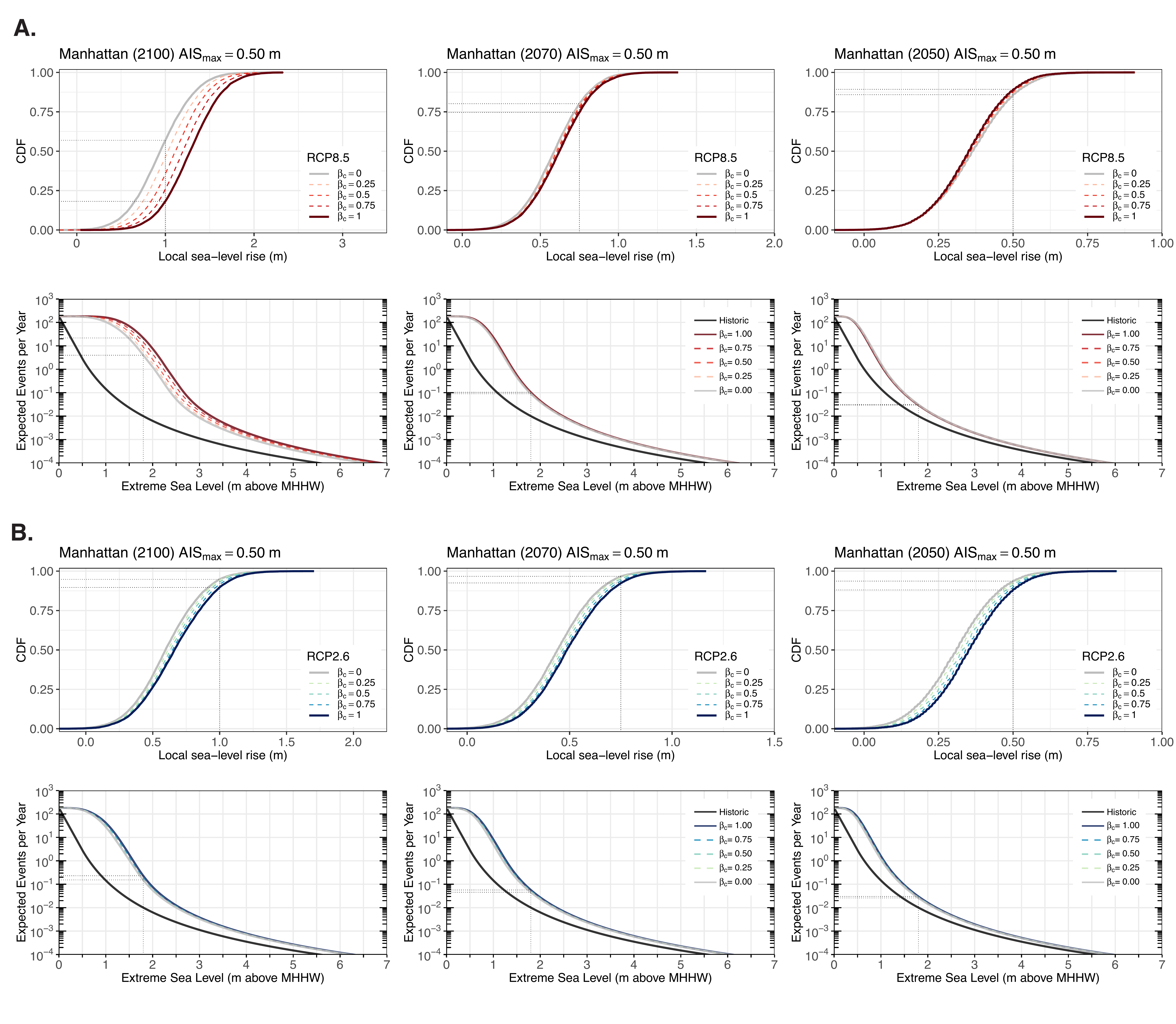}
\caption{\textbf{A.} Top: Probability boxes ('p-boxes'; solid lines) for 2100 (left), 2070 (center), and 2050 (right) local sea-level rise (SLR) in Manhattan (located at the Battery tide gauge) under the representative concentration pathway (RCP) 8.5 climate forcing scenario. Effective CDFs of local SLR (dashed lines) are generated within each p-box by averaging the edges using weights ($\beta_{c} \in$ [0,1]) that reflect a user's belief of AIS collapse initiation within the 21st century (higher values reflect higher likelihood of collapse) and by constraining the maximum possible 2100 Antarctic Ice Sheet (AIS) melt (AIS$_{max}$, relative to 2000; here, 0.5 m; Sec. \ref{sec:ais}). The black dotted lines highlight the cumulative probability of 1.0 m or 0.5 m of local SLR under different assumptions of AIS collapse initiation (i.e., values of $\beta_{c}$). Bottom: extreme sea level (ESL) event return curves for Manhattan showing the relationship between the expected number of ESL events per year and ESL height (meters above mean higher high water [MHHW]) for: 1) historical sea levels (black curve) and 2) the year 2100, 2070, and 2050 (RCP8.5) for different values of $\beta_{c}$. All curves incorporate generalized Pareto distribution (GPD) parameter uncertainty (Sec. \ref{Ssec:ESL}) and the future return curves additionally incorporate local SLR projection uncertainty by integrating across the entire local SLR probability distribution. The black dotted lines highlight the annual expected number of historically experienced 100-yr ESL events under different values of $\beta_{c}$. \textbf{B.} As for A, but for RCP2.6.}\label{Sfig:pbox500}
\end{figure}

\begin{figure}[htb]
\centering
\includegraphics[width=.999\textwidth]{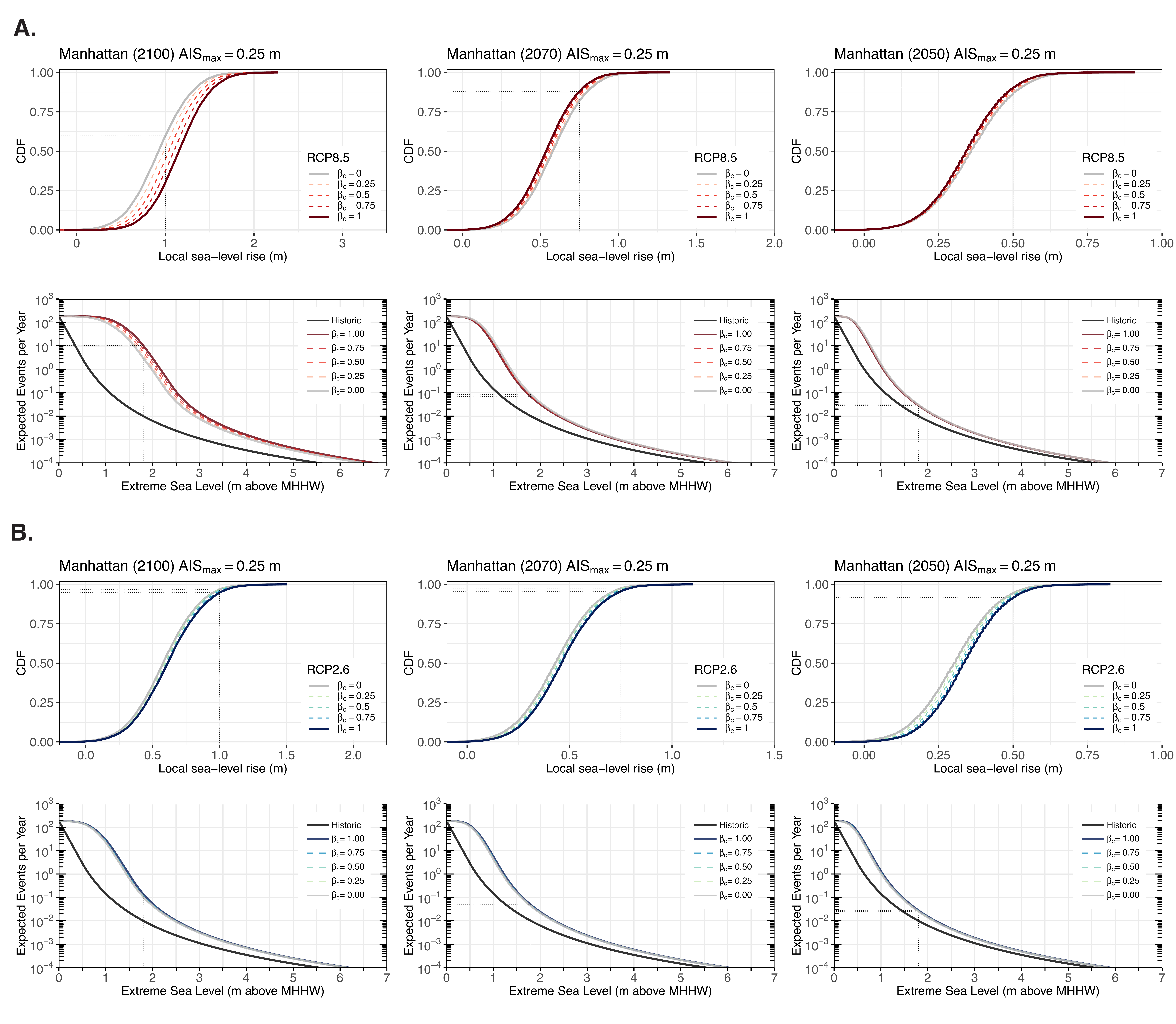}
\caption{\textbf{A.} Top: Probability boxes ('p-boxes'; solid lines) for 2100 (left), 2070 (center), and 2050 (right) local sea-level rise (SLR) in Manhattan (located at the Battery tide gauge) under the representative concentration pathway (RCP) 8.5 climate forcing scenario. Effective CDFs of local SLR (dashed lines) are generated within each p-box by averaging the edges using weights ($\beta_{c} \in$ [0,1]) that reflect a user's belief of AIS collapse initiation within the 21st century (higher values reflect higher likelihood of collapse) and by constraining the maximum possible 2100 Antarctic Ice Sheet (AIS) melt (AIS$_{max}$, relative to 2000; here, 0.25 m; Sec. \ref{sec:ais}). The black dotted lines highlight the cumulative probability of 1.0 m or 0.5 m of local SLR under different assumptions of AIS collapse initiation (i.e., values of $\beta_{c}$). Bottom: extreme sea level (ESL) event return curves for Manhattan showing the relationship between the expected number of ESL events per year and ESL height (meters above mean higher high water [MHHW]) for: 1) historical sea levels (black curve) and 2) the year 2100, 2070, and 2050 (RCP8.5) for different values of $\beta_{c}$. All curves incorporate generalized Pareto distribution (GPD) parameter uncertainty (Sec. \ref{Ssec:ESL}) and the future return curves additionally incorporate local SLR projection uncertainty by integrating across the entire local SLR probability distribution. The black dotted lines highlight the annual expected number of historically experienced 100-yr ESL events under different values of $\beta_{c}$. \textbf{B.} As for A, but for RCP2.6.}\label{Sfig:pbox250}
\end{figure}

\newpage
\clearpage

\begin{figure}[htb]
\centering
\includegraphics[width=.24\textwidth]{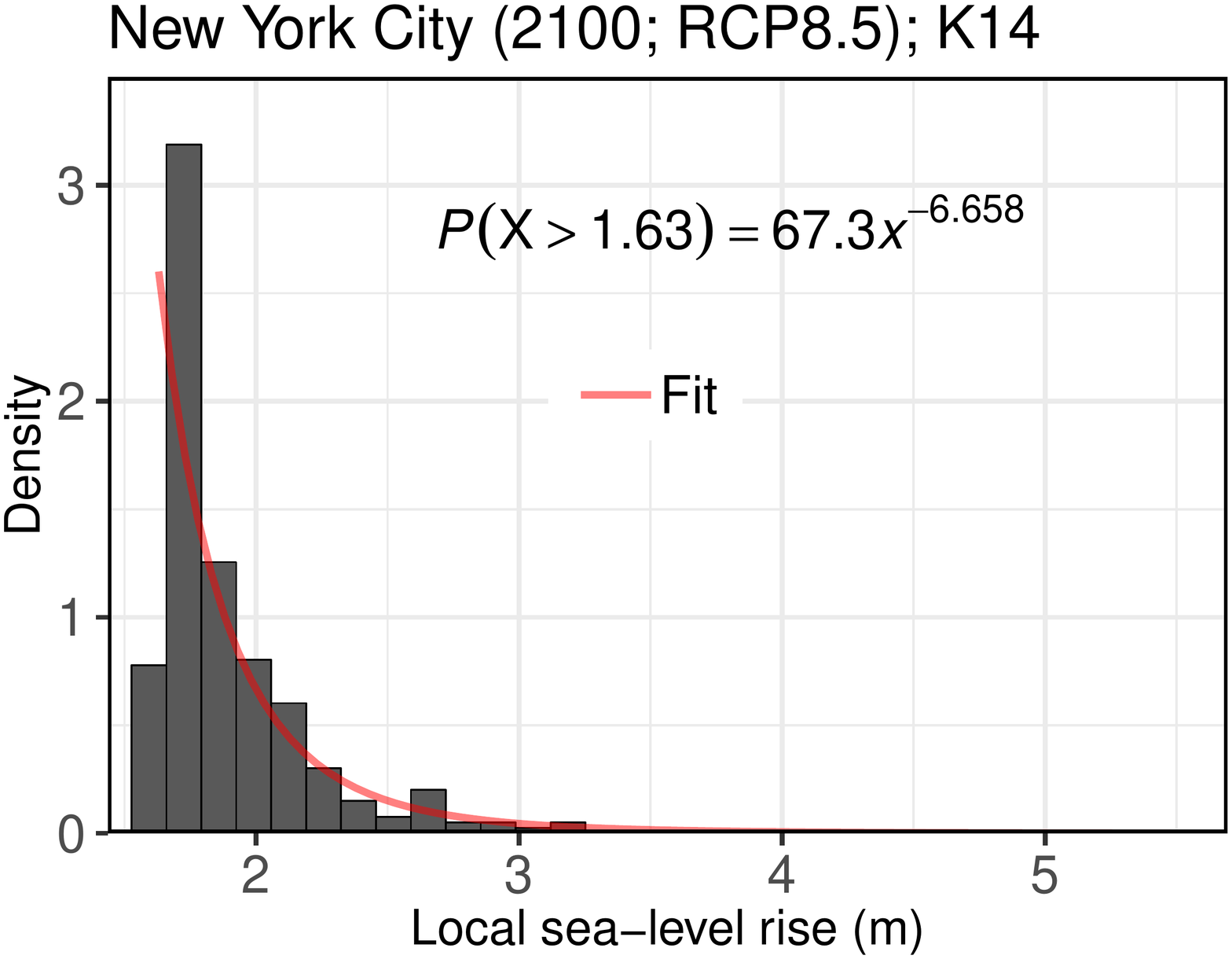}
\includegraphics[width=.24\textwidth]{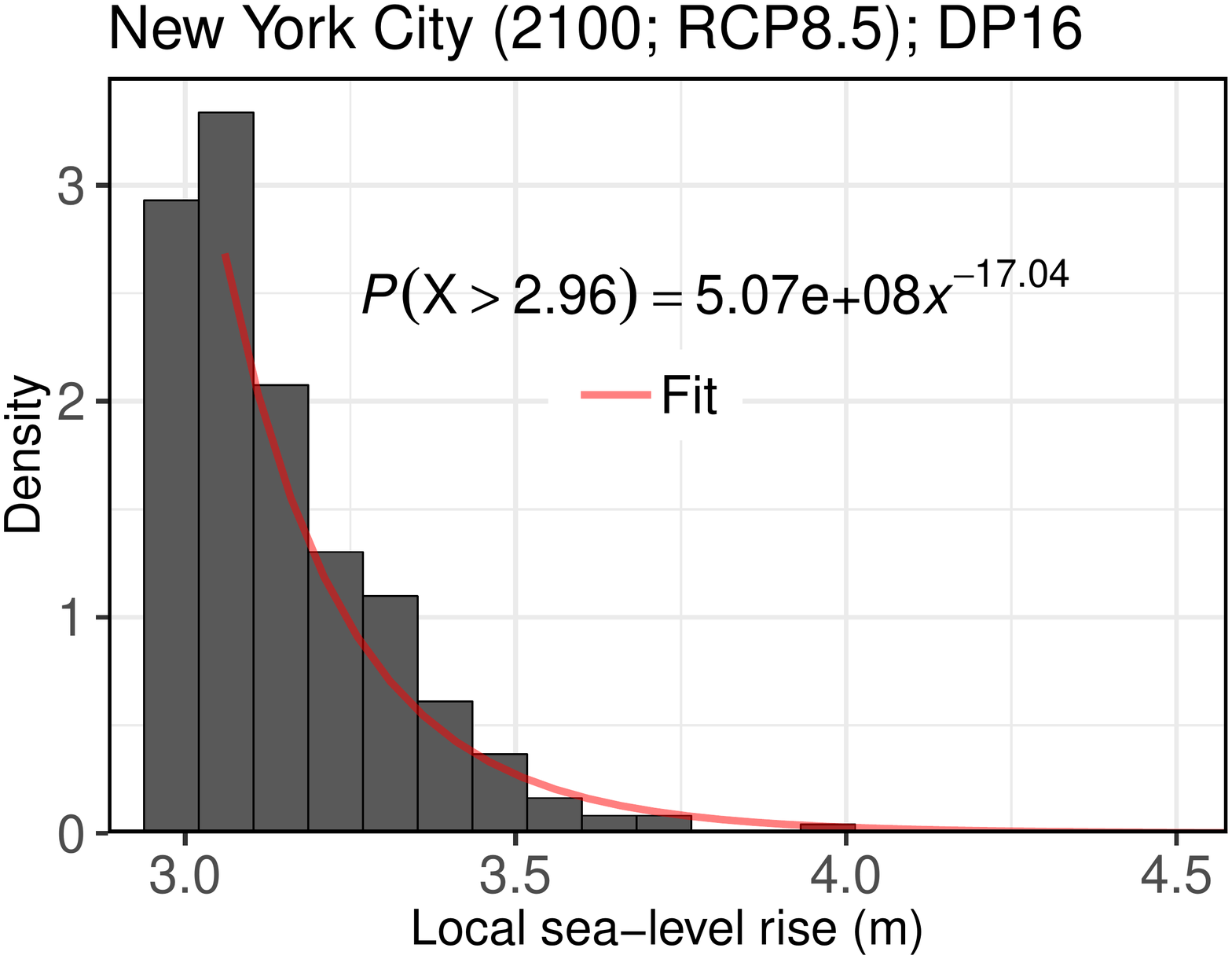}
\includegraphics[width=.24\textwidth]{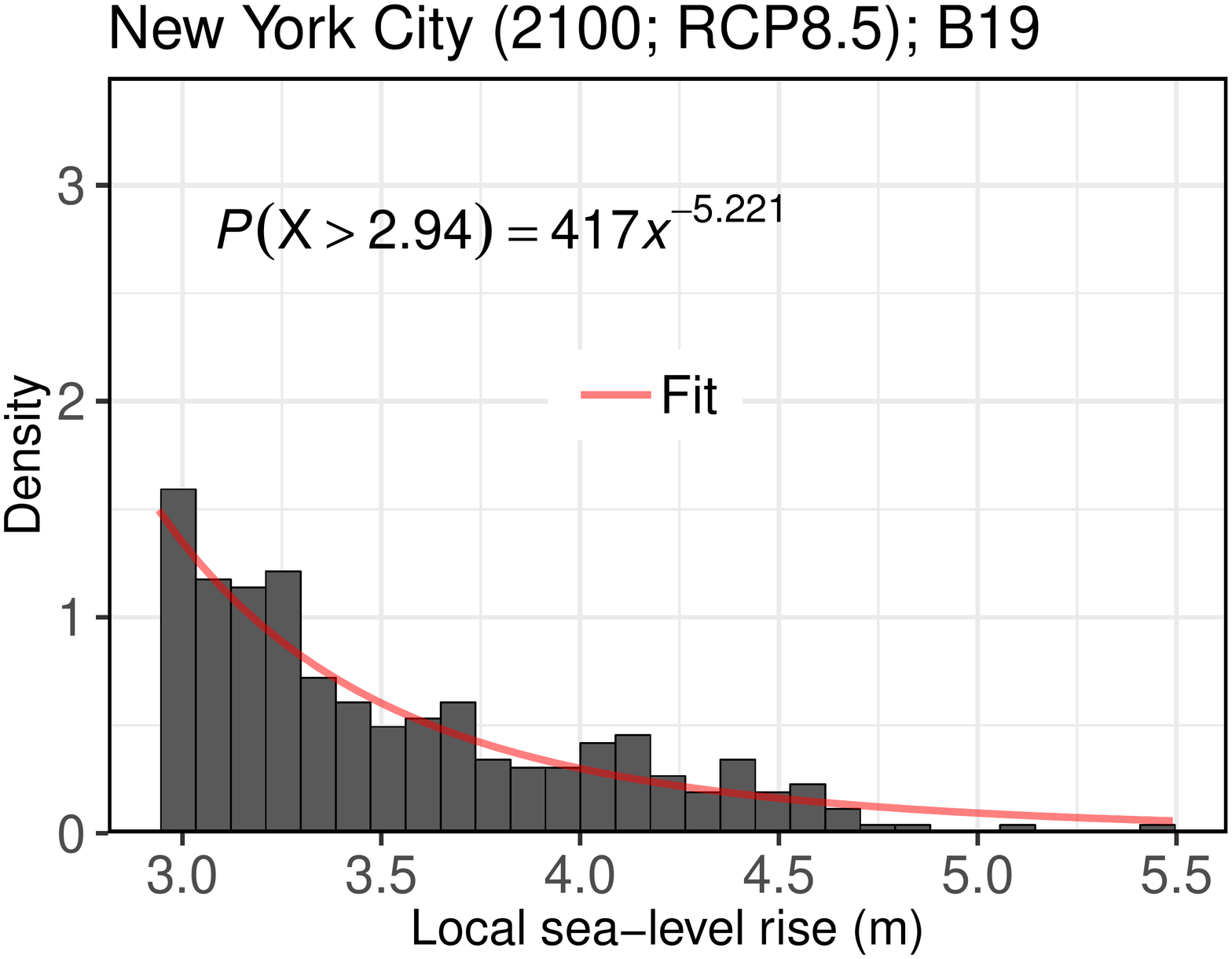}
\includegraphics[width=.24\textwidth]{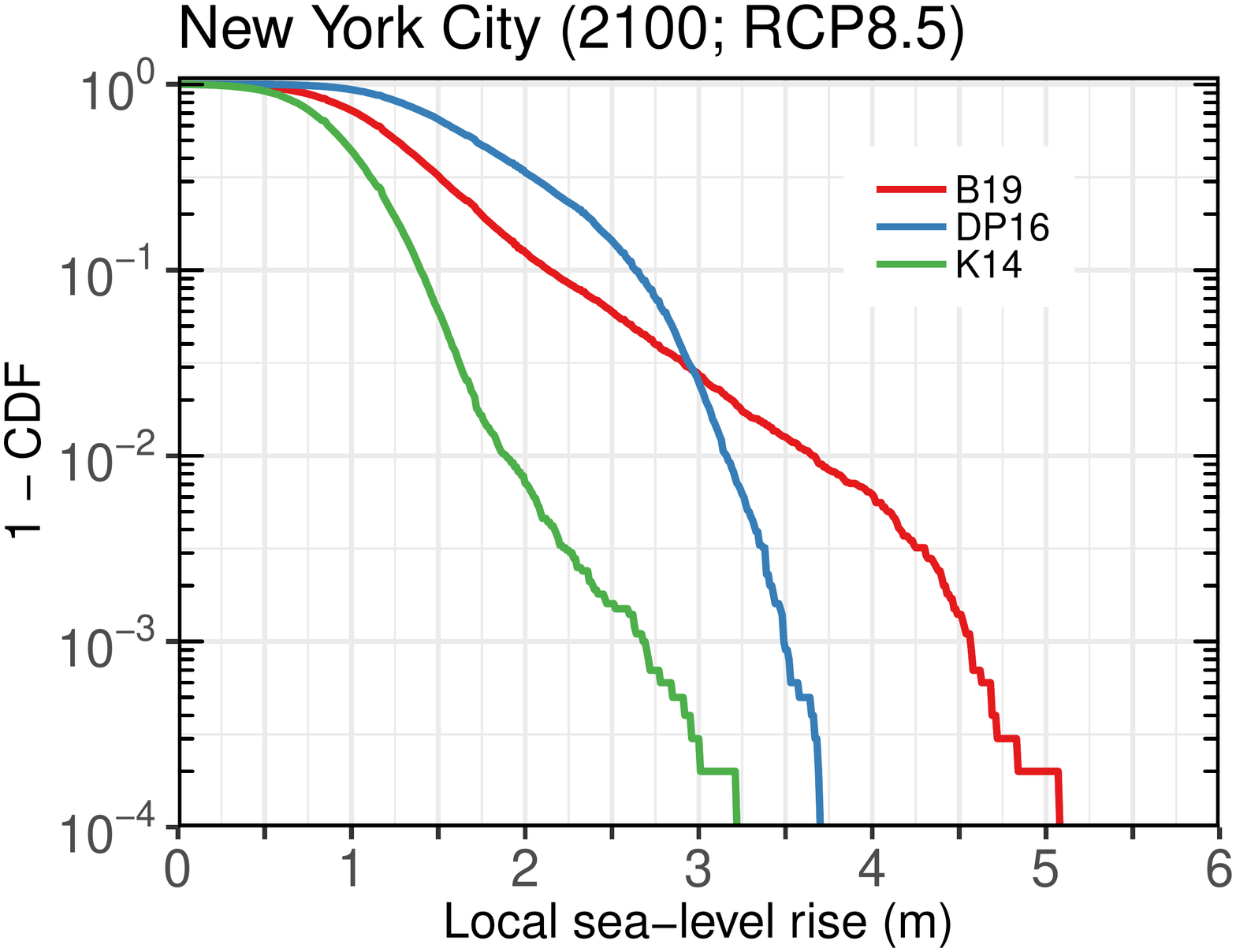}
\includegraphics[width=.24\textwidth]{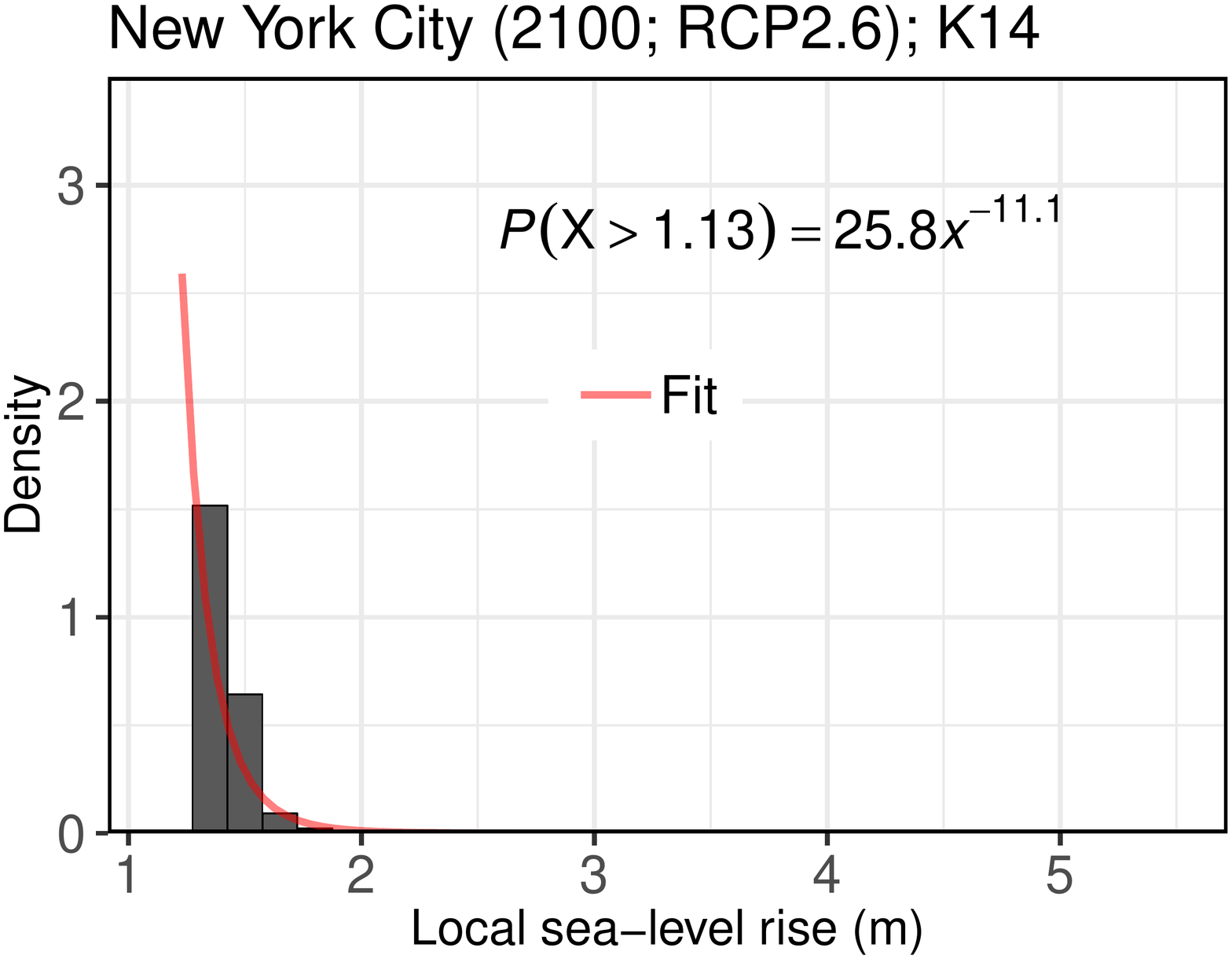}
\includegraphics[width=.24\textwidth]{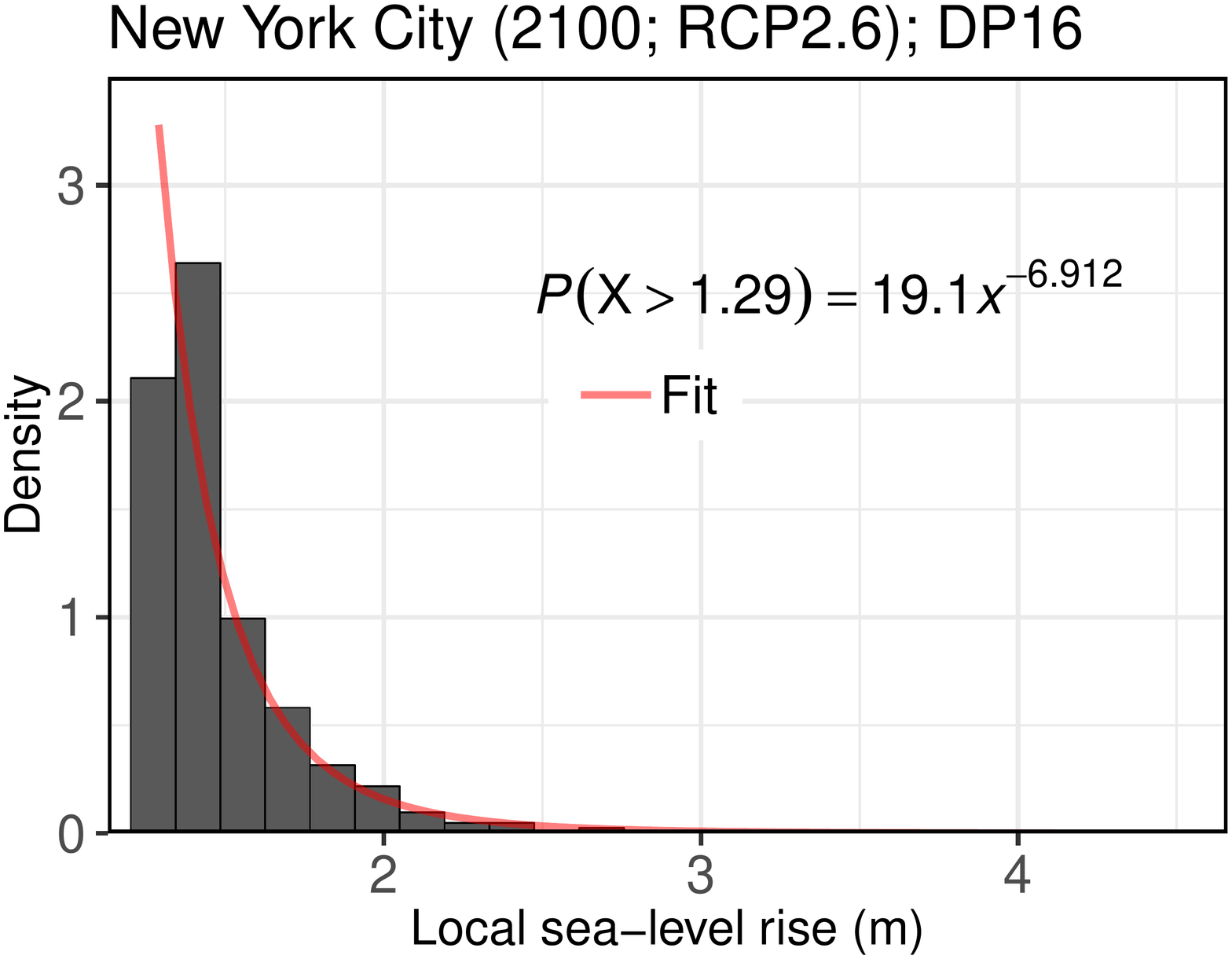}
\includegraphics[width=.24\textwidth]{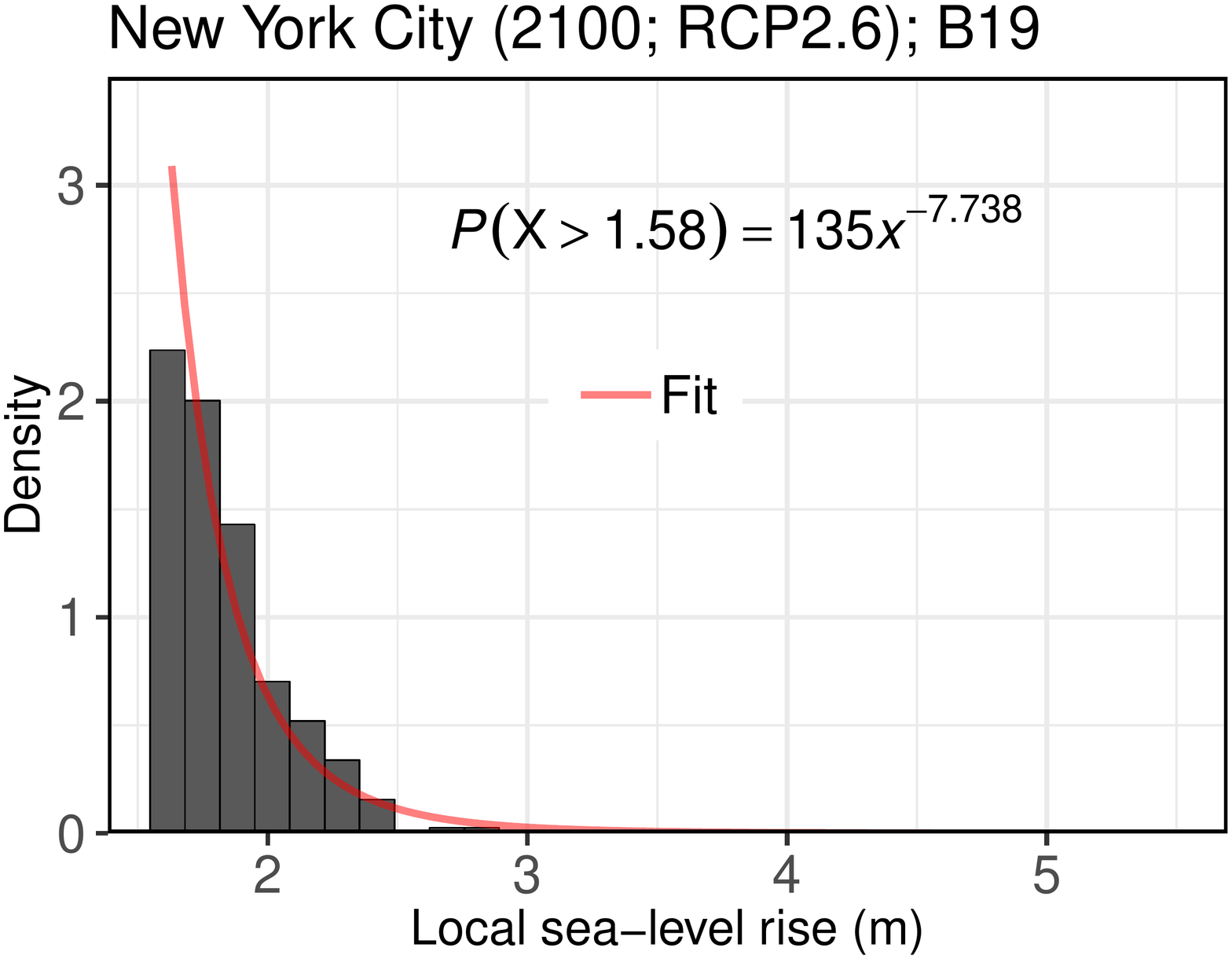}
\includegraphics[width=.24\textwidth]{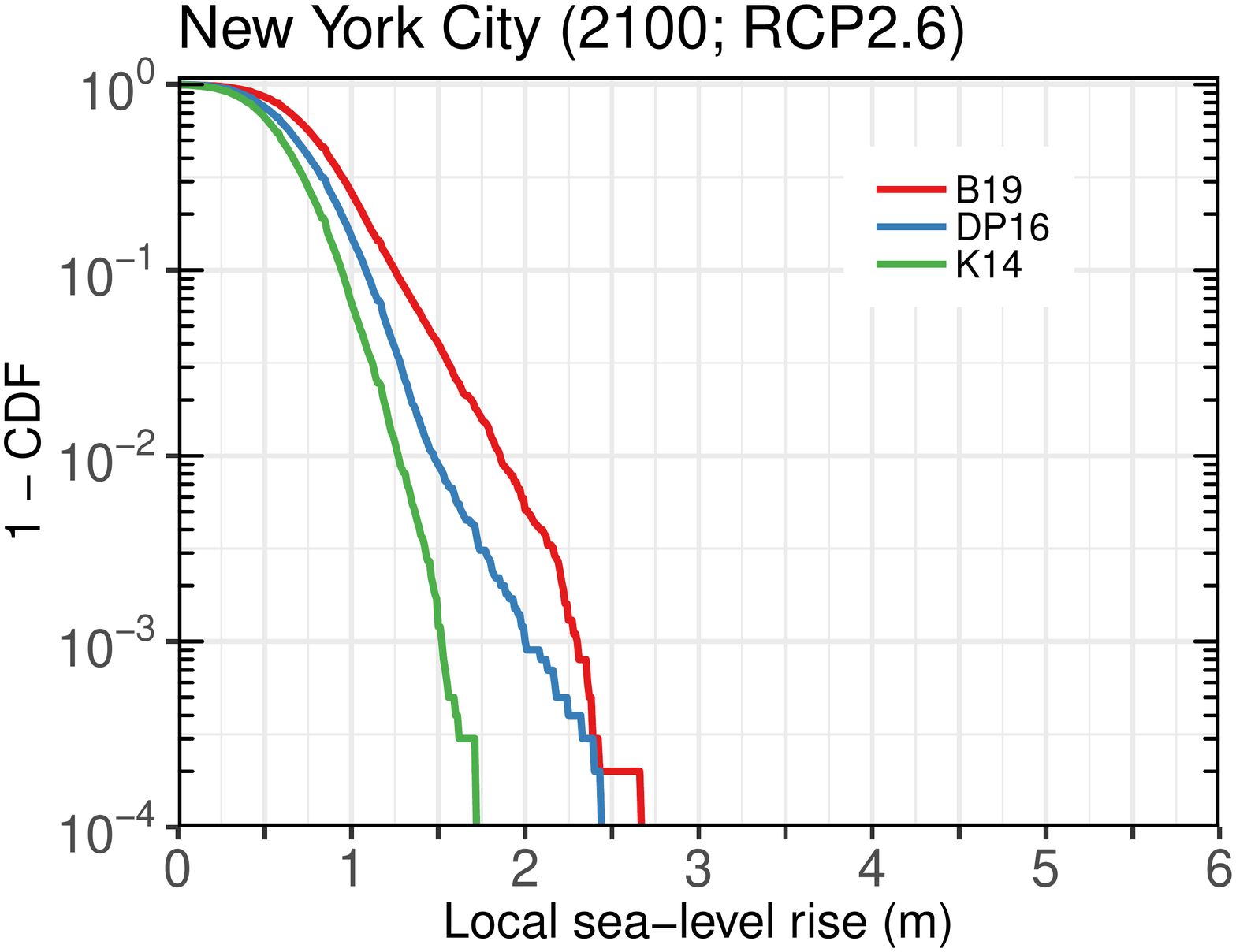}
\caption{\textbf{Top Row:} Density plot of local sea-level rise (SLR) samples (meters relative to 2000) above the 97th percentile for 2100 for New York City (RCP8.5) and a power function fit (red line). The smaller the absolute value of the exponent of the power function fit, the longer the tail of the SLR distribution. The SLR samples in each plot are from the probabilistic projections of \citet{Kopp2014a}(K14), \citet{Deconto2016a}(DP16), and \citet{Bamber2019a}(B19). Far right plot is a survival function for 2100 local SLR (meters) for New York City from \citet{Bamber2019a}(B19; red), \citet{Deconto2016a}(DP16; blue), and \citet{Kopp2014a}(K14; green). \textbf{Bottom Row:} As for Top Row, but for RCP2.6.} \label{Sfig:slrtail}
\end{figure}

\newpage
\clearpage

\begin{landscape} 
\begin{table}[htbp]
\centering
\setlength{\tabcolsep}{3pt}
\caption{Future local sea-level rise (SLR) projections (meters; relative to 2000) at a tide gauge located at the Battery in lower Manhattan (New York City) for 2100, 2070, and 2050 under representative concentration pathway (RCP) 8.5 and RCP2.6 and for different assumptions regarding future Antarctic ice sheet (AIS) behavior (e.g, likelihood of AIS collapse [$\beta_{c}$] and maximum 2100 AIS contribution [AIS$_{max}$]). Values given are: expected (5th percentile--95th percentile).}
{\small
\begin{tabular}{lcccccc|lccccccc}
\multicolumn{7}{l}{\textbf{Local Sea-Level  Rise}  (m; relative to 2000)}  \\
RCP8.5 &     &  \multicolumn{5}{c}{AIS$_{max}$}     &     &  RCP2.6   &     &  \multicolumn{5}{c}{AIS$_{max}$}  \\
 &    $\beta_{c}$     &  1.75  m   &  1.5  m   &  1.0  m     &  0.5  m     &  0.25  m     &     &     &  $\beta_{c}$   &  1.75  m   &  1.5  m   &  1.0  m   &  0.5  m   &  0.25  m  \\
\hline
2100   &   &   &   &   &   &   &   &  2100   &   &   &   &   &   &  \\
 &  1   &1.8  (1.0--2.9)   &  1.7  (0.9--2.7)   &  1.5  (0.9--2.2)   &  1.3  (0.8--1.8)   &  1.2  (0.7--1.7)   &   &   &  1   &0.7  (0.3--1.2)   &  0.7  (0.3--1.2)   &  0.7  (0.3--1.2)   &  0.7  (0.3--1.1)   &  0.6  (0.2--1.0)  \\
 &  0.75   &1.6  (0.8--2.5)   &  1.5  (0.8--2.3)   &  1.4  (0.8--2.0)   &  1.2  (0.7--1.7)   &  1.1  (0.6--1.6)   &   &   &  0.75   &0.7  (0.3--1.2)   &  0.7  (0.3--1.2)   &  0.7  (0.3--1.2)   &  0.7  (0.3--1.1)   &  0.6  (0.2--1.0)  \\
 &  0.5   &1.4  (0.7--2.2)   &  1.3  (0.7--2.0)   &  1.2  (0.7--1.8)   &  1.1  (0.6--1.6)   &  1.0  (0.6--1.6)   &   &   &  0.5   &0.7  (0.2--1.1)   &  0.7  (0.2--1.1)   &  0.7  (0.2--1.1)   &  0.6  (0.2--1.0)   &  0.6  (0.2--1.0)  \\
 &  0.25   &1.2  (0.6--1.8)   &  1.2  (0.6--1.8)   &  1.1  (0.6--1.7)   &  1.0  (0.5--1.6)   &  1.0  (0.5--1.5)   &   &   &  0.25   &0.6  (0.2--1.1)   &  0.6  (0.2--1.1)   &  0.6  (0.2--1.1)   &  0.6  (0.2--1.0)   &  0.6  (0.2--1.0)  \\
 &  0   &1.0  (0.4--1.5)   &  1.0  (0.4--1.5)   &  1.0  (0.4--1.5)   &  0.9  (0.4--1.5)   &  0.9  (0.4--1.5)   &   &   &  0   &0.6  (0.2--1.1)   &  0.6  (0.2--1.1)   &  0.6  (0.2--1.1)   &  0.6  (0.2--1.0)   &  0.6  (0.2--1.0)  \\
2070   &   &   &   &   &   &   &   &  2070   &   &   &   &   &   &  \\
 &  1   &0.8  (0.4--1.3)   &  0.8  (0.4--1.2)   &  0.7  (0.4--1.1)   &  0.6  (0.3--0.9)   &  0.5  (0.2--0.8)   &   &   &  1   &0.5  (0.2--0.8)   &  0.5  (0.2--0.8)   &  0.5  (0.2--0.8)   &  0.5  (0.2--0.8)   &  0.5  (0.2--0.7)  \\
 &  0.75   &0.8  (0.4--1.2)   &  0.8  (0.4--1.1)   &  0.7  (0.4--1.0)   &  0.6  (0.3--0.9)   &  0.6  (0.3--0.8)   &   &   &  0.75   &0.5  (0.2--0.8)   &  0.5  (0.2--0.8)   &  0.5  (0.2--0.8)   &  0.5  (0.2--0.8)   &  0.5  (0.2--0.7)  \\
 &  0.5   &0.7  (0.4--1.1)   &  0.7  (0.4--1.1)   &  0.7  (0.3--1.0)   &  0.6  (0.3--0.9)   &  0.6  (0.3--0.9)   &   &   &  0.5   &0.5  (0.2--0.8)   &  0.5  (0.2--0.8)   &  0.5  (0.2--0.8)   &  0.5  (0.2--0.8)   &  0.5  (0.2--0.7)  \\
 &  0.25   &0.7  (0.3--1.0)   &  0.7  (0.3--1.0)   &  0.6  (0.3--0.9)   &  0.6  (0.3--0.9)   &  0.6  (0.3--0.9)   &   &   &  0.25   &0.5  (0.2--0.8)   &  0.5  (0.2--0.8)   &  0.5  (0.2--0.8)   &  0.5  (0.2--0.7)   &  0.4  (0.2--0.7)  \\
 &  0   &0.6  (0.3--0.9)   &  0.6  (0.3--0.9)   &  0.6  (0.3--0.9)   &  0.6  (0.3--0.9)   &  0.6  (0.3--0.9)   &   &   &  0   &0.5  (0.2--0.7)   &  0.5  (0.2--0.7)   &  0.5  (0.2--0.7)   &  0.4  (0.2--0.7)   &  0.4  (0.2--0.7)  \\
2050   &   &   &   &   &   &   &   &  2050   &   &   &   &   &   &  \\
 &  1   &0.4  (0.2--0.7)   &  0.4  (0.2--0.6)   &  0.4  (0.2--0.6)   &  0.3  (0.2--0.6)   &  0.3  (0.1--0.5)   &   &   &  1   &0.4  (0.2--0.6)   &  0.4  (0.2--0.6)   &  0.4  (0.2--0.6)   &  0.4  (0.2--0.6)   &  0.3  (0.1--0.5)  \\
 &  0.75   &0.4  (0.2--0.6)   &  0.4  (0.2--0.6)   &  0.4  (0.2--0.6)   &  0.4  (0.2--0.6)   &  0.3  (0.1--0.6)   &   &   &  0.75   &0.4  (0.1--0.6)   &  0.4  (0.1--0.6)   &  0.3  (0.1--0.6)   &  0.3  (0.1--0.6)   &  0.3  (0.1--0.5)  \\
 &  0.5   &0.4  (0.2--0.6)   &  0.4  (0.2--0.6)   &  0.4  (0.2--0.6)   &  0.4  (0.2--0.6)   &  0.3  (0.1--0.6)   &   &   &  0.5   &0.3  (0.1--0.6)   &  0.3  (0.1--0.6)   &  0.3  (0.1--0.6)   &  0.3  (0.1--0.5)   &  0.3  (0.1--0.5)  \\
 &  0.25   &0.4  (0.2--0.6)   &  0.4  (0.2--0.6)   &  0.4  (0.2--0.6)   &  0.4  (0.2--0.6)   &  0.4  (0.2--0.6)   &   &   &  0.25   &0.3  (0.1--0.5)   &  0.3  (0.1--0.5)   &  0.3  (0.1--0.5)   &  0.3  (0.1--0.5)   &  0.3  (0.1--0.5)  \\
 &  0   &0.4  (0.2--0.6)   &  0.4  (0.2--0.6)   &  0.4  (0.2--0.6)   &  0.4  (0.2--0.6)   &  0.4  (0.1--0.6)   &   &   &  0   &0.3  (0.1--0.5)   &  0.3  (0.1--0.5)   &  0.3  (0.1--0.5)   &  0.3  (0.1--0.5)   &  0.3  (0.1--0.5)  \\
\hline
\end{tabular}}
\label{Stab:slr}
\end{table}
\end{landscape}

\newpage
\clearpage

\begin{figure}[htb]
\centering
\includegraphics[width=.9\textwidth]{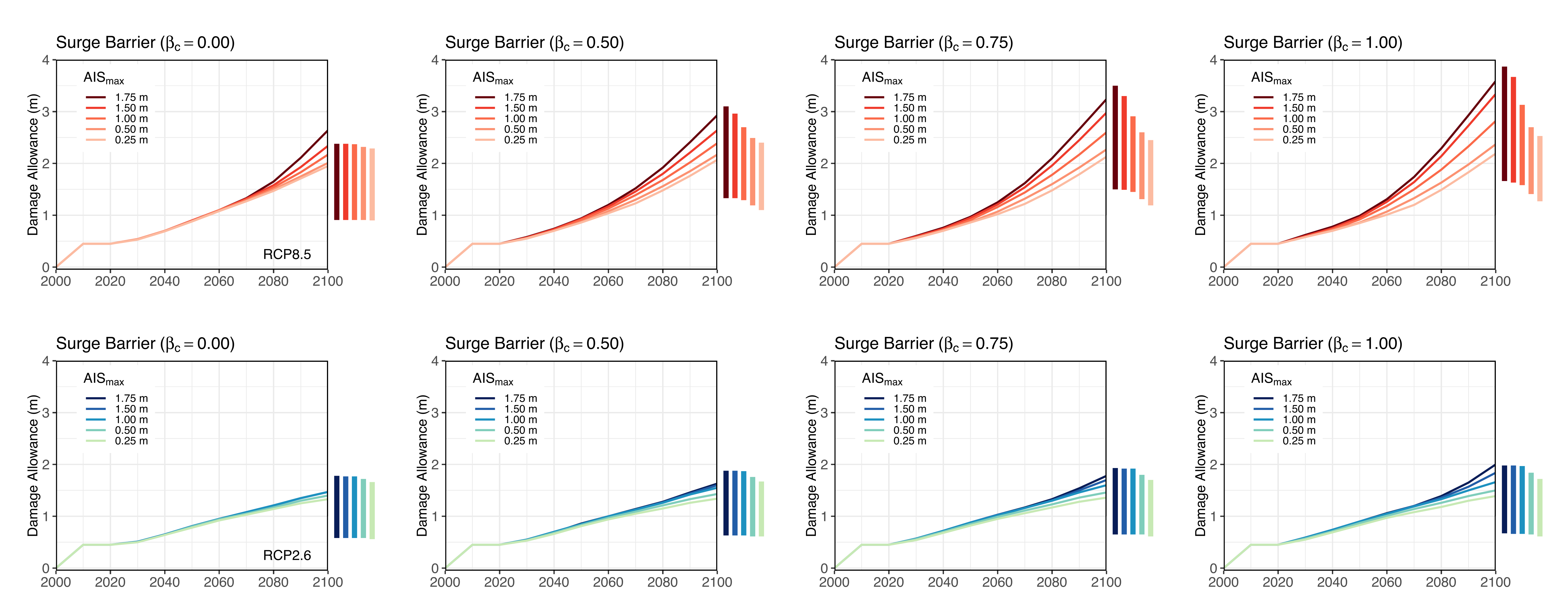}
\caption{\textbf{Top Row:} Instantaneous flood damage allowances (meters above mean higher high water [MHHW]) over time (2000--2100) for a storm surge barrier protecting Manhattan under different maximum 2100 Antarctic Ice Sheet (AIS) contribution thresholds (AIS$_{max}$, relative to 2000), different subjectively perceived likelihoods of AIS collapse ($\beta_{c}$; 0 being `most unlikely' and 1 being `most likely'), and for the representative concentration pathway (RCP) 8.5 climate forcing scenario. The colored bars in the margins of each plot show the 2100 damage allowances using only the 5/95th percentile local sea-level rise projections. The storm surge barrier allowances include 0.5 m of freeboard, have a 10\% probability of failure at the design height, and the barrier gates close when water levels are $>$ 1.0 m above MHHW. \textbf{Bottom Row:} As for Top Row, but for RCP2.6.}\label{Sfig:timeseries_surge}
\end{figure}


\begin{figure}[htb]
\centering
\includegraphics[width=.9\textwidth]{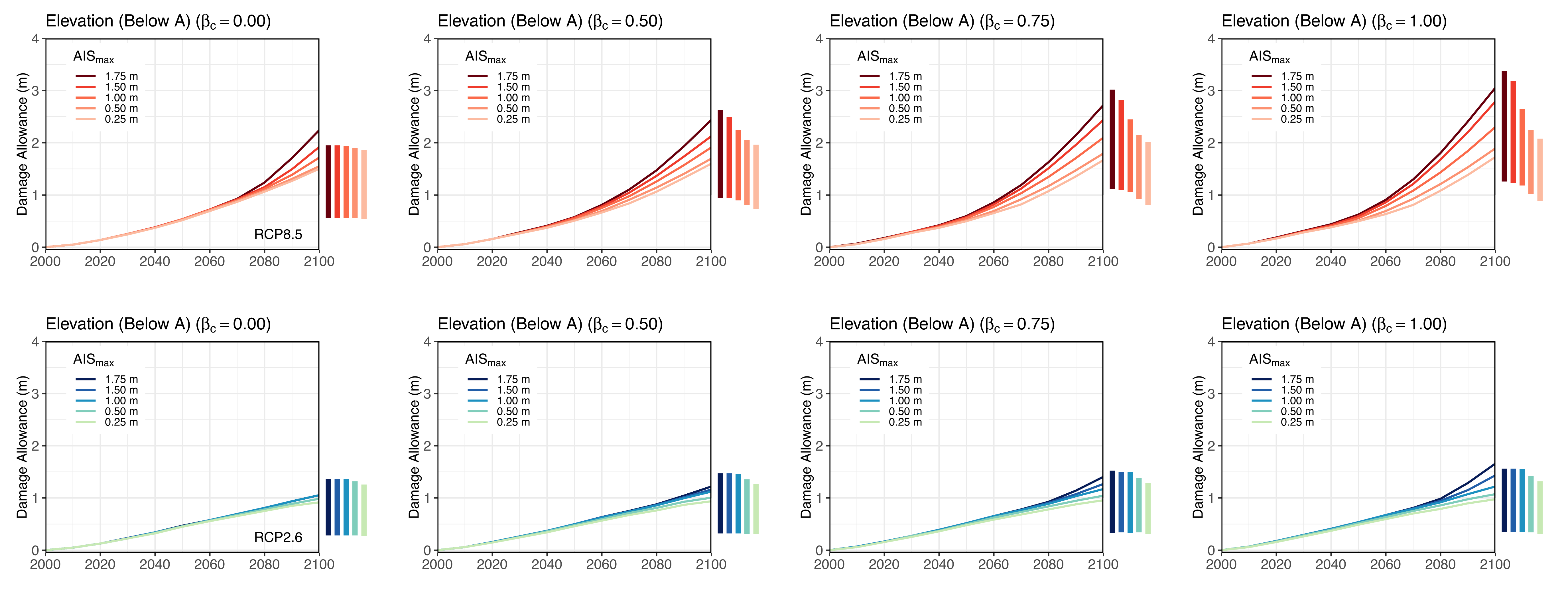}
\caption{\textbf{Top Row:} Instantaneous flood damage allowances (meters above mean higher high water [MHHW]) over time (2000--2100) for an elevation strategy below the damage allowance $A$ for Manhattan under different maximum 2100 Antarctic Ice Sheet (AIS) contribution thresholds (AIS$_{max}$, relative to 2000), different subjectively perceived likelihoods of AIS collapse ($\beta_{c}$; 0 being `most unlikely' and 1 being `most likely'), and for the representative concentration pathway (RCP) 8.5 climate forcing scenario. The colored bars in the margins of each plot show the 2100 damage allowances using only the 5/95th percentile local sea-level rise projections. The damage allowances assume perfect compliance (i.e., all structures below $A$ elevate). \textbf{Bottom Row:} As for Top Row, but for RCP2.6.}\label{Sfig:timeseries_elev2}
\end{figure}

\clearpage
\newpage

\begin{table}[htb]
\centering
\setlength{\tabcolsep}{3pt}

\caption{Damage allowances (meters) for a storm surge barrier for 2100, 2070, and 2050 under representative concentration pathway (RCP) 8.5 and RCP2.6 and for different assumptions regarding future Antarctic ice sheet (AIS) behavior (e.g, likelihood of AIS collapse [$\beta_{c}$] and maximum 2100 AIS contribution [AIS$_{max}$]). The allowances are relative to the current protection height around Manhattan (assumed to be a bulkhead 1.0 m above mean higher high water [MHHW]). The storm surge barrier allowances include 0.5 m of freeboard, have a 10\% probability of failure at the design height, and the barrier gates close when water levels are $>$ 1.0 m above MHHW (approximately once every 10 years based on observations from the recent past; Fig. \ref{fig:process}A).}
{\small
\begin{tabular}{lcccccc|lccccccc}
\multicolumn{7}{l}{\textbf{Storm  Surge  Barrier  Damage  Allowances}  (m)}  \\
RCP8.5   &     &  \multicolumn{5}{c}{AIS$_{max}$}     &     &  RCP2.6   &     &  \multicolumn{5}{c}{AIS$_{max}$}  \\
 &    $\beta_{c}$     &  1.75 m   &  1.5  m   &  1.0  m     &  0.5  m     &  0.25  m     &     &     &  $\beta_{c}$   &  1.75 m   &  1.5  m   &  1.0  m   &  0.5  m   &  0.25  m  \\
\hline
2100   &   &   &   &   &   &  &  & 2100   &   &   &   &   &   &  \\
 &  1.0   &3.6   &  3.3   &  2.8   &  2.4   &  2.2   &   &   &  1.0   &2.0   &  1.8   &  1.7   &  1.5   &  1.4  \\
 &  0.75   &3.2   &  3.0   &  2.6   &  2.3   &  2.1   &   &   &  0.75   &1.8   &  1.7   &  1.6   &  1.5   &  1.4  \\
 &  0.50   &2.9   &  2.6   &  2.4   &  2.2   &  2.1   &   & &  0.50   &1.6   &  1.6   &  1.6   &  1.4   &  1.3  \\
 &  0.25   &2.7   &  2.4   &  2.3   &  2.1   &  2.0   &   & &  0.25   &1.5   &  1.5   &  1.5   &  1.4   &  1.3  \\
 &  0.0   &2.6   &  2.3   &  2.2   &  2.0   &  1.9   &   & &  0.0   &1.5   &  1.5   &  1.5   &  1.4   &  1.3  \\
2070   &   &   &   &   &   &   &   &  2070   &   &   &   &   &   &  \\
 &  1.0   &1.7   &  1.6   &  1.5   &  1.3   &  1.2   &   &   &  1.0   &1.2   &  1.2   &  1.2   &  1.1   &  1.1  \\
 &  0.75   &1.6   &  1.5   &  1.4   &  1.3   &  1.2   &   &   &  0.75   &1.2   &  1.2   &  1.2   &  1.1   &  1.1  \\
 &  0.50   &1.5   &  1.5   &  1.4   &  1.3   &  1.2   &   & &  0.50   &1.1   &  1.1   &  1.1   &  1.1   &  1.1  \\
 &  0.25   &1.4   &  1.4   &  1.3   &  1.3   &  1.2   &   & &  0.25   &1.1   &  1.1   &  1.1   &  1.1   &  1.0  \\
 &  0.0   &1.3   &  1.3   &  1.3   &  1.3   &  1.3   &   & &  0.0   &1.1   &  1.1   &  1.1   &  1.1   &  1.0  \\
2050   &   &   &   &   &   &   &   &  2050   &   &   &   &   &   &  \\
 &  1.0   &1.0   &  1.0   &  0.9   &  0.9   &  0.8   &   &   &  1.0   &0.9   &  0.9   &  0.9   &  0.9   &  0.8  \\
 &  0.75   &1.0   &  0.9   &  0.9   &  0.9   &  0.9   &   &   &  0.75   &0.9   &  0.9   &  0.9   &  0.8   &  0.8  \\
 &  0.50   &0.9   &  0.9   &  0.9   &  0.9   &  0.9   &   & &  0.50   &0.9   &  0.8   &  0.8   &  0.8   &  0.8  \\
 &  0.25   &0.9   &  0.9   &  0.9   &  0.9   &  0.9   &   & &  0.25   &0.8   &  0.8   &  0.8   &  0.8   &  0.8  \\
 &  0.0   &0.9   &  0.9   &  0.9   &  0.9   &  0.9   &   & &  0.0   &0.8   &  0.8   &  0.8   &  0.8   &  0.8  \\
\end{tabular}}
\label{Stab:DmgAllwncSurgeB}
\end{table}
   
\clearpage
\newpage

\begin{table}[htb]
\centering
\setlength{\tabcolsep}{3pt}
\caption{Damage allowances (meters) for an elevation strategy in which structures elevate below the allowance height (i.e., Fig. \ref{fig:dmgfunc}A) for 2100, 2070, and 2050 under representative concentration pathway (RCP) 8.5 and RCP2.6 and for different assumptions regarding future Antarctic ice sheet (AIS) behavior (e.g, likelihood of AIS collapse [$\beta_{c}$] and maximum 2100 AIS contribution [AIS$_{max}$]). The allowances are relative to the current protection height around Manhattan (assumed to be a bulkhead 1.0 m above mean higher high water [MHHW]). The elevation strategy assumes perfect compliance (i.e., all structures elevate).}
{\small
\begin{tabular}{lcccccc|lccccccc}
\multicolumn{7}{l}{\textbf{Elevation  (below  A)  Damage  Allowances}  (m)}  \\
RCP8.5 &     &  \multicolumn{5}{c}{AIS$_{max}$}     &     &  RCP2.6   &     &  \multicolumn{5}{c}{AIS$_{max}$}  \\
 &    $\beta_{c}$     &  1.75  m   &  1.5  m   &  1.0  m     &  0.5  m     &  0.25  m     &     &     &  $\beta_{c}$   &  1.75  m   &  1.5  m   &  1.0  m   &  0.5  m   &  0.25  m  \\
\hline
2100   &   &   &   &   &   &   &   &  2100   &   &   &   &   &   &  \\
 &  1.0   &3.1   &  2.8   &  2.3   &  1.9   &  1.7   &   &   &  1.0   &1.7   &  1.4   &  1.2   &  1.1   &  1.0  \\
 &  0.75   &2.7   &  2.4   &  2.1   &  1.8   &  1.7   &   &   &  0.75   &1.4   &  1.3   &  1.2   &  1.0   &  1.0  \\
 &  0.50   &2.4   &  2.1   &  1.9   &  1.7   &  1.6   &   & &  0.50   &1.2   &  1.2   &  1.1   &  1.0   &  0.9  \\
 &  0.25   &2.3   &  1.9   &  1.8   &  1.6   &  1.6   &   & &  0.25   &1.1   &  1.1   &  1.1   &  1.0   &  0.9  \\
 &  0.0   &2.2   &  1.9   &  1.7   &  1.6   &  1.5   &   & &  0.0   &1.1   &  1.1   &  1.1   &  1.0   &  0.9  \\
2070   &   &   &   &   &   &   &   &  2070   &   &   &   &   &   &  \\
 &  1.0   &1.3   &  1.2   &  1.1   &  0.9   &  0.8   &   &   &  1.0   &0.8   &  0.8   &  0.8   &  0.8   &  0.7  \\
 &  0.75   &1.2   &  1.1   &  1.0   &  0.9   &  0.8   &   &   &  0.75   &0.8   &  0.8   &  0.8   &  0.7   &  0.7  \\
 &  0.50   &1.1   &  1.0   &  1.0   &  0.9   &  0.8   &   & &  0.50   &0.8   &  0.7   &  0.7   &  0.7   &  0.7  \\
 &  0.25   &1.0   &  1.0   &  0.9   &  0.9   &  0.9   &   & &  0.25   &0.7   &  0.7   &  0.7   &  0.7   &  0.7  \\
 &  0.0   &0.9   &  0.9   &  0.9   &  0.9   &  0.9   &   & &  0.0   &0.7   &  0.7   &  0.7   &  0.7   &  0.7  \\
2050   &   &   &   &   &   &   &   &  2050   &   &   &   &   &   &  \\
 &  1.0   &0.6   &  0.6   &  0.6   &  0.5   &  0.5   &   &   &  1.0   &0.5   &  0.5   &  0.5   &  0.5   &  0.5  \\
 &  0.75   &0.6   &  0.6   &  0.6   &  0.5   &  0.5   &   &   &  0.75   &0.5   &  0.5   &  0.5   &  0.5   &  0.5  \\
 &  0.50   &0.6   &  0.6   &  0.5   &  0.5   &  0.5   &   & &  0.50   &0.5   &  0.5   &  0.5   &  0.5   &  0.5  \\
 &  0.25   &0.6   &  0.5   &  0.5   &  0.5   &  0.5   &   & &  0.25   &0.5   &  0.5   &  0.5   &  0.5   &  0.5  \\
 &  0.0   &0.5   &  0.5   &  0.5   &  0.5   &  0.5   &   & &  0.0   &0.5   &  0.5   &  0.5   &  0.5   &  0.4  \\
\hline
\end{tabular}}
\label{Stab:DmgAllwncElevBelowA}
\end{table}

\subsection{Flood damage allowances: multi-strategy approach}

\begin{figure}[htb]
\centering
\includegraphics[width=.99\textwidth]{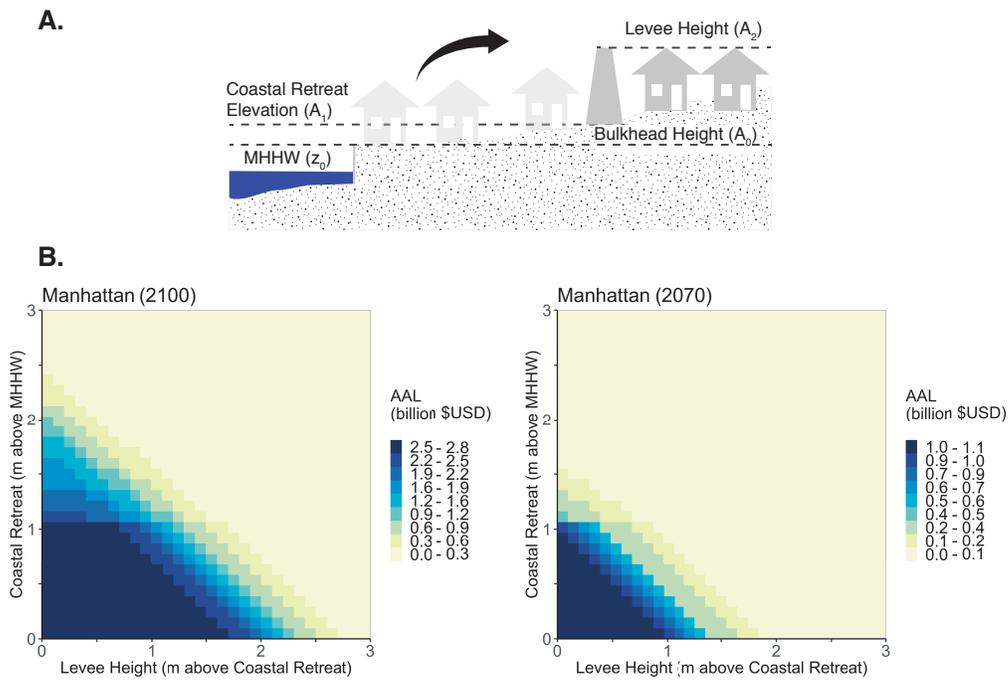}
\caption{ \textbf{A.} Schematic depicting the combined implementation of coastal retreat and a levee. \textbf{B.} Heat maps of annual average loss due to flood damages for 2100 (Left) and 2070 (Right) using the flood protection strategies of coastal retreat (y-axis) and a levee (x-axis) with design heights between 0 and 3 m for each strategy.  An additional 0.5 m of freeboard for the levee is not included in the depicted design heights along the x-axis. The levee has a 10\% probability of structural failure at the design height.} \label{Sfig:multi}
\end{figure}

We present an example of how a multi-strategy approach could be designed using both coastal retreat and a levee for Manhattan (Fig. \ref{Sfig:multi}A). For both 2070 and 2100, the flood damage risk (AAL) is mapped out for varying elevations below which coastal retreat occurs ($A_{1}$) and for the height of a levee ($A_{2}$, levee height is relative to $A_{1}$) assuming $\text{AIS}_{max}$ = 1.50 m and $\beta_{c}$ = 0.0 (Fig. \ref{Sfig:multi}B). A user could select a preferred AAL and subsequently set the coastal retreat and levee heights. For example, to maintain the current AAL of \$0.1 billion/yr in 2100 with coastal retreat below 2 m of elevation, a levee of roughly 0.75 m would need to be constructed. Additional heat maps could be used to depict alternative assumptions of future AIS behavior. 

\newpage
\clearpage

%
%
%
%
%
%
%
%

\acknowledgments

D.J.R was supported by the Science, Technology, and Environmental Policy (STEP) Program at Princeton University. R.E.K. was supported by grants from the National Science Foundation (ICER-1663807) and National Aeronautics and Space Administration (80NSSC17K0698). M.O. acknowledges support from the National Science Foundation, Award Number 1520683. Code for generating sea-level projections is available in the ProjectSL (https://github.com/bobkopp/ProjectSL) and LocalizeSL (https://github.com/bobkopp/LocalizeSL) repositories on Github. Code for generating extreme sea level projections is available in the hawaiiSL\_process (https://github.com/dmr2/hawaiiSL\_process), GPDfit (https://github.com/dmr2/GPDfit), and return\_curves (https://github.com/dmr2/return\_curves) repositories on Github. \hl{[ADD OTHERS]}



 \bibliography{damage_allowance_bib}

\begin{thebibliography}{}

\bibitem [\protect \citeauthoryear {%
Aerts%
}{%
Aerts%
}{%
{\protect \APACyear {2018}}%
}]{%
Aerts2018a}
\APACinsertmetastar {%
Aerts2018a}%
\begin{APACrefauthors}%
Aerts, J\BPBI C\BPBI J\BPBI H.%
\end{APACrefauthors}%
\unskip\
\newblock
\APACrefYearMonthDay{2018}{}{}.
\newblock
{\BBOQ}\APACrefatitle {A Review of Cost Estimates for Flood Adaptation} {A
  review of cost estimates for flood adaptation}.{\BBCQ}
\newblock
\APACjournalVolNumPages{Water}{10}{11}{}.
\newblock
\begin{APACrefDOI} \doi{10.3390/w10111646} \end{APACrefDOI}
\PrintBackRefs{\CurrentBib}

\bibitem [\protect \citeauthoryear {%
Aerts%
\ \protect \BOthers {.}}{%
Aerts%
\ \protect \BOthers {.}}{%
{\protect \APACyear {2014}}%
}]{%
Aerts2014a}
\APACinsertmetastar {%
Aerts2014a}%
\begin{APACrefauthors}%
Aerts, J\BPBI C\BPBI J\BPBI H.%
, Botzen, W\BPBI J.%
, Emanuel, K.%
, Lin, N.%
, {De Moel}, H.%
\BCBL {}\ \BBA {} Michel-Kerjan, E\BPBI O.%
\end{APACrefauthors}%
\unskip\
\newblock
\APACrefYearMonthDay{2014}{}{}.
\newblock
{\BBOQ}\APACrefatitle {{Climate adaptation: Evaluating flood resilience
  strategies for coastal megacities}} {{Climate adaptation: Evaluating flood
  resilience strategies for coastal megacities}}.{\BBCQ}
\newblock
\APACjournalVolNumPages{Science}{344}{6183}{473--475}.
\newblock
\begin{APACrefDOI} \doi{10.1126/science.1248222} \end{APACrefDOI}
\PrintBackRefs{\CurrentBib}

\bibitem [\protect \citeauthoryear {%
Arrow%
\ \protect \BOthers {.}}{%
Arrow%
\ \protect \BOthers {.}}{%
{\protect \APACyear {1996}}%
}]{%
Arrow1996a}
\APACinsertmetastar {%
Arrow1996a}%
\begin{APACrefauthors}%
Arrow, K\BPBI J.%
, Cropper, M\BPBI L.%
, Eads, G\BPBI C.%
, Hahn, R\BPBI W.%
, Lave, L\BPBI B.%
, Noll, R\BPBI G.%
\BDBL {}Stavins, R\BPBI N.%
\end{APACrefauthors}%
\unskip\
\newblock
\APACrefYearMonthDay{1996}{}{}.
\newblock
{\BBOQ}\APACrefatitle {{Is There a Role for Benefit-Cost Analysis in
  Environmental, Health, and Safety Regulation?}} {{Is There a Role for
  Benefit-Cost Analysis in Environmental, Health, and Safety
  Regulation?}}{\BBCQ}
\newblock
\APACjournalVolNumPages{Science}{272}{5259}{221--222}.
\newblock
\begin{APACrefDOI} \doi{10.1126/science.272.5259.221} \end{APACrefDOI}
\PrintBackRefs{\CurrentBib}

\bibitem [\protect \citeauthoryear {%
Bakker%
, Wong%
, Ruckert%
\BCBL {}\ \BBA {} Keller%
}{%
Bakker%
\ \protect \BOthers {.}}{%
{\protect \APACyear {2017}}%
}]{%
Bakker2017a}
\APACinsertmetastar {%
Bakker2017a}%
\begin{APACrefauthors}%
Bakker, A\BPBI M\BPBI R.%
, Wong, T\BPBI E.%
, Ruckert, K\BPBI L.%
\BCBL {}\ \BBA {} Keller, K.%
\end{APACrefauthors}%
\unskip\
\newblock
\APACrefYearMonthDay{2017}{}{}.
\newblock
{\BBOQ}\APACrefatitle {{Sea-level projections representing the deeply uncertain
  contribution of the West Antarctic ice sheet}} {{Sea-level projections
  representing the deeply uncertain contribution of the West Antarctic ice
  sheet}}.{\BBCQ}
\newblock
\APACjournalVolNumPages{Scientific Reports}{7}{1}{3880}.
\newblock
\begin{APACrefDOI} \doi{10.1038/s41598-017-04134-5} \end{APACrefDOI}
\PrintBackRefs{\CurrentBib}

\bibitem [\protect \citeauthoryear {%
Bamber%
\ \BBA {} Aspinall%
}{%
Bamber%
\ \BBA {} Aspinall%
}{%
{\protect \APACyear {2013}}%
}]{%
Bamber2013a}
\APACinsertmetastar {%
Bamber2013a}%
\begin{APACrefauthors}%
Bamber, J\BPBI L.%
\BCBT {}\ \BBA {} Aspinall, W\BPBI P.%
\end{APACrefauthors}%
\unskip\
\newblock
\APACrefYearMonthDay{2013}{}{}.
\newblock
{\BBOQ}\APACrefatitle {An expert judgement assessment of future sea level rise
  from the ice sheets} {An expert judgement assessment of future sea level rise
  from the ice sheets}.{\BBCQ}
\newblock
\APACjournalVolNumPages{Nature Climate Change}{3}{}{424--427}.
\newblock
\begin{APACrefDOI} \doi{10.1038/nclimate1778} \end{APACrefDOI}
\PrintBackRefs{\CurrentBib}

\bibitem [\protect \citeauthoryear {%
Bamber%
, Oppenheimer%
, Kopp%
, Aspinall%
\BCBL {}\ \BBA {} Cooke%
}{%
Bamber%
\ \protect \BOthers {.}}{%
{\protect \APACyear {2019}}%
}]{%
Bamber2019a}
\APACinsertmetastar {%
Bamber2019a}%
\begin{APACrefauthors}%
Bamber, J\BPBI L.%
, Oppenheimer, M.%
, Kopp, R\BPBI E.%
, Aspinall, W\BPBI P.%
\BCBL {}\ \BBA {} Cooke, R\BPBI M.%
\end{APACrefauthors}%
\unskip\
\newblock
\APACrefYearMonthDay{2019}{}{}.
\newblock
{\BBOQ}\APACrefatitle {Ice sheet contributions to future sea-level rise from
  structured expert judgment} {Ice sheet contributions to future sea-level rise
  from structured expert judgment}.{\BBCQ}
\newblock
\APACjournalVolNumPages{Proceedings of the National Academy of
  Sciences}{116}{23}{11195--11200}.
\newblock
\begin{APACrefDOI} \doi{10.1073/pnas.1817205116} \end{APACrefDOI}
\PrintBackRefs{\CurrentBib}

\bibitem [\protect \citeauthoryear {%
Baudrit%
, Guyonnet%
\BCBL {}\ \BBA {} Dubois%
}{%
Baudrit%
\ \protect \BOthers {.}}{%
{\protect \APACyear {2007}}%
}]{%
Baudrit2007a}
\APACinsertmetastar {%
Baudrit2007a}%
\begin{APACrefauthors}%
Baudrit, C.%
, Guyonnet, D.%
\BCBL {}\ \BBA {} Dubois, D.%
\end{APACrefauthors}%
\unskip\
\newblock
\APACrefYearMonthDay{2007}{}{}.
\newblock
{\BBOQ}\APACrefatitle {{Joint propagation of variability and imprecision in
  assessing the risk of groundwater contamination}} {{Joint propagation of
  variability and imprecision in assessing the risk of groundwater
  contamination}}.{\BBCQ}
\newblock
\APACjournalVolNumPages{Journal of Contaminant Hydrology}{93}{1-4}{72--84}.
\newblock
\begin{APACrefDOI} \doi{10.1016/j.jconhyd.2007.01.015} \end{APACrefDOI}
\PrintBackRefs{\CurrentBib}

\bibitem [\protect \citeauthoryear {%
{Buchanan}%
, {Kopp}%
, {Oppenheimer}%
\BCBL {}\ \BBA {} {Tebaldi}%
}{%
{Buchanan}%
\ \protect \BOthers {.}}{%
{\protect \APACyear {2016}}%
}]{%
Buchanan2016a}
\APACinsertmetastar {%
Buchanan2016a}%
\begin{APACrefauthors}%
{Buchanan}, M\BPBI K.%
, {Kopp}, R\BPBI E.%
, {Oppenheimer}, M.%
\BCBL {}\ \BBA {} {Tebaldi}, C.%
\end{APACrefauthors}%
\unskip\
\newblock
\APACrefYearMonthDay{2016}{}{}.
\newblock
{\BBOQ}\APACrefatitle {{Allowances for evolving coastal flood risk under
  uncertain local sea-level rise}} {{Allowances for evolving coastal flood risk
  under uncertain local sea-level rise}}.{\BBCQ}
\newblock
\APACjournalVolNumPages{Climatic Change}{}{}{}.
\newblock
\begin{APACrefDOI} \doi{10.1007/s10584-016-1664-7} \end{APACrefDOI}
\PrintBackRefs{\CurrentBib}

\bibitem [\protect \citeauthoryear {%
Buchanan%
, Oppenheimer%
\BCBL {}\ \BBA {} Kopp%
}{%
Buchanan%
\ \protect \BOthers {.}}{%
{\protect \APACyear {2017}}%
}]{%
Buchanan2017a}
\APACinsertmetastar {%
Buchanan2017a}%
\begin{APACrefauthors}%
Buchanan, M\BPBI K.%
, Oppenheimer, M.%
\BCBL {}\ \BBA {} Kopp, R\BPBI E.%
\end{APACrefauthors}%
\unskip\
\newblock
\APACrefYearMonthDay{2017}{}{}.
\newblock
{\BBOQ}\APACrefatitle {{Amplification of flood frequencies with local sea level
  rise and emerging flood regimes}} {{Amplification of flood frequencies with
  local sea level rise and emerging flood regimes}}.{\BBCQ}
\newblock
\APACjournalVolNumPages{Environmental Research Letters}{12}{6}{}.
\newblock
\begin{APACrefDOI} \doi{10.1088/1748-9326/aa6cb3} \end{APACrefDOI}
\PrintBackRefs{\CurrentBib}

\bibitem [\protect \citeauthoryear {%
Caldwell%
, Merrifield%
\BCBL {}\ \BBA {} Thompson%
}{%
Caldwell%
\ \protect \BOthers {.}}{%
{\protect \APACyear {2015}}%
}]{%
Caldwell2015}
\APACinsertmetastar {%
Caldwell2015}%
\begin{APACrefauthors}%
Caldwell, P\BPBI C.%
, Merrifield, M\BPBI A.%
\BCBL {}\ \BBA {} Thompson, P\BPBI R.%
\end{APACrefauthors}%
\unskip\
\newblock
\APACrefYearMonthDay{2015}{}{}.
\newblock
{\BBOQ}\APACrefatitle {{Sea level measured by tide gauges from global oceans
  --- the Joint Archive for Sea Level holdings (NCEI Accession 0019568)}} {{Sea
  level measured by tide gauges from global oceans --- the Joint Archive for
  Sea Level holdings (NCEI Accession 0019568)}}.{\BBCQ}
\newblock
\APACjournalVolNumPages{NOAA National Centers for Environmental
  Information}{Dataset}{}{}.
\PrintBackRefs{\CurrentBib}

\bibitem [\protect \citeauthoryear {%
Church%
, Clark%
\BCBL {}\ \protect \BOthers {.}}{%
Church%
\ \protect \BOthers {.}}{%
{\protect \APACyear {2013}}%
}]{%
Church2013a}
\APACinsertmetastar {%
Church2013a}%
\begin{APACrefauthors}%
Church, J\BPBI A.%
, Clark, P\BPBI U.%
\BCBL {}\ \BOthersPeriod {.}\end{APACrefauthors}%
\unskip\
\newblock
\APACrefYearMonthDay{2013}{}{}.
\newblock
{\BBOQ}\APACrefatitle {Chapter 13: Sea Level Change} {Chapter 13: Sea level
  change}.{\BBCQ}
\newblock
\BIn{} T\BPBI F.~Stocker\ \BOthers {.}\ (\BEDS), \APACrefbtitle {{Climate
  Change 2013: the Physical Science Basis}.} {{Climate Change 2013: the
  Physical Science Basis}.}
\newblock
\APACaddressPublisher{}{Cambridge University Press}.
\PrintBackRefs{\CurrentBib}

\bibitem [\protect \citeauthoryear {%
Coles%
}{%
Coles%
}{%
{\protect \APACyear {2001}}%
{\protect \APACexlab {{\protect \BCnt {1}}}}}]{%
Coles2001b}
\APACinsertmetastar {%
Coles2001b}%
\begin{APACrefauthors}%
Coles, S.%
\end{APACrefauthors}%
\unskip\
\newblock
\APACrefYearMonthDay{2001{\protect \BCnt {1}}}{}{}.
\newblock
{\BBOQ}\APACrefatitle {{An Introduction to Statistical Modeling of Extreme
  Values}} {{An Introduction to Statistical Modeling of Extreme
  Values}}.{\BBCQ}
\newblock
\BIn{} (\BCHAP\ {Classical Extreme Value Theory and Models}).
\newblock
\APACaddressPublisher{}{Springer}.
\PrintBackRefs{\CurrentBib}

\bibitem [\protect \citeauthoryear {%
Coles%
}{%
Coles%
}{%
{\protect \APACyear {2001}}%
{\protect \APACexlab {{\protect \BCnt {2}}}}}]{%
Coles2001a}
\APACinsertmetastar {%
Coles2001a}%
\begin{APACrefauthors}%
Coles, S.%
\end{APACrefauthors}%
\unskip\
\newblock
\APACrefYearMonthDay{2001{\protect \BCnt {2}}}{August}{}.
\newblock
{\BBOQ}\APACrefatitle {{An Introduction to Statistical Modeling of Extreme
  Values}} {{An Introduction to Statistical Modeling of Extreme
  Values}}.{\BBCQ}
\newblock
\BIn{} (\BPG~208).
\newblock
\APACaddressPublisher{}{Springer}.
\PrintBackRefs{\CurrentBib}

\bibitem [\protect \citeauthoryear {%
Deconto%
\ \BBA {} Pollard%
}{%
Deconto%
\ \BBA {} Pollard%
}{%
{\protect \APACyear {2016}}%
}]{%
Deconto2016a}
\APACinsertmetastar {%
Deconto2016a}%
\begin{APACrefauthors}%
Deconto, R\BPBI M.%
\BCBT {}\ \BBA {} Pollard, D.%
\end{APACrefauthors}%
\unskip\
\newblock
\APACrefYearMonthDay{2016}{}{}.
\newblock
{\BBOQ}\APACrefatitle {{Contribution of Antarctica to past and future sea-level
  rise}} {{Contribution of Antarctica to past and future sea-level
  rise}}.{\BBCQ}
\newblock
\APACjournalVolNumPages{Nature}{531}{7596}{591--597}.
\newblock
\begin{APACrefDOI} \doi{10.1038/nature17145} \end{APACrefDOI}
\PrintBackRefs{\CurrentBib}

\bibitem [\protect \citeauthoryear {%
Diaz%
}{%
Diaz%
}{%
{\protect \APACyear {2016}}%
}]{%
Diaz2016a}
\APACinsertmetastar {%
Diaz2016a}%
\begin{APACrefauthors}%
Diaz, D\BPBI B.%
\end{APACrefauthors}%
\unskip\
\newblock
\APACrefYearMonthDay{2016}{}{}.
\newblock
{\BBOQ}\APACrefatitle {{Estimating global damages from sea level rise with the
  Coastal Impact and Adaptation Model (CIAM)}} {{Estimating global damages from
  sea level rise with the Coastal Impact and Adaptation Model (CIAM)}}.{\BBCQ}
\newblock
\APACjournalVolNumPages{Climatic Change}{137}{1-2}{143--156}.
\newblock
\begin{APACrefDOI} \doi{10.1007/s10584-016-1675-4} \end{APACrefDOI}
\PrintBackRefs{\CurrentBib}

\bibitem [\protect \citeauthoryear {%
Dronkers%
\ \protect \BOthers {.}}{%
Dronkers%
\ \protect \BOthers {.}}{%
{\protect \APACyear {1990}}%
}]{%
CZMS1990a}
\APACinsertmetastar {%
CZMS1990a}%
\begin{APACrefauthors}%
Dronkers, J.%
, Gilbert, J\BPBI T\BPBI E.%
, Butler, L.%
, Carey, J.%
, Campbell, J.%
, James, E.%
\BDBL {}von Dadelszen, J.%
\end{APACrefauthors}%
\unskip\
\newblock
\APACrefYear{1990}.
\newblock
\APACrefbtitle {{Strategies for Adaptation to Sea Level Rise}} {{Strategies for
  Adaptation to Sea Level Rise}}.
\newblock
\APACaddressPublisher{Geneva}{{Intergovernmental Panel on Climate Change}}.
\PrintBackRefs{\CurrentBib}

\bibitem [\protect \citeauthoryear {%
Edwards%
\ \protect \BOthers {.}}{%
Edwards%
\ \protect \BOthers {.}}{%
{\protect \APACyear {2019}}%
}]{%
Edwards2019a}
\APACinsertmetastar {%
Edwards2019a}%
\begin{APACrefauthors}%
Edwards, T\BPBI L.%
, Brandon, M\BPBI A.%
, Durand, G.%
, Edwards, N\BPBI R.%
, Golledge, N\BPBI R.%
, Holden, P\BPBI B.%
\BDBL {}Wernecke, A.%
\end{APACrefauthors}%
\unskip\
\newblock
\APACrefYearMonthDay{2019}{}{}.
\newblock
{\BBOQ}\APACrefatitle {Revisiting Antarctic ice loss due to marine ice-cliff
  instability} {Revisiting antarctic ice loss due to marine ice-cliff
  instability}.{\BBCQ}
\newblock
\APACjournalVolNumPages{Nature}{566}{7742}{58--64}.
\newblock
\begin{APACrefDOI} \doi{10.1038/s41586-019-0901-4} \end{APACrefDOI}
\PrintBackRefs{\CurrentBib}

\bibitem [\protect \citeauthoryear {%
Fankhauser%
}{%
Fankhauser%
}{%
{\protect \APACyear {1995}}%
}]{%
Fankhauser1995a}
\APACinsertmetastar {%
Fankhauser1995a}%
\begin{APACrefauthors}%
Fankhauser, S.%
\end{APACrefauthors}%
\unskip\
\newblock
\APACrefYearMonthDay{1995}{}{}.
\newblock
{\BBOQ}\APACrefatitle {{Protection versus Retreat: The Economic Costs of
  Sea-Level Rise}} {{Protection versus Retreat: The Economic Costs of Sea-Level
  Rise}}.{\BBCQ}
\newblock
\APACjournalVolNumPages{Environment and Planning A}{27}{2}{299--319}.
\newblock
\begin{APACrefDOI} \doi{10.1068/a270299} \end{APACrefDOI}
\PrintBackRefs{\CurrentBib}

\bibitem [\protect \citeauthoryear {%
Fankhauser%
, Smith%
\BCBL {}\ \BBA {} Tol%
}{%
Fankhauser%
\ \protect \BOthers {.}}{%
{\protect \APACyear {1999}}%
}]{%
Fankhauser1999a}
\APACinsertmetastar {%
Fankhauser1999a}%
\begin{APACrefauthors}%
Fankhauser, S.%
, Smith, J\BPBI B.%
\BCBL {}\ \BBA {} Tol, R\BPBI S.%
\end{APACrefauthors}%
\unskip\
\newblock
\APACrefYearMonthDay{1999}{}{}.
\newblock
{\BBOQ}\APACrefatitle {{Weathering climate change: some simple rules to guide
  adaptation decisions}} {{Weathering climate change: some simple rules to
  guide adaptation decisions}}.{\BBCQ}
\newblock
\APACjournalVolNumPages{Ecological Economics}{30}{1}{67 - 78}.
\newblock
\begin{APACrefDOI} \doi{https://doi.org/10.1016/S0921-8009(98)00117-7}
  \end{APACrefDOI}
\PrintBackRefs{\CurrentBib}

\bibitem [\protect \citeauthoryear {%
Fischbach%
, Johnson%
\BCBL {}\ \BBA {} Molina-Perez%
}{%
Fischbach%
\ \protect \BOthers {.}}{%
{\protect \APACyear {2017}}%
}]{%
Fischbach2017a}
\APACinsertmetastar {%
Fischbach2017a}%
\begin{APACrefauthors}%
Fischbach, J.%
, Johnson, D.%
\BCBL {}\ \BBA {} Molina-Perez, E.%
\end{APACrefauthors}%
\unskip\
\newblock
\APACrefYearMonthDay{2017}{}{}.
\newblock
\APACrefbtitle {{Reducing Coastal Flood Risk with a Lake Pontchartrain
  Barrier}} {{Reducing Coastal Flood Risk with a Lake Pontchartrain Barrier}}\
  (\BNUM\ RR-1988-CPRA).
\newblock
\APAChowpublished {Web-only}.
\newblock
\APACaddressPublisher{Santa Monica, CA}{}.
\newblock
\begin{APACrefDOI} \doi{10.7249/rr1988} \end{APACrefDOI}
\PrintBackRefs{\CurrentBib}

\bibitem [\protect \citeauthoryear {%
Ghanbari%
, Arabi%
, Obeysekera%
\BCBL {}\ \BBA {} Sweet%
}{%
Ghanbari%
\ \protect \BOthers {.}}{%
{\protect \APACyear {2019}}%
}]{%
Ghanbari2019a}
\APACinsertmetastar {%
Ghanbari2019a}%
\begin{APACrefauthors}%
Ghanbari, M.%
, Arabi, M.%
, Obeysekera, J.%
\BCBL {}\ \BBA {} Sweet, W.%
\end{APACrefauthors}%
\unskip\
\newblock
\APACrefYearMonthDay{2019}{}{}.
\newblock
{\BBOQ}\APACrefatitle {{A Coherent Statistical Model for Coastal Flood
  Frequency Analysis Under Nonstationary Sea Level Conditions}} {{A Coherent
  Statistical Model for Coastal Flood Frequency Analysis Under Nonstationary
  Sea Level Conditions}}.{\BBCQ}
\newblock
\APACjournalVolNumPages{Earth's Future}{}{}{}.
\newblock
\begin{APACrefDOI} \doi{10.1029/2018EF001089} \end{APACrefDOI}
\PrintBackRefs{\CurrentBib}

\bibitem [\protect \citeauthoryear {%
Golledge%
\ \protect \BOthers {.}}{%
Golledge%
\ \protect \BOthers {.}}{%
{\protect \APACyear {2019}}%
}]{%
Golledge2019a}
\APACinsertmetastar {%
Golledge2019a}%
\begin{APACrefauthors}%
Golledge, N\BPBI R.%
, Keller, E\BPBI D.%
, Gomez, N.%
, Naughten, K\BPBI A.%
, Bernales, J.%
, Trusel, L\BPBI D.%
\BCBL {}\ \BBA {} Edwards, T\BPBI L.%
\end{APACrefauthors}%
\unskip\
\newblock
\APACrefYearMonthDay{2019}{}{}.
\newblock
{\BBOQ}\APACrefatitle {Global environmental consequences of
  twenty-first-century ice-sheet melt} {Global environmental consequences of
  twenty-first-century ice-sheet melt}.{\BBCQ}
\newblock
\APACjournalVolNumPages{Nature}{566}{7742}{65--72}.
\newblock
\begin{APACrefDOI} \doi{10.1038/s41586-019-0889-9} \end{APACrefDOI}
\PrintBackRefs{\CurrentBib}

\bibitem [\protect \citeauthoryear {%
Haasnoot%
, Kwakkel%
, Walker%
\BCBL {}\ \BBA {} ter Maat%
}{%
Haasnoot%
\ \protect \BOthers {.}}{%
{\protect \APACyear {2013}}%
}]{%
Haasnoot2013a}
\APACinsertmetastar {%
Haasnoot2013a}%
\begin{APACrefauthors}%
Haasnoot, M.%
, Kwakkel, J\BPBI H.%
, Walker, W\BPBI E.%
\BCBL {}\ \BBA {} ter Maat, J.%
\end{APACrefauthors}%
\unskip\
\newblock
\APACrefYearMonthDay{2013}{}{}.
\newblock
{\BBOQ}\APACrefatitle {{Dynamic adaptive policy pathways: A method for crafting
  robust decisions for a deeply uncertain world}} {{Dynamic adaptive policy
  pathways: A method for crafting robust decisions for a deeply uncertain
  world}}.{\BBCQ}
\newblock
\APACjournalVolNumPages{Global Environmental Change}{23}{2}{485 - 498}.
\newblock
\begin{APACrefDOI} \doi{https://doi.org/10.1016/j.gloenvcha.2012.12.006}
  \end{APACrefDOI}
\PrintBackRefs{\CurrentBib}

\bibitem [\protect \citeauthoryear {%
Haasnoot%
\ \protect \BOthers {.}}{%
Haasnoot%
\ \protect \BOthers {.}}{%
{\protect \APACyear {2019}}%
}]{%
Haasnoot2019a}
\APACinsertmetastar {%
Haasnoot2019a}%
\begin{APACrefauthors}%
Haasnoot, M.%
, van Aalst, M.%
, Rozenberg, J.%
, Dominique, K.%
, Matthews, J.%
, Bouwer, L\BPBI M.%
\BDBL {}Poff, N\BPBI L.%
\end{APACrefauthors}%
\unskip\
\newblock
\APACrefYearMonthDay{2019}{Apr}{04}.
\newblock
{\BBOQ}\APACrefatitle {Investments under non-stationarity: economic evaluation
  of adaptation pathways} {Investments under non-stationarity: economic
  evaluation of adaptation pathways}.{\BBCQ}
\newblock
\APACjournalVolNumPages{Climatic Change}{}{}{}.
\newblock
\begin{APACrefDOI} \doi{10.1007/s10584-019-02409-6} \end{APACrefDOI}
\PrintBackRefs{\CurrentBib}

\bibitem [\protect \citeauthoryear {%
Hall%
}{%
Hall%
}{%
{\protect \APACyear {2007}}%
}]{%
Hall2007a}
\APACinsertmetastar {%
Hall2007a}%
\begin{APACrefauthors}%
Hall, J.%
\end{APACrefauthors}%
\unskip\
\newblock
\APACrefYearMonthDay{2007}{}{}.
\newblock
{\BBOQ}\APACrefatitle {Probabilistic climate scenarios may misrepresent
  uncertainty and lead to bad adaptation decisions} {Probabilistic climate
  scenarios may misrepresent uncertainty and lead to bad adaptation
  decisions}.{\BBCQ}
\newblock
\APACjournalVolNumPages{Hydrological Processes}{21}{8}{1127-1129}.
\newblock
\begin{APACrefDOI} \doi{10.1002/hyp.6573} \end{APACrefDOI}
\PrintBackRefs{\CurrentBib}

\bibitem [\protect \citeauthoryear {%
Hallegatte%
, Green%
, Nicholls%
\BCBL {}\ \BBA {} Corfee-Morlot%
}{%
Hallegatte%
\ \protect \BOthers {.}}{%
{\protect \APACyear {2013}}%
}]{%
Hallegatte2013a}
\APACinsertmetastar {%
Hallegatte2013a}%
\begin{APACrefauthors}%
Hallegatte, S.%
, Green, C.%
, Nicholls, R\BPBI J.%
\BCBL {}\ \BBA {} Corfee-Morlot, J.%
\end{APACrefauthors}%
\unskip\
\newblock
\APACrefYearMonthDay{2013}{}{}.
\newblock
{\BBOQ}\APACrefatitle {{Future flood losses in major coastal cities}} {{Future
  flood losses in major coastal cities}}.{\BBCQ}
\newblock
\APACjournalVolNumPages{Nature Climate Change}{3}{9}{802--806}.
\newblock
\begin{APACrefDOI} \doi{10.1038/nclimate1979} \end{APACrefDOI}
\PrintBackRefs{\CurrentBib}

\bibitem [\protect \citeauthoryear {%
Hanson%
\ \protect \BOthers {.}}{%
Hanson%
\ \protect \BOthers {.}}{%
{\protect \APACyear {2011}}%
}]{%
Hanson2011a}
\APACinsertmetastar {%
Hanson2011a}%
\begin{APACrefauthors}%
Hanson, S.%
, Nicholls, R.%
, Ranger, N.%
, Hallegatte, S.%
, Corfee-Morlot, J.%
, Herweijer, C.%
\BCBL {}\ \BBA {} Chateau, J.%
\end{APACrefauthors}%
\unskip\
\newblock
\APACrefYearMonthDay{2011}{}{}.
\newblock
{\BBOQ}\APACrefatitle {{A global ranking of port cities with high exposure to
  climate extremes}} {{A global ranking of port cities with high exposure to
  climate extremes}}.{\BBCQ}
\newblock
\APACjournalVolNumPages{Climatic Change}{104}{1}{89--111}.
\newblock
\begin{APACrefDOI} \doi{10.1007/s10584-010-9977-4} \end{APACrefDOI}
\PrintBackRefs{\CurrentBib}

\bibitem [\protect \citeauthoryear {%
Healy%
\ \BBA {} Malhotra%
}{%
Healy%
\ \BBA {} Malhotra%
}{%
{\protect \APACyear {2009}}%
}]{%
Healy2009a}
\APACinsertmetastar {%
Healy2009a}%
\begin{APACrefauthors}%
Healy, A.%
\BCBT {}\ \BBA {} Malhotra, N.%
\end{APACrefauthors}%
\unskip\
\newblock
\APACrefYearMonthDay{2009}{}{}.
\newblock
{\BBOQ}\APACrefatitle {{Myopic Voters and Natural Disaster Policy}} {{Myopic
  Voters and Natural Disaster Policy}}.{\BBCQ}
\newblock
\APACjournalVolNumPages{American Political Science Review}{103}{3}{387--406}.
\newblock
\begin{APACrefDOI} \doi{10.1017/S0003055409990104} \end{APACrefDOI}
\PrintBackRefs{\CurrentBib}

\bibitem [\protect \citeauthoryear {%
Hinkel%
\ \protect \BOthers {.}}{%
Hinkel%
\ \protect \BOthers {.}}{%
{\protect \APACyear {2018}}%
}]{%
Hinkel2018a}
\APACinsertmetastar {%
Hinkel2018a}%
\begin{APACrefauthors}%
Hinkel, J.%
, Aerts, J\BPBI C\BPBI J\BPBI H.%
, Brown, S.%
, Jim{\'e}nez, J\BPBI A.%
, Lincke, D.%
, Nicholls, R\BPBI J.%
\BDBL {}Addo, K\BPBI A.%
\end{APACrefauthors}%
\unskip\
\newblock
\APACrefYearMonthDay{2018}{}{}.
\newblock
{\BBOQ}\APACrefatitle {The ability of societies to adapt to
  twenty-first-century sea-level rise} {The ability of societies to adapt to
  twenty-first-century sea-level rise}.{\BBCQ}
\newblock
\APACjournalVolNumPages{Nature Climate Change}{8}{7}{570--578}.
\newblock
\begin{APACrefDOI} \doi{10.1038/s41558-018-0176-z} \end{APACrefDOI}
\PrintBackRefs{\CurrentBib}

\bibitem [\protect \citeauthoryear {%
Hinkel%
\ \protect \BOthers {.}}{%
Hinkel%
\ \protect \BOthers {.}}{%
{\protect \APACyear {2014}}%
}]{%
Hinkel2014a}
\APACinsertmetastar {%
Hinkel2014a}%
\begin{APACrefauthors}%
Hinkel, J.%
, Lincke, D.%
, Vafeidis, A\BPBI T.%
, Perrette, M.%
, Nicholls, R\BPBI J.%
, Tol, R\BPBI S\BPBI J.%
\BDBL {}Levermann, A.%
\end{APACrefauthors}%
\unskip\
\newblock
\APACrefYearMonthDay{2014}{}{}.
\newblock
{\BBOQ}\APACrefatitle {{Coastal flood damage and adaptation costs under 21st
  century sea-level rise.}} {{Coastal flood damage and adaptation costs under
  21st century sea-level rise.}}{\BBCQ}
\newblock
\APACjournalVolNumPages{Proceedings of the National Academy of Sciences of the
  United States of America}{111}{9}{3292--7}.
\newblock
\begin{APACrefDOI} \doi{10.1073/pnas.1222469111} \end{APACrefDOI}
\PrintBackRefs{\CurrentBib}

\bibitem [\protect \citeauthoryear {%
Horton%
\ \protect \BOthers {.}}{%
Horton%
\ \protect \BOthers {.}}{%
{\protect \APACyear {2018}}%
}]{%
Horton2018a}
\APACinsertmetastar {%
Horton2018a}%
\begin{APACrefauthors}%
Horton, B\BPBI P.%
, Kopp, R\BPBI E.%
, Garner, A\BPBI J.%
, Hay, C\BPBI C.%
, Khan, N\BPBI S.%
, Roy, K.%
\BCBL {}\ \BBA {} Shaw, T\BPBI A.%
\end{APACrefauthors}%
\unskip\
\newblock
\APACrefYearMonthDay{2018}{}{}.
\newblock
{\BBOQ}\APACrefatitle {Mapping Sea-Level Change in Time, Space, and
  Probability} {Mapping sea-level change in time, space, and
  probability}.{\BBCQ}
\newblock
\APACjournalVolNumPages{Annual Review of Environment and
  Resources}{43}{1}{481-521}.
\newblock
\begin{APACrefDOI} \doi{10.1146/annurev-environ-102017-025826} \end{APACrefDOI}
\PrintBackRefs{\CurrentBib}

\bibitem [\protect \citeauthoryear {%
Hunter%
}{%
Hunter%
}{%
{\protect \APACyear {2012}}%
}]{%
Hunter2012a}
\APACinsertmetastar {%
Hunter2012a}%
\begin{APACrefauthors}%
Hunter, J.%
\end{APACrefauthors}%
\unskip\
\newblock
\APACrefYearMonthDay{2012}{}{}.
\newblock
{\BBOQ}\APACrefatitle {A simple technique for estimating an allowance for
  uncertain sea-level rise} {A simple technique for estimating an allowance for
  uncertain sea-level rise}.{\BBCQ}
\newblock
\APACjournalVolNumPages{Climatic Change}{113}{}{239--252}.
\newblock
\begin{APACrefDOI} \doi{10.1007/s10584-011-0332-1} \end{APACrefDOI}
\PrintBackRefs{\CurrentBib}

\bibitem [\protect \citeauthoryear {%
Jevrejeva%
, Jackson%
, Grinsted%
, Lincke%
\BCBL {}\ \BBA {} Marzeion%
}{%
Jevrejeva%
\ \protect \BOthers {.}}{%
{\protect \APACyear {2018}}%
}]{%
Jevrejeva2018a}
\APACinsertmetastar {%
Jevrejeva2018a}%
\begin{APACrefauthors}%
Jevrejeva, S.%
, Jackson, L\BPBI P.%
, Grinsted, A.%
, Lincke, D.%
\BCBL {}\ \BBA {} Marzeion, B.%
\end{APACrefauthors}%
\unskip\
\newblock
\APACrefYearMonthDay{2018}{}{}.
\newblock
{\BBOQ}\APACrefatitle {{Flood damage costs under the sea level rise with
  warming of 1.5 ${^\circ}$C and 2.0 ${^\circ}$C}} {{Flood damage costs under
  the sea level rise with warming of 1.5 ${^\circ}$C and 2.0
  ${^\circ}$C}}.{\BBCQ}
\newblock
\APACjournalVolNumPages{Environ. Res. Lett.}{13}{074014}{11}.
\newblock
\begin{APACrefDOI} \doi{10.1088/1748-9326/aacc76} \end{APACrefDOI}
\PrintBackRefs{\CurrentBib}

\bibitem [\protect \citeauthoryear {%
Jonkman%
, Hillen%
, Nicholls%
, Kanning%
\BCBL {}\ \BBA {} van Ledden%
}{%
Jonkman%
\ \protect \BOthers {.}}{%
{\protect \APACyear {2013}}%
}]{%
Jonkman2013a}
\APACinsertmetastar {%
Jonkman2013a}%
\begin{APACrefauthors}%
Jonkman, S\BPBI N.%
, Hillen, M\BPBI M.%
, Nicholls, R\BPBI J.%
, Kanning, W.%
\BCBL {}\ \BBA {} van Ledden, M.%
\end{APACrefauthors}%
\unskip\
\newblock
\APACrefYearMonthDay{2013}{}{}.
\newblock
{\BBOQ}\APACrefatitle {Costs of Adapting Coastal Defences to Sea-Level Rise---
  New Estimates and Their Implications} {Costs of adapting coastal defences to
  sea-level rise--- new estimates and their implications}.{\BBCQ}
\newblock
\APACjournalVolNumPages{Journal of Coastal Research}{}{}{1212-1226}.
\newblock
\begin{APACrefDOI} \doi{10.2112/JCOASTRES-D-12-00230.1} \end{APACrefDOI}
\PrintBackRefs{\CurrentBib}

\bibitem [\protect \citeauthoryear {%
Kanyama%
, Svahn%
\BCBL {}\ \BBA {} Sonnek%
}{%
Kanyama%
\ \protect \BOthers {.}}{%
{\protect \APACyear {2019}}%
}]{%
Kanyama2019a}
\APACinsertmetastar {%
Kanyama2019a}%
\begin{APACrefauthors}%
Kanyama, A\BPBI C.%
, Svahn, P\BPBI W.%
\BCBL {}\ \BBA {} Sonnek, K\BPBI M.%
\end{APACrefauthors}%
\unskip\
\newblock
\APACrefYearMonthDay{2019}{}{}.
\newblock
{\BBOQ}\APACrefatitle {{``We want to know where the line is'': comparing
  current planning for future sea-level rise with three core principles of
  robust decision support approaches}} {{``We want to know where the line is'':
  comparing current planning for future sea-level rise with three core
  principles of robust decision support approaches}}.{\BBCQ}
\newblock
\APACjournalVolNumPages{Journal of Environmental Planning and
  Management}{0}{0}{1--20}.
\newblock
\begin{APACrefDOI} \doi{10.1080/09640568.2018.1496070} \end{APACrefDOI}
\PrintBackRefs{\CurrentBib}

\bibitem [\protect \citeauthoryear {%
Kaplan%
\ \BBA {} Garrick%
}{%
Kaplan%
\ \BBA {} Garrick%
}{%
{\protect \APACyear {1981}}%
}]{%
Kaplan1981a}
\APACinsertmetastar {%
Kaplan1981a}%
\begin{APACrefauthors}%
Kaplan, S.%
\BCBT {}\ \BBA {} Garrick, B\BPBI J.%
\end{APACrefauthors}%
\unskip\
\newblock
\APACrefYearMonthDay{1981}{}{}.
\newblock
{\BBOQ}\APACrefatitle {{On The Quantitative Definition of Risk}} {{On The
  Quantitative Definition of Risk}}.{\BBCQ}
\newblock
\APACjournalVolNumPages{Risk Analysis}{1}{1}{11--27}.
\newblock
\begin{APACrefDOI} \doi{10.1111/j.1539-6924.1981.tb01350.x} \end{APACrefDOI}
\PrintBackRefs{\CurrentBib}

\bibitem [\protect \citeauthoryear {%
Klein%
\ \protect \BOthers {.}}{%
Klein%
\ \protect \BOthers {.}}{%
{\protect \APACyear {2001}}%
}]{%
Klein2001a}
\APACinsertmetastar {%
Klein2001a}%
\begin{APACrefauthors}%
Klein, R\BPBI J\BPBI T.%
, Nicholls, R\BPBI J.%
, Ragoonaden, S.%
, Capobianco, M.%
, Aston, J.%
, Buckley, E\BPBI N.%
\BDBL {}Buckley{\#}t, E\BPBI N.%
\end{APACrefauthors}%
\unskip\
\newblock
\APACrefYearMonthDay{2001}{}{}.
\newblock
{\BBOQ}\APACrefatitle {{Technological Options for Adaptation to Climate Change
  in Coastal Zones}} {{Technological Options for Adaptation to Climate Change
  in Coastal Zones}}.{\BBCQ}
\newblock
\APACjournalVolNumPages{Journal of Coastal Research Journal of Coastal
  Research}{17}{3}{531--543}.
\PrintBackRefs{\CurrentBib}

\bibitem [\protect \citeauthoryear {%
Knopman%
, Wachs%
, Miller%
, Davis%
\BCBL {}\ \BBA {} Pfrommer%
}{%
Knopman%
\ \protect \BOthers {.}}{%
{\protect \APACyear {2017}}%
}]{%
Knopman2017a}
\APACinsertmetastar {%
Knopman2017a}%
\begin{APACrefauthors}%
Knopman, D.%
, Wachs, M.%
, Miller, B\BPBI M.%
, Davis, S\BPBI G.%
\BCBL {}\ \BBA {} Pfrommer, K.%
\end{APACrefauthors}%
\unskip\
\newblock
\APACrefYearMonthDay{2017}{}{}.
\newblock
\APACrefbtitle {{Not Everything Is Broken: The Future of U.S. Transportation
  and Water Infrastructure Funding and Finance}} {{Not Everything Is Broken:
  The Future of U.S. Transportation and Water Infrastructure Funding and
  Finance}}\ \APACbVolEdTR {}{{Research Report}\ \BNUM\ RR-1739-RC}.
\newblock
\APACaddressInstitution{}{{RAND Corporation}}.
\PrintBackRefs{\CurrentBib}

\bibitem [\protect \citeauthoryear {%
Kopp%
\ \protect \BOthers {.}}{%
Kopp%
\ \protect \BOthers {.}}{%
{\protect \APACyear {2017}}%
}]{%
Kopp2017a}
\APACinsertmetastar {%
Kopp2017a}%
\begin{APACrefauthors}%
Kopp, R\BPBI E.%
, DeConto, R\BPBI M.%
, Bader, D\BPBI A.%
, Hay, C\BPBI C.%
, Horton, R\BPBI M.%
, Kulp, S.%
\BDBL {}Strauss, B\BPBI H.%
\end{APACrefauthors}%
\unskip\
\newblock
\APACrefYearMonthDay{2017}{}{}.
\newblock
{\BBOQ}\APACrefatitle {Evolving Understanding of Antarctic Ice-Sheet Physics
  and Ambiguity in Probabilistic Sea-Level Projections} {Evolving understanding
  of antarctic ice-sheet physics and ambiguity in probabilistic sea-level
  projections}.{\BBCQ}
\newblock
\APACjournalVolNumPages{Earth's Future}{}{}{n/a--n/a}.
\newblock
\APACrefnote{2017EF000663}
\newblock
\begin{APACrefDOI} \doi{10.1002/2017EF000663} \end{APACrefDOI}
\PrintBackRefs{\CurrentBib}

\bibitem [\protect \citeauthoryear {%
Kopp%
\ \protect \BOthers {.}}{%
Kopp%
\ \protect \BOthers {.}}{%
{\protect \APACyear {2014}}%
}]{%
Kopp2014a}
\APACinsertmetastar {%
Kopp2014a}%
\begin{APACrefauthors}%
Kopp, R\BPBI E.%
, Horton, R\BPBI M.%
, Little, C\BPBI M.%
, Mitrovica, J\BPBI X.%
, Oppenheimer, M.%
, Rasmussen, D\BPBI J.%
\BDBL {}Tebaldi, C.%
\end{APACrefauthors}%
\unskip\
\newblock
\APACrefYearMonthDay{2014}{}{}.
\newblock
{\BBOQ}\APACrefatitle {Probabilistic 21st and 22nd century sea-level
  projections at a global network of tide gauge sites} {Probabilistic 21st and
  22nd century sea-level projections at a global network of tide gauge
  sites}.{\BBCQ}
\newblock
\APACjournalVolNumPages{Earth's Future}{2}{}{383--406}.
\newblock
\begin{APACrefDOI} \doi{10.1002/2014EF000239} \end{APACrefDOI}
\PrintBackRefs{\CurrentBib}

\bibitem [\protect \citeauthoryear {%
{Le Bars}%
, Drijfhout%
\BCBL {}\ \BBA {} de Vries%
}{%
{Le Bars}%
\ \protect \BOthers {.}}{%
{\protect \APACyear {2017}}%
}]{%
LeBars2017a}
\APACinsertmetastar {%
LeBars2017a}%
\begin{APACrefauthors}%
{Le Bars}, D.%
, Drijfhout, S.%
\BCBL {}\ \BBA {} de Vries, H.%
\end{APACrefauthors}%
\unskip\
\newblock
\APACrefYearMonthDay{2017}{apr}{}.
\newblock
{\BBOQ}\APACrefatitle {{A high-end sea level rise probabilistic projection
  including rapid Antarctic ice sheet mass loss}} {{A high-end sea level rise
  probabilistic projection including rapid Antarctic ice sheet mass
  loss}}.{\BBCQ}
\newblock
\APACjournalVolNumPages{Environmental Research Letters}{12}{4}{044013}.
\newblock
\begin{APACrefDOI} \doi{10.1088/1748-9326/aa6512} \end{APACrefDOI}
\PrintBackRefs{\CurrentBib}

\bibitem [\protect \citeauthoryear {%
{Le Cozannet}%
, Manceau%
\BCBL {}\ \BBA {} Rohmer%
}{%
{Le Cozannet}%
\ \protect \BOthers {.}}{%
{\protect \APACyear {2017}}%
}]{%
LeCozannet2017a}
\APACinsertmetastar {%
LeCozannet2017a}%
\begin{APACrefauthors}%
{Le Cozannet}, G.%
, Manceau, J\BHBI C.%
\BCBL {}\ \BBA {} Rohmer, J.%
\end{APACrefauthors}%
\unskip\
\newblock
\APACrefYearMonthDay{2017}{}{}.
\newblock
{\BBOQ}\APACrefatitle {{Bounding probabilistic sea-level projections within the
  framework of the possibility theory}} {{Bounding probabilistic sea-level
  projections within the framework of the possibility theory}}.{\BBCQ}
\newblock
\APACjournalVolNumPages{Environ. Res. Lett.}{12}{1}{014012}.
\PrintBackRefs{\CurrentBib}

\bibitem [\protect \citeauthoryear {%
Lempert%
}{%
Lempert%
}{%
{\protect \APACyear {2002}}%
}]{%
Lempert2002a}
\APACinsertmetastar {%
Lempert2002a}%
\begin{APACrefauthors}%
Lempert, R\BPBI J.%
\end{APACrefauthors}%
\unskip\
\newblock
\APACrefYearMonthDay{2002}{}{}.
\newblock
{\BBOQ}\APACrefatitle {A new decision sciences for complex systems} {A new
  decision sciences for complex systems}.{\BBCQ}
\newblock
\APACjournalVolNumPages{PNAS}{99}{suppl 3}{7309--7313}.
\newblock
\begin{APACrefDOI} \doi{10.1073/pnas.082081699} \end{APACrefDOI}
\PrintBackRefs{\CurrentBib}

\bibitem [\protect \citeauthoryear {%
Lempert%
, Popper%
\BCBL {}\ \BBA {} Bankes%
}{%
Lempert%
\ \protect \BOthers {.}}{%
{\protect \APACyear {2003}}%
}]{%
Lempert2003a}
\APACinsertmetastar {%
Lempert2003a}%
\begin{APACrefauthors}%
Lempert, R\BPBI J.%
, Popper, S\BPBI W.%
\BCBL {}\ \BBA {} Bankes, S\BPBI C.%
\end{APACrefauthors}%
\unskip\
\newblock
\APACrefYear{2003}.
\newblock
\APACrefbtitle {Shaping the Next One Hundred Years: New Methods for
  Quantitative, Long-Term Policy Analysis} {Shaping the next one hundred years:
  New methods for quantitative, long-term policy analysis}\ (\PrintOrdinal{1}\
  \BEd).
\newblock
\APACaddressPublisher{}{RAND Corporation}.
\PrintBackRefs{\CurrentBib}

\bibitem [\protect \citeauthoryear {%
Lenk%
, Rybski%
, Heidrich%
, Dawson%
\BCBL {}\ \BBA {} Kropp%
}{%
Lenk%
\ \protect \BOthers {.}}{%
{\protect \APACyear {2017}}%
}]{%
Lenk2017a}
\APACinsertmetastar {%
Lenk2017a}%
\begin{APACrefauthors}%
Lenk, S.%
, Rybski, D.%
, Heidrich, O.%
, Dawson, R\BPBI J.%
\BCBL {}\ \BBA {} Kropp, J\BPBI P.%
\end{APACrefauthors}%
\unskip\
\newblock
\APACrefYearMonthDay{2017}{}{}.
\newblock
{\BBOQ}\APACrefatitle {Costs of sea dikes -- regressions and uncertainty
  estimates} {Costs of sea dikes -- regressions and uncertainty
  estimates}.{\BBCQ}
\newblock
\APACjournalVolNumPages{Natural Hazards and Earth System
  Sciences}{17}{5}{765--779}.
\newblock
\begin{APACrefDOI} \doi{10.5194/nhess-17-765-2017} \end{APACrefDOI}
\PrintBackRefs{\CurrentBib}

\bibitem [\protect \citeauthoryear {%
Levermann%
\ \protect \BOthers {.}}{%
Levermann%
\ \protect \BOthers {.}}{%
{\protect \APACyear {2014}}%
}]{%
Levermann2014a}
\APACinsertmetastar {%
Levermann2014a}%
\begin{APACrefauthors}%
Levermann, A.%
, Winkelmann, R.%
, Nowicki, S.%
, Fastook, J\BPBI L.%
, Frieler, K.%
, Greve, R.%
\BDBL {}Bindschadler, R\BPBI A.%
\end{APACrefauthors}%
\unskip\
\newblock
\APACrefYearMonthDay{2014}{}{}.
\newblock
{\BBOQ}\APACrefatitle {{Projecting Antarctic ice discharge using response
  functions from SeaRISE ice-sheet models}} {{Projecting Antarctic ice
  discharge using response functions from SeaRISE ice-sheet models}}.{\BBCQ}
\newblock
\APACjournalVolNumPages{Earth System Dynamics}{5}{2}{271--293}.
\newblock
\begin{APACrefDOI} \doi{10.5194/esd-5-271-2014} \end{APACrefDOI}
\PrintBackRefs{\CurrentBib}

\bibitem [\protect \citeauthoryear {%
Lincke%
\ \BBA {} Hinkel%
}{%
Lincke%
\ \BBA {} Hinkel%
}{%
{\protect \APACyear {2018}}%
}]{%
Lincke2018a}
\APACinsertmetastar {%
Lincke2018a}%
\begin{APACrefauthors}%
Lincke, D.%
\BCBT {}\ \BBA {} Hinkel, J.%
\end{APACrefauthors}%
\unskip\
\newblock
\APACrefYearMonthDay{2018}{}{}.
\newblock
{\BBOQ}\APACrefatitle {{Economically robust protection against 21st century
  sea-level rise}} {{Economically robust protection against 21st century
  sea-level rise}}.{\BBCQ}
\newblock
\APACjournalVolNumPages{Global Environmental Change}{51}{April}{67--73}.
\newblock
\begin{APACrefDOI} \doi{10.1016/j.gloenvcha.2018.05.003} \end{APACrefDOI}
\PrintBackRefs{\CurrentBib}

\bibitem [\protect \citeauthoryear {%
Little%
, Oppenheimer%
\BCBL {}\ \BBA {} Urban%
}{%
Little%
, Oppenheimer%
\BCBL {}\ \BBA {} Urban%
}{%
{\protect \APACyear {2013}}%
}]{%
Little2013b}
\APACinsertmetastar {%
Little2013b}%
\begin{APACrefauthors}%
Little, C\BPBI M.%
, Oppenheimer, M.%
\BCBL {}\ \BBA {} Urban, N\BPBI M.%
\end{APACrefauthors}%
\unskip\
\newblock
\APACrefYearMonthDay{2013}{{\APACmonth{03}}}{}.
\newblock
{\BBOQ}\APACrefatitle {{Upper bounds on twenty-first-century Antarctic ice loss
  assessed using a probabilistic framework}} {{Upper bounds on
  twenty-first-century Antarctic ice loss assessed using a probabilistic
  framework}}.{\BBCQ}
\newblock
\APACjournalVolNumPages{Nature Climate Change}{3}{}{654-659}.
\newblock
\begin{APACrefDOI} \doi{10.1038/nclimate1845} \end{APACrefDOI}
\PrintBackRefs{\CurrentBib}

\bibitem [\protect \citeauthoryear {%
Little%
, Urban%
\BCBL {}\ \BBA {} Oppenheimer%
}{%
Little%
, Urban%
\BCBL {}\ \BBA {} Oppenheimer%
}{%
{\protect \APACyear {2013}}%
}]{%
Little2013a}
\APACinsertmetastar {%
Little2013a}%
\begin{APACrefauthors}%
Little, C\BPBI M.%
, Urban, N\BPBI M.%
\BCBL {}\ \BBA {} Oppenheimer, M.%
\end{APACrefauthors}%
\unskip\
\newblock
\APACrefYearMonthDay{2013}{}{}.
\newblock
{\BBOQ}\APACrefatitle {Probabilistic framework for assessing the ice sheet
  contribution to sea level change} {Probabilistic framework for assessing the
  ice sheet contribution to sea level change}.{\BBCQ}
\newblock
\APACjournalVolNumPages{Proceedings of the National Academy of
  Sciences}{110}{}{3264--3269}.
\newblock
\begin{APACrefDOI} \doi{10.1073/pnas.1214457110} \end{APACrefDOI}
\PrintBackRefs{\CurrentBib}

\bibitem [\protect \citeauthoryear {%
McInerney%
, Lempert%
\BCBL {}\ \BBA {} Keller%
}{%
McInerney%
\ \protect \BOthers {.}}{%
{\protect \APACyear {2012}}%
}]{%
McInerney2012a}
\APACinsertmetastar {%
McInerney2012a}%
\begin{APACrefauthors}%
McInerney, D.%
, Lempert, R.%
\BCBL {}\ \BBA {} Keller, K.%
\end{APACrefauthors}%
\unskip\
\newblock
\APACrefYearMonthDay{2012}{}{}.
\newblock
{\BBOQ}\APACrefatitle {What are robust strategies in the face of uncertain
  climate threshold responses?} {What are robust strategies in the face of
  uncertain climate threshold responses?}{\BBCQ}
\newblock
\APACjournalVolNumPages{Climatic change}{112}{3-4}{547--568}.
\PrintBackRefs{\CurrentBib}

\bibitem [\protect \citeauthoryear {%
Merrell%
, Reynolds%
, Cardenas%
, Gunn%
\BCBL {}\ \BBA {} Hufton%
}{%
Merrell%
\ \protect \BOthers {.}}{%
{\protect \APACyear {2011}}%
}]{%
Merrell2011a}
\APACinsertmetastar {%
Merrell2011a}%
\begin{APACrefauthors}%
Merrell, W\BPBI J.%
, Reynolds, L\BPBI G.%
, Cardenas, A.%
, Gunn, J\BPBI R.%
\BCBL {}\ \BBA {} Hufton, A\BPBI J.%
\end{APACrefauthors}%
\unskip\
\newblock
\APACrefYearMonthDay{2011}{}{}.
\newblock
{\BBOQ}\APACrefatitle {The Ike Dike: A Coastal Barrier Protecting the
  Houston/Galveston Region from Hurricane Storm Surge} {The ike dike: A coastal
  barrier protecting the houston/galveston region from hurricane storm
  surge}.{\BBCQ}
\newblock
\BIn{} V.~Badescu\ \BBA {} R\BPBI B.~Cathcart\ (\BEDS), \APACrefbtitle
  {Macro-engineering Seawater in Unique Environments: Arid Lowlands and Water
  Bodies Rehabilitation} {Macro-engineering seawater in unique environments:
  Arid lowlands and water bodies rehabilitation}\ (\BPGS\ 691--716).
\newblock
\APACaddressPublisher{Berlin, Heidelberg}{Springer Berlin Heidelberg}.
\newblock
\begin{APACrefDOI} \doi{10.1007/978-3-642-14779-1_31} \end{APACrefDOI}
\PrintBackRefs{\CurrentBib}

\bibitem [\protect \citeauthoryear {%
Merz%
, Kreibich%
, Schwarze%
\BCBL {}\ \BBA {} Thieken%
}{%
Merz%
\ \protect \BOthers {.}}{%
{\protect \APACyear {2010}}%
}]{%
Merz2010a}
\APACinsertmetastar {%
Merz2010a}%
\begin{APACrefauthors}%
Merz, B.%
, Kreibich, H.%
, Schwarze, R.%
\BCBL {}\ \BBA {} Thieken, A.%
\end{APACrefauthors}%
\unskip\
\newblock
\APACrefYearMonthDay{2010}{}{}.
\newblock
{\BBOQ}\APACrefatitle {Review article "Assessment of economic flood damage"}
  {Review article "assessment of economic flood damage"}.{\BBCQ}
\newblock
\APACjournalVolNumPages{Natural Hazards and Earth System
  Sciences}{10}{8}{1697--1724}.
\newblock
\begin{APACrefDOI} \doi{10.5194/nhess-10-1697-2010} \end{APACrefDOI}
\PrintBackRefs{\CurrentBib}

\bibitem [\protect \citeauthoryear {%
Mooyaart%
\ \BBA {} Jonkman%
}{%
Mooyaart%
\ \BBA {} Jonkman%
}{%
{\protect \APACyear {2017}}%
}]{%
Mooyaart2017a}
\APACinsertmetastar {%
Mooyaart2017a}%
\begin{APACrefauthors}%
Mooyaart, L\BPBI F.%
\BCBT {}\ \BBA {} Jonkman, S\BPBI N.%
\end{APACrefauthors}%
\unskip\
\newblock
\APACrefYearMonthDay{2017}{2019/05/23}{}.
\newblock
{\BBOQ}\APACrefatitle {{Overview and Design Considerations of Storm Surge
  Barriers}} {{Overview and Design Considerations of Storm Surge
  Barriers}}.{\BBCQ}
\newblock
\APACjournalVolNumPages{Journal of Waterway, Port, Coastal, and Ocean
  Engineering}{143}{4}{}.
\newblock
\begin{APACrefDOI} \doi{10.1061/(ASCE)WW.1943-5460.0000383} \end{APACrefDOI}
\PrintBackRefs{\CurrentBib}

\bibitem [\protect \citeauthoryear {%
Morang%
}{%
Morang%
}{%
{\protect \APACyear {2016}}%
}]{%
Morang2016a}
\APACinsertmetastar {%
Morang2016a}%
\begin{APACrefauthors}%
Morang, A.%
\end{APACrefauthors}%
\unskip\
\newblock
\APACrefYearMonthDay{2016}{}{}.
\newblock
{\BBOQ}\APACrefatitle {{Hurricane Barriers in New England and New Jersey:
  History and Status after Five Decades}} {{Hurricane Barriers in New England
  and New Jersey: History and Status after Five Decades}}.{\BBCQ}
\newblock
\APACjournalVolNumPages{Journal of Coastal Research}{317}{}{181--205}.
\newblock
\begin{APACrefDOI} \doi{10.2112/JCOASTRES-D-14-00074.1} \end{APACrefDOI}
\PrintBackRefs{\CurrentBib}

\bibitem [\protect \citeauthoryear {%
{New York City Department of Finance}%
}{%
{New York City Department of Finance}%
}{%
{\protect \APACyear {2018}}%
}]{%
NYC2018a}
\APACinsertmetastar {%
NYC2018a}%
\begin{APACrefauthors}%
{New York City Department of Finance}.%
\end{APACrefauthors}%
\unskip\
\newblock
\APACrefYearMonthDay{2018}{April}{}.
\newblock
\APACrefbtitle {Annual Report of the New York City Property Tax: Fiscal Year
  2018} {Annual report of the new york city property tax: Fiscal year 2018}\
  \APACbVolEdTR{}{\BTR{}}.
\newblock
\APACaddressInstitution{}{{NYC Department of Finance}}.
\PrintBackRefs{\CurrentBib}

\bibitem [\protect \citeauthoryear {%
{NRC}%
}{%
{NRC}%
}{%
{\protect \APACyear {2014}}%
}]{%
NRC2014a}
\APACinsertmetastar {%
NRC2014a}%
\begin{APACrefauthors}%
{NRC}.%
\end{APACrefauthors}%
\unskip\
\newblock
\APACrefYearMonthDay{2014}{}{}.
\newblock
{\BBOQ}\APACrefatitle {{Institutional Landscape for Coastal Risk Management
  Responsibilities}} {{Institutional Landscape for Coastal Risk Management
  Responsibilities}}.{\BBCQ}
\newblock
\BIn{} \APACrefbtitle {{Reducing Coastal Risk on the East and Gulf Coasts}}
  {{Reducing Coastal Risk on the East and Gulf Coasts}}\ (\BCHAP~2).
\newblock
\APACaddressPublisher{Washington, {D.C.}}{The National Academies Press}.
\PrintBackRefs{\CurrentBib}

\bibitem [\protect \citeauthoryear {%
{NYC}%
}{%
{NYC}%
}{%
{\protect \APACyear {2019}}%
{\protect \APACexlab {{\protect \BCnt {1}}}}}]{%
NYC2019c}
\APACinsertmetastar {%
NYC2019c}%
\begin{APACrefauthors}%
{NYC}.%
\end{APACrefauthors}%
\unskip\
\newblock
\APACrefYearMonthDay{2019{\protect \BCnt {1}}}{March}{}.
\newblock
\APACrefbtitle {{Climate Resiliency Design Guidelines}} {{Climate Resiliency
  Design Guidelines}}\ \APACbVolEdTR{}{\BTR{}\ \BNUM\ version 3.0}.
\newblock
\APACaddressInstitution{}{New York City Mayor's Office of Resiliency and
  Recovery}.
\PrintBackRefs{\CurrentBib}

\bibitem [\protect \citeauthoryear {%
{NYC}%
}{%
{NYC}%
}{%
{\protect \APACyear {2019}}%
{\protect \APACexlab {{\protect \BCnt {2}}}}}]{%
NYC2019b}
\APACinsertmetastar {%
NYC2019b}%
\begin{APACrefauthors}%
{NYC}.%
\end{APACrefauthors}%
\unskip\
\newblock
\APACrefYearMonthDay{2019{\protect \BCnt {2}}}{April}{}.
\newblock
\APACrefbtitle {{East Side Coastal Resiliency Project}} {{East Side Coastal
  Resiliency Project}}\ \APACbVolEdTR {}{{Draft Environmental Impact Statement
  (EIS)}\ \BNUM\ {CEQR No: 15DPR013M}}.
\newblock
\APACaddressInstitution{}{{New York City Office of Management and Budget}}.
\PrintBackRefs{\CurrentBib}

\bibitem [\protect \citeauthoryear {%
{NYC}%
}{%
{NYC}%
}{%
{\protect \APACyear {2019}}%
{\protect \APACexlab {{\protect \BCnt {3}}}}}]{%
NYC2019a}
\APACinsertmetastar {%
NYC2019a}%
\begin{APACrefauthors}%
{NYC}.%
\end{APACrefauthors}%
\unskip\
\newblock
\APACrefYearMonthDay{2019{\protect \BCnt {3}}}{March}{}.
\newblock
\APACrefbtitle {{Lower Manhattan Climate Resilience Study}} {{Lower Manhattan
  Climate Resilience Study}}\ \APACbVolEdTR{}{\BTR{}}.
\newblock
\APACaddressInstitution{}{{New York City Mayor's Office of Recovery \&
  Resilience}}.
\PrintBackRefs{\CurrentBib}

\bibitem [\protect \citeauthoryear {%
{NYC Planning}%
}{%
{NYC Planning}%
}{%
{\protect \APACyear {2018}}%
}]{%
NYCpluto2018a}
\APACinsertmetastar {%
NYCpluto2018a}%
\begin{APACrefauthors}%
{NYC Planning}.%
\end{APACrefauthors}%
\unskip\
\newblock
\APACrefYearMonthDay{2018}{}{}.
\newblock
\APACrefbtitle {{MapPLUTO database}.} {{MapPLUTO database}.}
\newblock
\APAChowpublished {data retrived from the NYC Planning Department:
  https://www1.nyc.gov/site/planning/data-maps/open-data/dwn-pluto-mappluto.page}.
\PrintBackRefs{\CurrentBib}

\bibitem [\protect \citeauthoryear {%
Pirazzoli%
\ \BBA {} Umgiesser%
}{%
Pirazzoli%
\ \BBA {} Umgiesser%
}{%
{\protect \APACyear {2006}}%
}]{%
Pirazzoli2006a}
\APACinsertmetastar {%
Pirazzoli2006a}%
\begin{APACrefauthors}%
Pirazzoli, P\BPBI A.%
\BCBT {}\ \BBA {} Umgiesser, G.%
\end{APACrefauthors}%
\unskip\
\newblock
\APACrefYearMonthDay{2006}{}{}.
\newblock
{\BBOQ}\APACrefatitle {{The Projected "MOSE" Barriers Against Flooding in
  Venice (Italy) and the Expected Global Sea-level Rise}} {{The Projected
  "MOSE" Barriers Against Flooding in Venice (Italy) and the Expected Global
  Sea-level Rise}}.{\BBCQ}
\newblock
\APACjournalVolNumPages{J. of Marine Env. Eng.}{8}{}{0--0}.
\PrintBackRefs{\CurrentBib}

\bibitem [\protect \citeauthoryear {%
Pollard%
, DeConto%
\BCBL {}\ \BBA {} Alley%
}{%
Pollard%
\ \protect \BOthers {.}}{%
{\protect \APACyear {2015}}%
}]{%
Pollard2015a}
\APACinsertmetastar {%
Pollard2015a}%
\begin{APACrefauthors}%
Pollard, D.%
, DeConto, R\BPBI M.%
\BCBL {}\ \BBA {} Alley, R\BPBI B.%
\end{APACrefauthors}%
\unskip\
\newblock
\APACrefYearMonthDay{2015}{}{}.
\newblock
{\BBOQ}\APACrefatitle {Potential Antarctic Ice Sheet retreat driven by
  hydrofracturing and ice cliff failure} {Potential antarctic ice sheet retreat
  driven by hydrofracturing and ice cliff failure}.{\BBCQ}
\newblock
\APACjournalVolNumPages{Earth and Planetary Science Letters}{412}{}{112 - 121}.
\newblock
\begin{APACrefDOI} \doi{https://doi.org/10.1016/j.epsl.2014.12.035}
  \end{APACrefDOI}
\PrintBackRefs{\CurrentBib}

\bibitem [\protect \citeauthoryear {%
{Public Law}%
}{%
{Public Law}%
}{%
{\protect \APACyear {1936}}%
}]{%
Flood1936a}
\APACinsertmetastar {%
Flood1936a}%
\begin{APACrefauthors}%
{Public Law}.%
\end{APACrefauthors}%
\unskip\
\newblock
\APACrefYearMonthDay{1936}{}{}.
\newblock
\APACrefbtitle {{Flood Control Act of 1936}.} {{Flood Control Act of 1936}.}
\newblock
\APAChowpublished {{Pub. L. 74--738}}.
\PrintBackRefs{\CurrentBib}

\bibitem [\protect \citeauthoryear {%
Ramm%
, White%
, Chan%
\BCBL {}\ \BBA {} Watson%
}{%
Ramm%
\ \protect \BOthers {.}}{%
{\protect \APACyear {2017}}%
}]{%
Ramm2017a}
\APACinsertmetastar {%
Ramm2017a}%
\begin{APACrefauthors}%
Ramm, T\BPBI D.%
, White, C\BPBI J.%
, Chan, A\BPBI H\BPBI C.%
\BCBL {}\ \BBA {} Watson, C\BPBI S.%
\end{APACrefauthors}%
\unskip\
\newblock
\APACrefYearMonthDay{2017}{}{}.
\newblock
{\BBOQ}\APACrefatitle {{A review of methodologies applied in Australian
  practice to evaluate long-term coastal adaptation options}} {{A review of
  methodologies applied in Australian practice to evaluate long-term coastal
  adaptation options}}.{\BBCQ}
\newblock
\APACjournalVolNumPages{Climate Risk Management}{17}{}{35 - 51}.
\newblock
\begin{APACrefDOI} \doi{https://doi.org/10.1016/j.crm.2017.06.005}
  \end{APACrefDOI}
\PrintBackRefs{\CurrentBib}

\bibitem [\protect \citeauthoryear {%
Ranger%
, Reeder%
\BCBL {}\ \BBA {} Lowe%
}{%
Ranger%
\ \protect \BOthers {.}}{%
{\protect \APACyear {2013}}%
}]{%
Ranger2013a}
\APACinsertmetastar {%
Ranger2013a}%
\begin{APACrefauthors}%
Ranger, N.%
, Reeder, T.%
\BCBL {}\ \BBA {} Lowe, J.%
\end{APACrefauthors}%
\unskip\
\newblock
\APACrefYearMonthDay{2013}{}{}.
\newblock
{\BBOQ}\APACrefatitle {{Addressing 'deep' uncertainty over long-term climate in
  major infrastructure projects: four innovations of the Thames Estuary 2100
  Project}} {{Addressing 'deep' uncertainty over long-term climate in major
  infrastructure projects: four innovations of the Thames Estuary 2100
  Project}}.{\BBCQ}
\newblock
\APACjournalVolNumPages{{EURO Journal on Decision Processes}}{}{}{233--262}.
\newblock
\begin{APACrefDOI} \doi{10.1007/s40070-013-0014-5} \end{APACrefDOI}
\PrintBackRefs{\CurrentBib}

\bibitem [\protect \citeauthoryear {%
Rasmussen%
\ \protect \BOthers {.}}{%
Rasmussen%
\ \protect \BOthers {.}}{%
{\protect \APACyear {2018}}%
}]{%
Rasmussen2018a}
\APACinsertmetastar {%
Rasmussen2018a}%
\begin{APACrefauthors}%
Rasmussen, D\BPBI J.%
, Bittermann, K.%
, Buchanan, M\BPBI K.%
, Kulp, S.%
, Strauss, B\BPBI H.%
, Kopp, R\BPBI E.%
\BCBL {}\ \BBA {} Oppenheimer, M.%
\end{APACrefauthors}%
\unskip\
\newblock
\APACrefYearMonthDay{2018}{}{}.
\newblock
{\BBOQ}\APACrefatitle {{Extreme sea level implications of 1.5 ${^\circ}$C, 2.0
  ${^\circ}$C, and 2.5 ${^\circ}$C temperature stabilization targets in the
  21st and 22nd centuries}} {{Extreme sea level implications of 1.5
  ${^\circ}$C, 2.0 ${^\circ}$C, and 2.5 ${^\circ}$C temperature stabilization
  targets in the 21st and 22nd centuries}}.{\BBCQ}
\newblock
\APACjournalVolNumPages{Environmental Research Letters}{13}{3}{034040}.
\newblock
\begin{APACrefDOI} \doi{10.1088/1748-9326/aaac87} \end{APACrefDOI}
\PrintBackRefs{\CurrentBib}

\bibitem [\protect \citeauthoryear {%
Ritz%
\ \protect \BOthers {.}}{%
Ritz%
\ \protect \BOthers {.}}{%
{\protect \APACyear {2015}}%
}]{%
Ritz2015a}
\APACinsertmetastar {%
Ritz2015a}%
\begin{APACrefauthors}%
Ritz, C.%
, Edwards, T\BPBI L.%
, Durand, G.%
, Payne, A\BPBI J.%
, Peyaud, V.%
\BCBL {}\ \BBA {} Hindmarsh, R\BPBI C\BPBI A.%
\end{APACrefauthors}%
\unskip\
\newblock
\APACrefYearMonthDay{2015}{11}{18}.
\newblock
{\BBOQ}\APACrefatitle {Potential sea-level rise from Antarctic ice-sheet
  instability constrained by observations} {Potential sea-level rise from
  antarctic ice-sheet instability constrained by observations}.{\BBCQ}
\newblock
\APACjournalVolNumPages{Nature}{528}{}{115 EP -}.
\PrintBackRefs{\CurrentBib}

\bibitem [\protect \citeauthoryear {%
Scussolini%
\ \protect \BOthers {.}}{%
Scussolini%
\ \protect \BOthers {.}}{%
{\protect \APACyear {2017}}%
}]{%
Scussolini2017a}
\APACinsertmetastar {%
Scussolini2017a}%
\begin{APACrefauthors}%
Scussolini, P.%
, Tran, T\BPBI V\BPBI T.%
, Koks, E.%
, Diaz-Loaiza, A.%
, Ho, P\BPBI L.%
\BCBL {}\ \BBA {} Lasage, R.%
\end{APACrefauthors}%
\unskip\
\newblock
\APACrefYearMonthDay{2017}{2019/05/23}{}.
\newblock
{\BBOQ}\APACrefatitle {{Adaptation to Sea Level Rise: A Multidisciplinary
  Analysis for Ho Chi Minh City, Vietnam}} {{Adaptation to Sea Level Rise: A
  Multidisciplinary Analysis for Ho Chi Minh City, Vietnam}}.{\BBCQ}
\newblock
\APACjournalVolNumPages{Water Resources Research}{53}{12}{10841--10857}.
\newblock
\begin{APACrefDOI} \doi{10.1002/2017WR021344} \end{APACrefDOI}
\PrintBackRefs{\CurrentBib}

\bibitem [\protect \citeauthoryear {%
SIRR%
}{%
SIRR%
}{%
{\protect \APACyear {2013}}%
}]{%
SIRR2013a}
\APACinsertmetastar {%
SIRR2013a}%
\begin{APACrefauthors}%
SIRR.%
\end{APACrefauthors}%
\unskip\
\newblock
\APACrefYearMonthDay{2013}{}{}.
\newblock
\APACrefbtitle {{A Stronger, More Resilient New York}} {{A Stronger, More
  Resilient New York}}\ \APACbVolEdTR{}{\BTR{}}.
\newblock
\APACaddressInstitution{New York, NY, USA}{{New York City Special Initiative
  for Rebuilding and Resiliency}}.
\PrintBackRefs{\CurrentBib}

\bibitem [\protect \citeauthoryear {%
Slangen%
\ \protect \BOthers {.}}{%
Slangen%
\ \protect \BOthers {.}}{%
{\protect \APACyear {2017}}%
}]{%
Slangen2017a}
\APACinsertmetastar {%
Slangen2017a}%
\begin{APACrefauthors}%
Slangen, A.%
, Hunter, J.%
, Woodworth, P.%
, de Winter, R.%
, Edwards, T.%
, Reerink, T.%
\BCBL {}\ \BBA {} van~de Wal, R.%
\end{APACrefauthors}%
\unskip\
\newblock
\APACrefYearMonthDay{2017}{}{}.
\newblock
{\BBOQ}\APACrefatitle {{The Impact of Uncertainties in Ice Sheet Dynamics on
  Sea-Level Allowances at Tide Gauge Locations}} {{The Impact of Uncertainties
  in Ice Sheet Dynamics on Sea-Level Allowances at Tide Gauge
  Locations}}.{\BBCQ}
\newblock
\APACjournalVolNumPages{Journal of Marine Science and Engineering}{5}{2}{21}.
\newblock
\begin{APACrefDOI} \doi{10.3390/jmse5020021} \end{APACrefDOI}
\PrintBackRefs{\CurrentBib}

\bibitem [\protect \citeauthoryear {%
Sriver%
, Lempert%
, Wikman-Svahn%
\BCBL {}\ \BBA {} Keller%
}{%
Sriver%
\ \protect \BOthers {.}}{%
{\protect \APACyear {2018}}%
}]{%
Sriver2018a}
\APACinsertmetastar {%
Sriver2018a}%
\begin{APACrefauthors}%
Sriver, R\BPBI L.%
, Lempert, R\BPBI J.%
, Wikman-Svahn, P.%
\BCBL {}\ \BBA {} Keller, K.%
\end{APACrefauthors}%
\unskip\
\newblock
\APACrefYearMonthDay{2018}{02}{}.
\newblock
{\BBOQ}\APACrefatitle {Characterizing uncertain sea-level rise projections to
  support investment decisions} {Characterizing uncertain sea-level rise
  projections to support investment decisions}.{\BBCQ}
\newblock
\APACjournalVolNumPages{PLOS ONE}{13}{2}{1-35}.
\newblock
\begin{APACrefDOI} \doi{10.1371/journal.pone.0190641} \end{APACrefDOI}
\PrintBackRefs{\CurrentBib}

\bibitem [\protect \citeauthoryear {%
{Sustainable Solutions Lab}%
}{%
{Sustainable Solutions Lab}%
}{%
{\protect \APACyear {2018}}%
{\protect \APACexlab {{\protect \BCnt {1}}}}}]{%
SustainableSolutionsLab2018a}
\APACinsertmetastar {%
SustainableSolutionsLab2018a}%
\begin{APACrefauthors}%
{Sustainable Solutions Lab}.%
\end{APACrefauthors}%
\unskip\
\newblock
\APACrefYearMonthDay{2018{\protect \BCnt {1}}}{}{}.
\newblock
\APACrefbtitle {{Feasibility of Harbor-wide Barrier Systems: Preliminary
  Analysis for Boston Harbor}} {{Feasibility of Harbor-wide Barrier Systems:
  Preliminary Analysis for Boston Harbor}}\ \APACbVolEdTR{}{\BTR{}}.
\newblock
\APACaddressInstitution{Boston, MA}{University of Massachusetts, Boston}.
\PrintBackRefs{\CurrentBib}

\bibitem [\protect \citeauthoryear {%
{Sustainable Solutions Lab}%
}{%
{Sustainable Solutions Lab}%
}{%
{\protect \APACyear {2018}}%
{\protect \APACexlab {{\protect \BCnt {2}}}}}]{%
SustainableSolutionsLab2018b}
\APACinsertmetastar {%
SustainableSolutionsLab2018b}%
\begin{APACrefauthors}%
{Sustainable Solutions Lab}.%
\end{APACrefauthors}%
\unskip\
\newblock
\APACrefYearMonthDay{2018{\protect \BCnt {2}}}{}{}.
\newblock
\APACrefbtitle {{Financing Climate Resilience: Mobilizing Resources and
  Incentives to Protect Boston from Climate Risks}} {{Financing Climate
  Resilience: Mobilizing Resources and Incentives to Protect Boston from
  Climate Risks}}\ \APACbVolEdTR{}{\BTR{}}.
\newblock
\APACaddressInstitution{Boston, MA}{University of Massachusetts, Boston}.
\PrintBackRefs{\CurrentBib}

\bibitem [\protect \citeauthoryear {%
Sweet%
\ \protect \BOthers {.}}{%
Sweet%
\ \protect \BOthers {.}}{%
{\protect \APACyear {2016}}%
}]{%
Sweet2016a}
\APACinsertmetastar {%
Sweet2016a}%
\begin{APACrefauthors}%
Sweet, W\BPBI V.%
, Menendez, M.%
, Genz, A.%
, Obeysekera, J.%
, Park, J.%
\BCBL {}\ \BBA {} Marra, J\BPBI J.%
\end{APACrefauthors}%
\unskip\
\newblock
\APACrefYearMonthDay{2016}{}{}.
\newblock
{\BBOQ}\APACrefatitle {{In Tide's Way: Southeast Florida's September 2015
  Sunny-day Flood}} {{In Tide's Way: Southeast Florida's September 2015
  Sunny-day Flood}}.{\BBCQ}
\newblock
\APACjournalVolNumPages{Bull. Am. Meteorol. Soc.}{97}{12}{S25-S30}.
\newblock
\begin{APACrefDOI} \doi{10.1175/BAMS-D-16-0117.1} \end{APACrefDOI}
\PrintBackRefs{\CurrentBib}

\bibitem [\protect \citeauthoryear {%
{TAW}%
}{%
{TAW}%
}{%
{\protect \APACyear {1998}}%
}]{%
TAW1998a}
\APACinsertmetastar {%
TAW1998a}%
\begin{APACrefauthors}%
{TAW}.%
\end{APACrefauthors}%
\unskip\
\newblock
\APACrefYearMonthDay{1998}{January}{}.
\newblock
\APACrefbtitle {{Fundamentals on Water Defences (English translation of
  `Grondslagen voor Waterkeren')}} {{Fundamentals on Water Defences (English
  translation of `Grondslagen voor Waterkeren')}}\ \APACbVolEdTR{}{\BTR{}}.
\newblock
\APACaddressInstitution{}{{Rijkswaterstaat (The Netherlands)}}.
\PrintBackRefs{\CurrentBib}

\bibitem [\protect \citeauthoryear {%
Tebaldi%
, Strauss%
\BCBL {}\ \BBA {} Zervas%
}{%
Tebaldi%
\ \protect \BOthers {.}}{%
{\protect \APACyear {2012}}%
}]{%
Tebaldi2012a}
\APACinsertmetastar {%
Tebaldi2012a}%
\begin{APACrefauthors}%
Tebaldi, C.%
, Strauss, B\BPBI H.%
\BCBL {}\ \BBA {} Zervas, C\BPBI E.%
\end{APACrefauthors}%
\unskip\
\newblock
\APACrefYearMonthDay{2012}{{\APACmonth{03}}}{}.
\newblock
{\BBOQ}\APACrefatitle {Modelling sea level rise impacts on storm surges along
  {US} coasts} {Modelling sea level rise impacts on storm surges along {US}
  coasts}.{\BBCQ}
\newblock
\APACjournalVolNumPages{Environmental Research Letters}{7}{}{014032}.
\newblock
\begin{APACrefDOI} \doi{10.1088/1748-9326/7/1/014032} \end{APACrefDOI}
\PrintBackRefs{\CurrentBib}

\bibitem [\protect \citeauthoryear {%
{UK Environment Agency}%
}{%
{UK Environment Agency}%
}{%
{\protect \APACyear {2012}}%
}]{%
EA2012a}
\APACinsertmetastar {%
EA2012a}%
\begin{APACrefauthors}%
{UK Environment Agency}.%
\end{APACrefauthors}%
\unskip\
\newblock
\APACrefYearMonthDay{2012}{}{}.
\newblock
\APACrefbtitle {{Thames Estuary 2100}} {{Thames Estuary 2100}}\
  \APACbVolEdTR{}{\BTR{}}.
\newblock
\APACaddressInstitution{}{{Environment Agency (UK)}}.
\PrintBackRefs{\CurrentBib}

\bibitem [\protect \citeauthoryear {%
{U.S. Water Resources Council}%
}{%
{U.S. Water Resources Council}%
}{%
{\protect \APACyear {1983}}%
}]{%
USWRC1983a}
\APACinsertmetastar {%
USWRC1983a}%
\begin{APACrefauthors}%
{U.S. Water Resources Council}.%
\end{APACrefauthors}%
\unskip\
\newblock
\APACrefYearMonthDay{1983}{March}{}.
\newblock
\APACrefbtitle {{Economic and Environmental Principles and Guidelines for Water
  Related Land Resources Implementation Studies}.} {{Economic and Environmental
  Principles and Guidelines for Water Related Land Resources Implementation
  Studies}.}
\PrintBackRefs{\CurrentBib}

\bibitem [\protect \citeauthoryear {%
{USACE}%
}{%
{USACE}%
}{%
{\protect \APACyear {2015}}%
}]{%
USACE2015a}
\APACinsertmetastar {%
USACE2015a}%
\begin{APACrefauthors}%
{USACE}.%
\end{APACrefauthors}%
\unskip\
\newblock
\APACrefYearMonthDay{2015}{January}{}.
\newblock
\APACrefbtitle {{North Atlantic Coast Comprehensive Study: Resilient Adaptation
  to Increasing Risk, Physical Depth-Damage Function Summary Report}} {{North
  Atlantic Coast Comprehensive Study: Resilient Adaptation to Increasing Risk,
  Physical Depth-Damage Function Summary Report}}\ \APACbVolEdTR {}{{Technical
  Report}}.
\newblock
\APACaddressInstitution{}{{U.S. Army Corps of Engineers, North Atlantic
  Division}}.
\PrintBackRefs{\CurrentBib}

\bibitem [\protect \citeauthoryear {%
{USACE}%
}{%
{USACE}%
}{%
{\protect \APACyear {2016}}%
}]{%
USACE2016a}
\APACinsertmetastar {%
USACE2016a}%
\begin{APACrefauthors}%
{USACE}.%
\end{APACrefauthors}%
\unskip\
\newblock
\APACrefYearMonthDay{2016}{December}{}.
\newblock
\APACrefbtitle {{South Shore of Staten Island Coastal Storm Risk Management}}
  {{South Shore of Staten Island Coastal Storm Risk Management}}\ \APACbVolEdTR
  {}{{Final Environmental Impact Statement (EIS)}}.
\newblock
\APACaddressInstitution{}{{U.S. Army Corps of Engineers, New York District}}.
\PrintBackRefs{\CurrentBib}

\bibitem [\protect \citeauthoryear {%
{USACE}%
}{%
{USACE}%
}{%
{\protect \APACyear {2019}}%
}]{%
USACE2019a}
\APACinsertmetastar {%
USACE2019a}%
\begin{APACrefauthors}%
{USACE}.%
\end{APACrefauthors}%
\unskip\
\newblock
\APACrefYearMonthDay{2019}{February}{}.
\newblock
\APACrefbtitle {{New York -- New Jersey Harbor and Tributaries Coastal Storm
  Risk Management Feasibility Study}} {{New York -- New Jersey Harbor and
  Tributaries Coastal Storm Risk Management Feasibility Study}}\ \APACbVolEdTR
  {}{{Interim Report}}.
\newblock
\APACaddressInstitution{}{{U.S. Army Corps of Engineers, New York District}}.
\PrintBackRefs{\CurrentBib}

\bibitem [\protect \citeauthoryear {%
van Dantzig%
}{%
van Dantzig%
}{%
{\protect \APACyear {1956}}%
}]{%
vanDantzig1956a}
\APACinsertmetastar {%
vanDantzig1956a}%
\begin{APACrefauthors}%
van Dantzig, D.%
\end{APACrefauthors}%
\unskip\
\newblock
\APACrefYearMonthDay{1956}{}{}.
\newblock
{\BBOQ}\APACrefatitle {{Economic Decision Problems for Flood Prevention}}
  {{Economic Decision Problems for Flood Prevention}}.{\BBCQ}
\newblock
\APACjournalVolNumPages{Econometrica}{24}{3}{276--287}.
\PrintBackRefs{\CurrentBib}

\bibitem [\protect \citeauthoryear {%
Van~Vuuren%
\ \protect \BOthers {.}}{%
Van~Vuuren%
\ \protect \BOthers {.}}{%
{\protect \APACyear {2011}}%
}]{%
VanVuuren2011a}
\APACinsertmetastar {%
VanVuuren2011a}%
\begin{APACrefauthors}%
Van~Vuuren, D\BPBI P.%
, Edmonds, J.%
, Kainuma, M.%
, Riahi, K.%
, Thomson, A.%
, Hibbard, K.%
\BDBL {}Lamarque, J\BHBI F.%
\end{APACrefauthors}%
\unskip\
\newblock
\APACrefYearMonthDay{2011}{}{}.
\newblock
{\BBOQ}\APACrefatitle {The representative concentration pathways: an overview}
  {The representative concentration pathways: an overview}.{\BBCQ}
\newblock
\APACjournalVolNumPages{Climatic Change}{109}{}{5--31}.
\newblock
\begin{APACrefDOI} \doi{10.1007/s10584-011-0148-z} \end{APACrefDOI}
\PrintBackRefs{\CurrentBib}

\bibitem [\protect \citeauthoryear {%
Vitousek%
\ \protect \BOthers {.}}{%
Vitousek%
\ \protect \BOthers {.}}{%
{\protect \APACyear {2017}}%
}]{%
Vitousek2017a}
\APACinsertmetastar {%
Vitousek2017a}%
\begin{APACrefauthors}%
Vitousek, S.%
, Barnard, P\BPBI L.%
, Fletcher, C\BPBI H.%
, Frazer, N.%
, Erikson, L.%
\BCBL {}\ \BBA {} Storlazzi, C\BPBI D.%
\end{APACrefauthors}%
\unskip\
\newblock
\APACrefYearMonthDay{2017}{}{}.
\newblock
{\BBOQ}\APACrefatitle {{Doubling of coastal flooding frequency within decades
  due to sea-level rise}} {{Doubling of coastal flooding frequency within
  decades due to sea-level rise}}.{\BBCQ}
\newblock
\APACjournalVolNumPages{Scientific Reports}{7}{1}{1--9}.
\newblock
\begin{APACrefDOI} \doi{10.1038/s41598-017-01362-7} \end{APACrefDOI}
\PrintBackRefs{\CurrentBib}

\bibitem [\protect \citeauthoryear {%
Vousdoukas%
\ \protect \BOthers {.}}{%
Vousdoukas%
\ \protect \BOthers {.}}{%
{\protect \APACyear {2018}}%
}]{%
Vousdoukas2018a}
\APACinsertmetastar {%
Vousdoukas2018a}%
\begin{APACrefauthors}%
Vousdoukas, M\BPBI I.%
, Mentaschi, L.%
, Voukouvalas, E.%
, Verlaan, M.%
, Jevrejeva, S.%
, Jackson, L\BPBI P.%
\BCBL {}\ \BBA {} Feyen, L.%
\end{APACrefauthors}%
\unskip\
\newblock
\APACrefYearMonthDay{2018}{}{}.
\newblock
{\BBOQ}\APACrefatitle {{Global probabilistic projections of extreme sea levels
  show intensification of coastal flood hazard}} {{Global probabilistic
  projections of extreme sea levels show intensification of coastal flood
  hazard}}.{\BBCQ}
\newblock
\APACjournalVolNumPages{Nature Communications}{9}{1}{1--12}.
\newblock
\begin{APACrefDOI} \doi{10.1038/s41467-018-04692-w} \end{APACrefDOI}
\PrintBackRefs{\CurrentBib}

\bibitem [\protect \citeauthoryear {%
Wahl%
\ \protect \BOthers {.}}{%
Wahl%
\ \protect \BOthers {.}}{%
{\protect \APACyear {2017}}%
}]{%
Wahl2017a}
\APACinsertmetastar {%
Wahl2017a}%
\begin{APACrefauthors}%
Wahl, T.%
, Haigh, I\BPBI D.%
, Nicholls, R\BPBI J.%
, Arns, A.%
, Dangendorf, S.%
, Hinkel, J.%
\BCBL {}\ \BBA {} Slangen, A\BPBI B\BPBI A.%
\end{APACrefauthors}%
\unskip\
\newblock
\APACrefYearMonthDay{2017}{}{}.
\newblock
{\BBOQ}\APACrefatitle {{Understanding extreme sea levels for broad-scale
  coastal impact and adaptation analysis}} {{Understanding extreme sea levels
  for broad-scale coastal impact and adaptation analysis}}.{\BBCQ}
\newblock
\APACjournalVolNumPages{Nature Communications}{8}{May}{16075}.
\newblock
\begin{APACrefDOI} \doi{10.1038/ncomms16075} \end{APACrefDOI}
\PrintBackRefs{\CurrentBib}

\bibitem [\protect \citeauthoryear {%
Walker%
, Haasnoot%
\BCBL {}\ \BBA {} Kwakkel%
}{%
Walker%
\ \protect \BOthers {.}}{%
{\protect \APACyear {2013}}%
}]{%
Walker2013a}
\APACinsertmetastar {%
Walker2013a}%
\begin{APACrefauthors}%
Walker, W\BPBI E.%
, Haasnoot, M.%
\BCBL {}\ \BBA {} Kwakkel, J\BPBI H.%
\end{APACrefauthors}%
\unskip\
\newblock
\APACrefYearMonthDay{2013}{}{}.
\newblock
{\BBOQ}\APACrefatitle {{Adapt or Perish: A Review of Planning Approaches for
  Adaptation under Deep Uncertainty}} {{Adapt or Perish: A Review of Planning
  Approaches for Adaptation under Deep Uncertainty}}.{\BBCQ}
\newblock
\APACjournalVolNumPages{Sustainability}{5}{3}{955--979}.
\newblock
\begin{APACrefDOI} \doi{10.3390/su5030955} \end{APACrefDOI}
\PrintBackRefs{\CurrentBib}

\bibitem [\protect \citeauthoryear {%
White%
}{%
White%
}{%
{\protect \APACyear {1945}}%
}]{%
White1945a}
\APACinsertmetastar {%
White1945a}%
\begin{APACrefauthors}%
White, G\BPBI F.%
\end{APACrefauthors}%
\unskip\
\newblock
\APACrefYear{1945}.
\newblock
\APACrefbtitle {{Human adjustment to floods: A geographical approach to the
  flood problem in the United States}} {{Human adjustment to floods: A
  geographical approach to the flood problem in the United States}}.
\newblock
\APACaddressPublisher{}{University of Chicago Press}.
\PrintBackRefs{\CurrentBib}

\bibitem [\protect \citeauthoryear {%
Wise%
\ \protect \BOthers {.}}{%
Wise%
\ \protect \BOthers {.}}{%
{\protect \APACyear {2014}}%
}]{%
Wise2014a}
\APACinsertmetastar {%
Wise2014a}%
\begin{APACrefauthors}%
Wise, R.%
, Fazey, I.%
, Smith, M\BPBI S.%
, Park, S.%
, Eakin, H.%
, Garderen, E\BPBI A\BPBI V.%
\BCBL {}\ \BBA {} Campbell, B.%
\end{APACrefauthors}%
\unskip\
\newblock
\APACrefYearMonthDay{2014}{}{}.
\newblock
{\BBOQ}\APACrefatitle {Reconceptualising adaptation to climate change as part
  of pathways of change and response} {Reconceptualising adaptation to climate
  change as part of pathways of change and response}.{\BBCQ}
\newblock
\APACjournalVolNumPages{Global Environmental Change}{28}{}{325 - 336}.
\newblock
\begin{APACrefDOI} \doi{https://doi.org/10.1016/j.gloenvcha.2013.12.002}
  \end{APACrefDOI}
\PrintBackRefs{\CurrentBib}

\bibitem [\protect \citeauthoryear {%
Wolff%
}{%
Wolff%
}{%
{\protect \APACyear {2008}}%
}]{%
Wolff2008a}
\APACinsertmetastar {%
Wolff2008a}%
\begin{APACrefauthors}%
Wolff, T\BPBI F.%
\end{APACrefauthors}%
\unskip\
\newblock
\APACrefYearMonthDay{2008}{}{}.
\newblock
{\BBOQ}\APACrefatitle {Reliability of Levee Systems} {Reliability of levee
  systems}.{\BBCQ}
\newblock
\BIn{} K\BHBI K.~Phoon\ (\BED), \APACrefbtitle {{Reliability-Based Design in
  Geotechnical Engineering: Computations and Applications}} {{Reliability-Based
  Design in Geotechnical Engineering: Computations and Applications}}\
  (\PrintOrdinal{First}\ \BEd, \BCHAP~12).
\newblock
\APACaddressPublisher{New York, NY, USA}{Taylor \& Francis}.
\PrintBackRefs{\CurrentBib}

\bibitem [\protect \citeauthoryear {%
Wong%
, Bakker%
\BCBL {}\ \BBA {} Keller%
}{%
Wong%
\ \protect \BOthers {.}}{%
{\protect \APACyear {2017}}%
}]{%
Wong2017a}
\APACinsertmetastar {%
Wong2017a}%
\begin{APACrefauthors}%
Wong, T\BPBI E.%
, Bakker, A\BPBI M\BPBI R.%
\BCBL {}\ \BBA {} Keller, K.%
\end{APACrefauthors}%
\unskip\
\newblock
\APACrefYearMonthDay{2017}{Sep}{01}.
\newblock
{\BBOQ}\APACrefatitle {Impacts of Antarctic fast dynamics on sea-level
  projections and coastal flood defense} {Impacts of antarctic fast dynamics on
  sea-level projections and coastal flood defense}.{\BBCQ}
\newblock
\APACjournalVolNumPages{Climatic Change}{144}{2}{347--364}.
\newblock
\begin{APACrefDOI} \doi{10.1007/s10584-017-2039-4} \end{APACrefDOI}
\PrintBackRefs{\CurrentBib}

\bibitem [\protect \citeauthoryear {%
Wong%
\ \BBA {} Keller%
}{%
Wong%
\ \BBA {} Keller%
}{%
{\protect \APACyear {2017}}%
}]{%
Wong2017b}
\APACinsertmetastar {%
Wong2017b}%
\begin{APACrefauthors}%
Wong, T\BPBI E.%
\BCBT {}\ \BBA {} Keller, K.%
\end{APACrefauthors}%
\unskip\
\newblock
\APACrefYearMonthDay{2017}{10}{}.
\newblock
{\BBOQ}\APACrefatitle {{Deep Uncertainty Surrounding Coastal Flood Risk
  Projections: A Case Study for New Orleans}} {{Deep Uncertainty Surrounding
  Coastal Flood Risk Projections: A Case Study for New Orleans}}.{\BBCQ}
\newblock
\APACjournalVolNumPages{Earth's Future}{5}{10}{1015--1026}.
\newblock
\begin{APACrefDOI} \doi{10.1002/2017EF000607} \end{APACrefDOI}
\PrintBackRefs{\CurrentBib}

\end{thebibliography}

\end{document}